\documentclass[useAMS,usenatbib,times,letter,amssymb]{mn2e}

\usepackage{graphics,latexsym,epsfig,graphicx,amssymb,lscape,color,soul,url}     

\renewcommand{\d}[0]{{\rm d}}
\newcommand{\e}[0]{{\rm e}}
\renewcommand{\i}[0]{{\rm i}}
\newcommand{\ave}[1]{\langle #1 \rangle}
\newcommand{\Ave}[1]{\Big\langle #1 \Big\rangle}
\newcommand{\Ref}[1]{(\ref{#1})}

\renewcommand{\vec}[1]{{\bmath{#1}}}
\newcommand{\mat}[1]{\mathbfss{#1}}

\newcommand{\msol}[0]{{\rm M}_\odot}

\newcommand{\verB}[1]{#1}
\newcommand{\lensfit}[0]{\emph{lens}fit\,}

\title[Galaxy-galaxy-galaxy lensing in CFHTLenS]{CFHTLenS:
  Higher-order galaxy-mass correlations probed by galaxy-galaxy-galaxy lensing}

\author[Simon et
al.]{P. Simon$^1$\thanks{Email:psimon@astro.uni-bonn.de},
  T. Erben$^1$, P. Schneider$^1$, C. Heymans$^2$, H.
  Hildebrandt$^{3,1}$, \newauthor H. Hoekstra$^{4,5}$, T.D. Kitching$^2$,
  Y. Mellier$^6$, L. Miller$^7$, L. Van Waerbeke$^3$, \newauthor
  C. Bonnett$^8$, J. Coupon$^{9}$,  L. Fu$^{10}$,
  M.J. Hudson$^{11,12}$, K.  Kuijken$^{4}$,
  B.T.P. Rowe$^{13,14}$, \newauthor T.  Schrabback$^{1,4,15}$, E.
  Semboloni$^{4}$,  and M. Velander$^{4,7}$\\
  $^1$Argelander Institute for Astronomy, University of Bonn, Auf
  dem H{\"u}gel 71, 53121 Bonn, Germany.\\
  $^2$Scottish Universities Physics Alliance, Institute for Astronomy,
  \\~~University of Edinburgh, Royal Observatory, Blackford
  Hill, Edinburgh, EH9 3HJ, UK. \\
  $^3$Department of Physics and Astronomy, University of British
  Columbia, 6224 Agricultural Road, Vancouver, V6T 1Z1, BC, Canada.\\
  $^4$Leiden Observatory, Leiden University, Niels Bohrweg 2, 2333
  CA Leiden, The Netherlands.\\
  $^5$Department of Physics and Astronomy, University of Victoria,
  Victoria, BC V8P 5C2, Canada.\\
  $^6$Institut d'Astrophysique de Paris, Université Pierre et Marie
  Curie - Paris 6, 98 bis Boulevard Arago, F-75014 Paris, France.\\
  $^7$Department of Physics, Oxford University, Keble Road, Oxford
  OX1 3RH, UK.\\
  $^8$Institut de Ciencies de l’Espai, CSIC/IEEC, F. de Ciencies,
  Torre C5 par-2, Barcelona 08193, Spain.\\
  $^9$Institute of Astronomy and Astrophysics, Academia Sinica,
  P.O. Box 23-141, Taipei 10617, Taiwan.\\
  $^{10}$Key Lab for Astrophysics, Shanghai Normal University, 100
  Guilin Road, 200234, Shanghai, China. \\
  $^{11}$Department of Physics and Astronomy, University of Waterloo,
  Waterloo, ON, N2L 3G1, Canada.\\
  $^{12}$Perimeter Institute for Theoretical Physics, 31 Caroline
  Street N, Waterloo, ON, N2L 1Y5, Canada.\\
  $^{13}$Department of Physics and Astronomy, University College
  London, Gower Street, London WC1E 6BT, UK.\\
  $^{14}$California Institute of Technology, 1200 E California
  Boulevard, Pasadena CA 91125, USA.\\
  $^{15}$Kavli Institute for Particle Astrophysics and Cosmology,
  Stanford University, 382 Via Pueblo Mall, Stanford, CA 94305-4060,
  USA.}

\date{Version of \today}

\begin{document}
\pagerange{\pageref{firstpage}--\pageref{lastpage}} \pubyear{2008}

\def\LaTeX{L\kern-.36em\raise.3ex\hbox{a}\kern-.15em
    T\kern-.1667em\lower.7ex\hbox{E}\kern-.125emX}

\maketitle
\label{firstpage}

\begin{abstract}

  We present the first direct measurement of the galaxy-matter
  bispectrum as a function of galaxy luminosity, stellar mass and SED
  type. Our analysis uses a galaxy-galaxy-galaxy lensing technique
  (G3L), on angular scales between 9 arcsec to 50 arcmin, to quantify
  (i) the excess surface mass density around galaxy pairs (excess mass
  hereafter) and (ii) the excess shear-shear correlations around
  single galaxies, both of which yield a measure of two types of
  galaxy-matter bispectra. We apply our method to the state-of-the-art
  Canada-France-Hawaii Telescope Lensing Survey (CFHTLenS), spanning
  154 square degrees. This survey allows us to detect a significant
  change of the bispectra with lens properties (stellar mass,
  luminosity and SED type). Measurements for lens populations with
  distinct redshift distributions become comparable by a newly devised
  normalisation technique. That will also aid future comparisons to
  other surveys or simulations. A significant dependence of the
  normalised G3L statistics on luminosity within \mbox{$-23\le
    M_r\le-18$} and stellar mass within \mbox{$5\times10^9M_\odot\le
    M_\ast\le2\times10^{11}M_\odot$} is found ($h=0.73$). Both
  bispectra exhibit a stronger signal for more luminous lenses or
  those with higher stellar mass (up to a factor 2-3). This is
  accompanied by a steeper equilateral bispectrum for more luminous or
  higher stellar mass lenses for the excess mass.  Importantly, we
  find the excess mass to be very sensitive to galaxy type as recently
  predicted with semi-analytic galaxy models: luminous
  (\mbox{$M_r<-21$}) late-type galaxies show no detectable signal,
  while all excess mass detected for luminous galaxies seems to be
  associated with early-type galaxies. We also present the first
  observational constraints on third-order stochastic galaxy biasing
  parameters.

\end{abstract}



\section{Introduction}

Over the course of the last two decades, the gravitational lensing
effect has allowed us to establish a new branch of science that
exploits the distortion of light bundles from distant galaxies
(``sources'') in order to probe the large-scale gravitational field
produced by intervening matter. Strong tidal gravitational fields
cause an obvious distortion of individual galaxy images
\citep[``strong lensing''; cf.][]{2006glsw.conf.....M}, whereas weak
deflections can only be inferred by statistical methods utilising many
galaxy images \citep[``weak lensing'';
cf.][]{2006glsw.conf..269S}. For the latter, usually shear image
distortions are harnessed, although the study of higher-order flexion
distortions may also be feasible in the near future
\citep[cf.][]{2002ApJ...564...65G,2005ApJ...619..741G,2011MNRAS.412.2665V}. Recently,
the lensing magnification effect has also moved into the focus of
research as new source of information on cosmological large-scale
structure \citep{2009A&A...507..683H}. As the gravitational field is
solely determined by the mass density of the objects under
examination, no further assumptions on their properties need to be
made when studying lensing. This makes it a unique tool for
cosmologists to examine the large-scale structure of the Universe, in
particular the relation between luminous components, such as galaxies,
and the dark component. Within the current $\Lambda\rm CDM$ standard
model of cosmology \citep{pea99,2003moco.book.....D}, the major
fraction of matter is so-called dark matter, whereas ordinary baryonic
matter is subdominant \citep{2011ApJS..192...18K}. Therefore, lensing
plays a key role in scrutinising the dominant matter component or in
testing the standard model.

Statistical methods have been developed that quantify the average mass
distribution around galaxies by cross-correlating tangential shear, as
observed from background sources, with foreground lens galaxy
positions. Galaxy-galaxy lensing (GGL), as the first highly successful
application, in effect measures the stacked projected surface mass
density profiles around galaxies
\citep{bbs96,hgd98,fms99,2001astro.ph..8013M,2003MNRAS.340..609H,2004AJ....127.2544S,2004MNRAS.355..129S,2004ApJ...606...67H,2006A&A...455..441K,2006MNRAS.368..715M,2007ApJ...669...21P,2011arXiv1107.4093V,2012arXiv1207.1120M,
  2012ApJ...744..159L}.  The GGL signal is thus a function of
lens-source separation (and their redshifts) only, i.e., a two-point
statistic that is based on a lens and the image ellipticity of a
source galaxy.  For a review see \citet{2006glsw.conf..269S} or
\citet{2008ARNPS..58...99H}. GGL studies revealed, e.g., a mass
distribution far exceeding the extension of visible light: lenses are
embedded in a dark matter halo of a size with at least $\sim
100\,h^{-1}\rm kpc$ \citep{2004ApJ...606...67H} and a mean density
profile consistent with those found in $\Lambda\rm CDM$ simulations
\citep{1996ApJ...462..563N,2005Natur.435..629S}. As extension of GGL,
the light distribution within the lens can be utilised to align the
stacked mass fields, which allows the measurement of the mean
ellipticity of the halo mass distribution in a coordinate frame
aligned with the stellar light distribution of a lens
\citep{2004ApJ...606...67H, 2006MNRAS.370.1008M, 2012arXiv1206.4304V,
  SETAL2012}. More generally, on larger spatial scales the technique
has been exploited to infer the spatial distribution of lenses with
respect to the matter distribution, the second-order galaxy biasing
\citep{2001ApJ...558L..11H, hvg02,
  2003MNRAS.346..994P,2004AJ....127.2544S,2005PhRvD..71d3511S,2007A&A...461..861S,
  2012ApJ...750...37J}. More recently, GGL in combination with galaxy
clustering in redshift surveys has been employed to test general
relativity \citep{2010Natur.464..256R}, or to successfully constrain
cosmological parameters \citep{2012arXiv1207.1120M}.

\citet[][SW05 hereafter]{2005A&A...432..783S} introduced two new GGL
correlation functions that involve three instead of two galaxies,
either two lenses and one source (``lens-lens-shear'') or two sources
and one lens (``lens-shear-shear''). Therefore, this new class of
correlators represents the third-order level of GGL or simply
``G3L''. Both correlators express new aspects of the average matter
distribution around lenses, which can be translated into third-order
galaxy biasing parameters (SW05), especially if represented in terms
of aperture statistics \citep{1998ApJ...498...43S}. This paper chooses
the aperture statistics to represent the G3L signal. Thereby we
essentially express the angular bispectrum of the (projected)
matter-galaxy three-point correlation. A rigorous mathematical
description of the aperture statistics is given in the following
section.

A more intuitive interpretation \citep{RNSIMON2011} of G3L is given by
the definition of the real-space correlation functions: the
lens-lens-shear correlation function measures the average excess shear
\citep[or excess mass,][]{2008A&A...479..655S} around clustered lens
pairs, i.e., in excess of the average shear pattern around pairs
formed from a hypothetical set of lenses that is uniformly randomly
distributed on the sky (unclustered) but exhibit the same GGL signal
as the lenses in the data. It is a probe for the joint matter
environment of galaxy pairs, not single galaxies. This correlator
promises to put additional constraints on galaxy models
\citep{2012arXiv1204.2232S} as it appears to be very sensitive to
galaxy types. On the other hand, the lens-shear-shear correlation
function measures the ``excess shear-shear correlation'': it
quantifies the shear-shear correlation function in the neighbourhood
of a lens in excess of shear-shear correlations as expected from
randomly scattered lenses. Thereby it picks up the (projected) matter
density two-point correlation function of matter physically associated
with lenses.  In a way this makes the lens-shear-shear correlator
similar to the traditional GGL, but now also probing the variance in
the surface matter density around lenses instead of merely the
average. The angular matter-galaxy bispectra are Fourier-transforms of
these correlators.

\citet{2008A&A...479..655S} have demonstrated with the Red-Sequence
Cluster Survey \citep[RCS1;][]{2005ApJS..157....1G} that both G3L
correlation functions can readily be measured with existing lensing
surveys. The RCS1 study aimed to obtain a high signal-to-noise ratio of
the lensing signal, for which all available lenses were combined into
one lens catalogue. Therefore, apart from this feasibility study in
existing data, little more is known on the dependence of the G3L
signal on galaxy properties. This paper is a first step to fill this
gap by systematically measuring G3L for a series of lens samples with
varying properties. The amount of data available through the CFHTLenS
analysis allows this to be done for the first time. An accompanying
paper by \citet{VETAL2012} explores the GGL signal of CFHTLenS in the
light of the halo model \citep{2002PhR...372....1C}.
 
The paper is laid out as follows. Sect. \ref{sect:theory} summarises
the aperture statistics that is devised to express the G3L signal,
gives their practical estimators and lists possible sources of
systematics.  In Sect. \ref{sect:data}, we outline the selection
criteria of our source and lens samples. Lenses are selected by
luminosity, stellar mass, redshift and two galaxy spectral types, all
to be analysed separately. Sect. \ref{sect:results} presents our G3L
results. For a large range of angular scales covered in this study,
the G3L signal is characterised by a simple power law whose parameters
are given. Sect. \ref{sect:interpretation} offers a physical
interpretation of the G3L statistics in terms of 3D galaxy-matter
bispectra. In this context, we also introduce a normalisation scheme
to remove, to lowest order, the impact of the exact shape of the lens
redshift distribution and the source redshift distribution from the
signal. Finally, the Sects. \ref{sect:discussion} and
\ref{sect:conclusions} present our discussion and conclusions.

Throughout the paper we adopt a WMAP7 \citep{2011ApJS..192...18K}
fiducial cosmology for the matter density $\Omega_{\rm m}=0.27$, the
cosmological constant $\Omega_\Lambda=1-\Omega_{\rm m}=0.73$ (both in
units of the critical density) and $H_0=100h\,\rm
km\,s^{-1}Mpc^{-1}$. These parameters are consistent with
gravitational lensing constraints obtained from CFHTLenS itself
\citep{KETAL2012,BETAL2012,HETAL2012}. If not stated otherwise, we
explicitly use $h=0.73$, in particular for the absolute galaxy
magnitudes and their stellar masses.


\section{Formalism}
\label{sect:theory}

This section summarises the theory and notation of G3L as detailed in
SW05, and lists possible G3L specific systematics.

\subsection{Galaxy-galaxy lensing preliminaries}

The weak gravitational lensing effect \citep[see][ and references
therein]{2006glsw.conf..269S} probes the three-dimensional relative
matter density fluctuations $\delta_{\rm
  m}(\vec{R}_\perp,\chi)=\Delta\rho_{\rm m}/\bar{\rho}_{\rm m}$ in
projection along the line-of-sight in terms of the lensing convergence
\begin{equation}
  \label{eq:kappa}
  \kappa(\vec{\theta})=
  \frac{3\Omega_{\rm m}}{2D_{\rm H}^2}
  \int_0^{\chi_{\rm h}}\d\chi\,\frac{g(\chi)f_{\rm K}(\chi)}{a(\chi)}\,
  \delta_{\rm m}\!\big(f_{\rm K}(\chi)\vec{\theta},\chi\big)\;.
\end{equation}
Here $\vec{R}_\perp=f_{\rm K}(\chi)\vec{\theta}$ is a 2D vector
perpendicular to a reference line-of-sight and $\vec{\theta}$ the
angular position on the sky. The comoving angular diameter distance
$f_{\rm K}(\chi)$ is written as a function of comoving radial distance
$\chi$. By $D_{\rm H}:=c/H_0$ we define the Hubble length, and
$a(\chi)$ is the cosmic scale factor at a distance $\chi$; we set
$a(0)=1$ by definition; $c$ is the vacuum speed of light. By
$\chi_{\rm h}$ we denote the comoving Hubble radius of today as the
theoretical maximum distance at which we can observe objects. The
lensing efficiency averaged over the probability density distribution
function (p.d.f.)  $p_{\rm b}(\chi)\d\chi$ of background galaxies
(``sources'') is expressed by
\begin{equation}
  g(\chi)=
  \int_\chi^{\chi_{\rm h}}\d\chi^\prime\,
  p_{\rm b}(\chi^\prime)\frac{f_{\rm K}(\chi^\prime-\chi)}
  {f_{\rm K}(\chi^\prime)}\;.
\end{equation}

Although the convergence in principle is observable through
magnification of galaxy images, past weak lensing analyses and this
paper focus on the related gravitational shear \citep{kas93}
\begin{equation}
  \label{eq:kappagamma}
  \gamma_{\rm c}(\vec{\theta})=
  \frac{1}{\pi}\int\d^2\vartheta\,{\cal
    D}(\vec{\vartheta}-\vec{\theta})\kappa(\vec{\vartheta})~;~
  {\cal D}(\vec{\theta}):=-\frac{1}{(\vec{\theta}^\ast)^2}\;.
\end{equation}
\verB{By $\vec{\theta}^\ast$ we denote the complex conjugate of $\vec{\theta}$.}
For this purpose, the complex ellipticity of the galaxy image
\begin{equation}
  \epsilon(\vec{\theta})\approx\gamma_{\rm
    c}(\vec{\theta})+\epsilon_{\rm s}~;~\ave{\epsilon_{\rm s}}=0
\end{equation}
serves as a noisy estimator of $\gamma_{\rm c}$; the noise term
originates from the unknown intrinsic shape $\epsilon_{\rm s}$. In
addition, due to the finite number of sources, one also experiences
sampling noise of the shear field. Note that we adopt the commonly
used complex notation of 2D vectors and spinors (in the case of shears
and ellipticities), where real and imaginary parts are the components
along two Cartesian axes in a tangential plane on the sky.

Galaxy-galaxy lensing techniques correlate the total matter
distribution $\kappa(\vec{\theta})$ with the relative number density
distribution $\kappa_{\rm g}(\vec{\theta})$ of lens galaxies
(``lenses'') on the sky by means of cross-correlating the lensing
signal with positions of foreground galaxies,
\begin{equation}
  \kappa_{\rm g}(\vec{\theta})=
  \frac{n_{\rm g}(\vec{\theta})-\overline{n}_{\rm
      g}}{\overline{n}_{\rm g}}=
  \int_0^{\chi_{\rm h}}\d\chi\,
  p_{\rm f}(\chi)\,
  \delta_{\rm g}\!\big(f_{\rm K}(\chi)\vec{\theta},\chi\big)\;,
\end{equation}
where $p_{\rm f}(\chi)\d\chi$ is the p.d.f. of the lens \verB{(foreground)}
comoving distances along the line-of-sight; $n_{\rm g}(\vec{\theta})$
is the projected number density of lenses and $\bar{n}_{\rm g}$ its
statistical mean. For the scope of this paper, $p_{\rm f}(\chi)$ is
estimated from a redshift p.d.f. $p_z(z)\d z=p_{\rm f}(\chi)\d\chi$ of
a selected lens sample.

\subsection{G3L aperture statistics}

\begin{figure}
  \begin{center}
    \epsfig{file=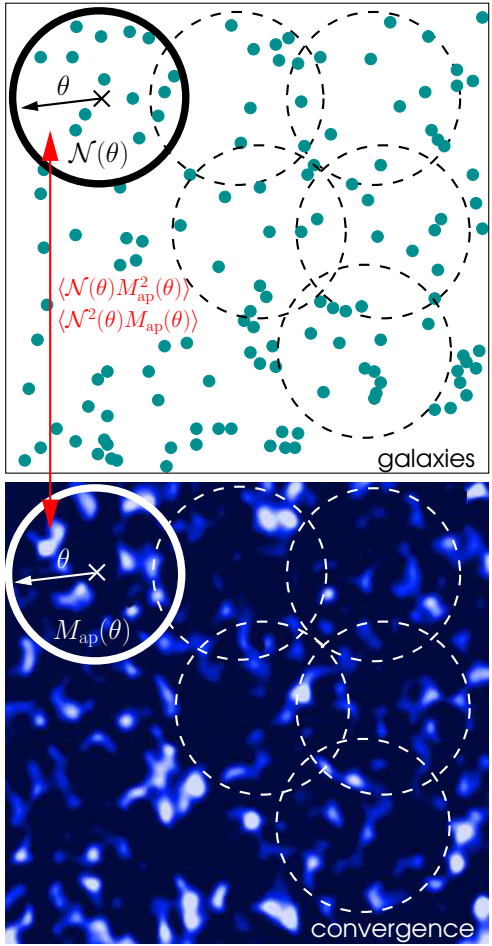,width=65mm,angle=0}
  \end{center}
  \caption{\label{fig:apstatsketch}
    \verB{Illustration of the aperture statistics. Fluctuations ${\cal
        N}(\theta)$ in the projected galaxy number density (\emph{top
      panel}), smoothed to the characteristic filter scale $\theta$,
      are statistically compared to the filtered projected matter
      fluctuations $M_{\rm ap}(\theta)$ (lensing convergence;
      \emph{bottom panel}). We take ${\cal N}^2(\theta)M_{\rm ap}(\theta)$ or
      ${\cal N}(\theta)M^2_{\rm ap}(\theta)$, and average these for
      different aperture centres (dashed circles) to estimate
      third-order moments of the joint probability distribution of
      ${\cal N}(\theta)$ and $M_{\rm ap}(\theta)$.}}
\end{figure}

For practical purposes, the aperture statistics are a convenient
measure for a lensing analysis \citep{svj98, 1998ApJ...498...43S,
  1998A&A...334....1V, cnp02}. They quantify moments of fluctuations
in $\kappa(\vec{\theta})$ and $\kappa_{\rm g}(\vec{\theta})$ within
apertures of a variable angular scale $\theta$. The moments are
determined from the smoothed fields $\kappa(\vec{\theta})$ and
$\kappa_{\rm g}(\vec{\theta})$,
\begin{eqnarray}
  M_{\rm ap}(\theta)&=&
  \int\frac{\d^2\vartheta}{\theta^2}\,u\left(|\vec{\vartheta}|\theta^{-1}\right)\,
  \kappa\left(\vec{\vartheta}\right)\;,\\
  {\cal N}(\theta)&=&
  \int\frac{\d^2\vartheta}{\theta^2}\,
  u\left(|\vec{\vartheta}|\theta^{-1}\right)\,
  \kappa_{\rm g}(\vec{\vartheta})\;,
\end{eqnarray}
where $u(\vartheta/\theta)\theta^{-2}$ is the
smoothing kernel. For mathematical convenience, we placed the aperture
centre at $\vec{\theta}_{\rm c}=0$ in the previous
definition. Third-order moments are defined by considering the
ensemble average of
\begin{eqnarray}
\label{eq:n2map}
\ave{{\cal N}^2M_{\rm ap}}
(\theta_1;\theta_2;\theta_3):=
\Ave{{\cal N}(\theta_1){\cal N}(\theta_2)M_{\rm ap}(\theta_3)}\;,\\
  \label{eq:nmap2}
\ave{{\cal N}M^2_{\rm ap}}
(\theta_1;\theta_2;\theta_3):=
\Ave{{\cal N}(\theta_1)M_{\rm ap}(\theta_2)M_{\rm ap}(\theta_3)}\;,
\end{eqnarray}
over all random realisations of the fields $\kappa(\vec{\theta})$ and
$\kappa_{\rm g}(\vec{\theta})$.
\verB{Due to the assumed statistical homogeneity of the fields, the
  averages do not depend on the aperture centre position.}
Therefore, in practice, where only one realisation or survey is
available, these quantities are estimated by averaging the products
${\cal N}(\theta_1){\cal N}(\theta_2)M_{\rm ap}(\theta_3)$ and ${\cal
  N}(\theta_1)M_{\rm ap}(\theta_2)M_{\rm ap}(\theta_3)$ for different
aperture centres covering the survey area. \verB{See
  Fig. \ref{fig:apstatsketch} for an illustration.}

For a compensated filter $u$, i.e., $\int_0^\infty\d\theta\,\theta
u(\theta)=0$, the aperture mass can in principle be obtained directly
from the observable shear through \citep{svj98}
\begin{equation}
  M_{\rm ap}(\theta)=
  \int_0^\infty\int_0^{2\pi}\frac{\d\varphi\,\d\vartheta\,\vartheta}{\theta^2}\,
  q\left(\vartheta\theta^{-1}\right)\,
  \Re{\left(\gamma(\vec{\vartheta};\varphi)\right)}\;,
\end{equation}
where $\gamma(\vec{\vartheta};\varphi):=-\e^{-2\i\varphi}\gamma_{\rm
  c}(\vec{\vartheta})$ denotes the Cartesian shear $\gamma_{\rm c}$ at
angular position $\vec{\vartheta}$ rotated by the polar angle
$\varphi$. The real part of $\gamma(\vec{\vartheta};\varphi)$ is the
tangential shear, the imaginary part the cross shear.  The relation
between the filters $u(x)$ and $q(x)$ is given by
\begin{equation}
  q(x)=\left(\frac{2}{x^2}\int_0^x\d s\,s\,u(s)\right)-u(x)\;.
\end{equation}

This paper uses the exponential aperture filter from
\citet{1998A&A...334....1V}, exponential filter hereafter,
\begin{equation}
  u(x)=\frac{1}{2\pi}\left(1-\frac{x^2}{2}\right)\e^{-x^2/2}\;,
\end{equation}
which effectively has a finite support because of the Gaussian factor
that suppresses the filter strongly to zero for
$\vartheta\gtrsim3\theta$ (SW05). The Fourier transform of the
aperture filter is
\begin{equation}
  \tilde{u}(\ell)=
  \int\d^2\theta\,u(\theta)\e^{+\i\vec{\ell}\cdot\vec{\theta}}
  =\frac{\ell^2}{2}\e^{-\ell^2/2}\;.
\end{equation}
We generally denote a Fourier transform of $f(\vec{\theta})$ by
$\tilde{f}(\vec{\ell})$ in the following.  The exponential filter
$\tilde{u}(\ell)$ peaks in Fourier space at an angular wave number of
$\ell=\sqrt{2}$, which determines a characteristic angular scale
selected by an aperture radius of $\theta$.

\begin{figure}
  \begin{center}
    \epsfig{file=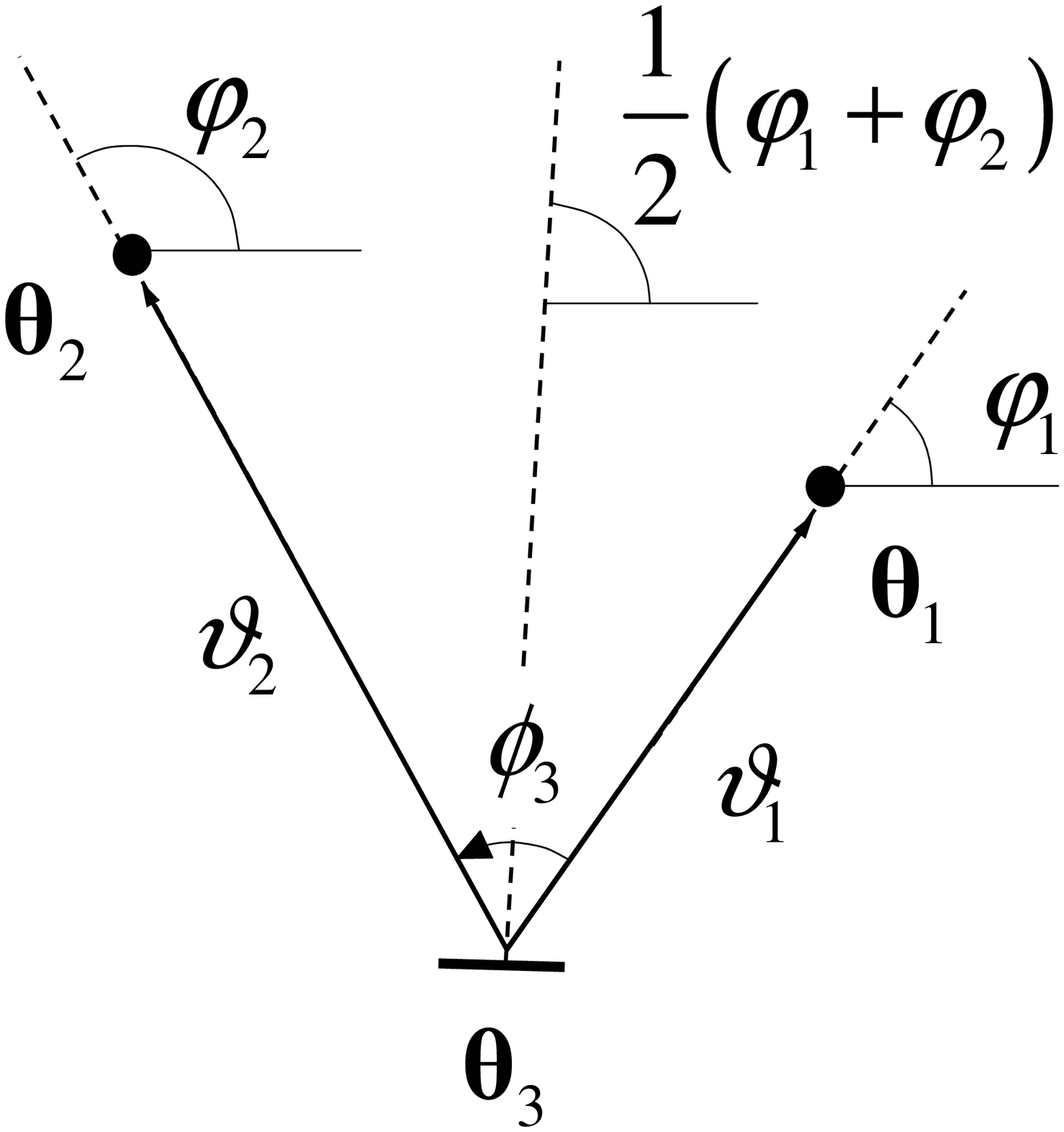,width=65mm,angle=0}

    \vspace{0.35cm}
    \epsfig{file=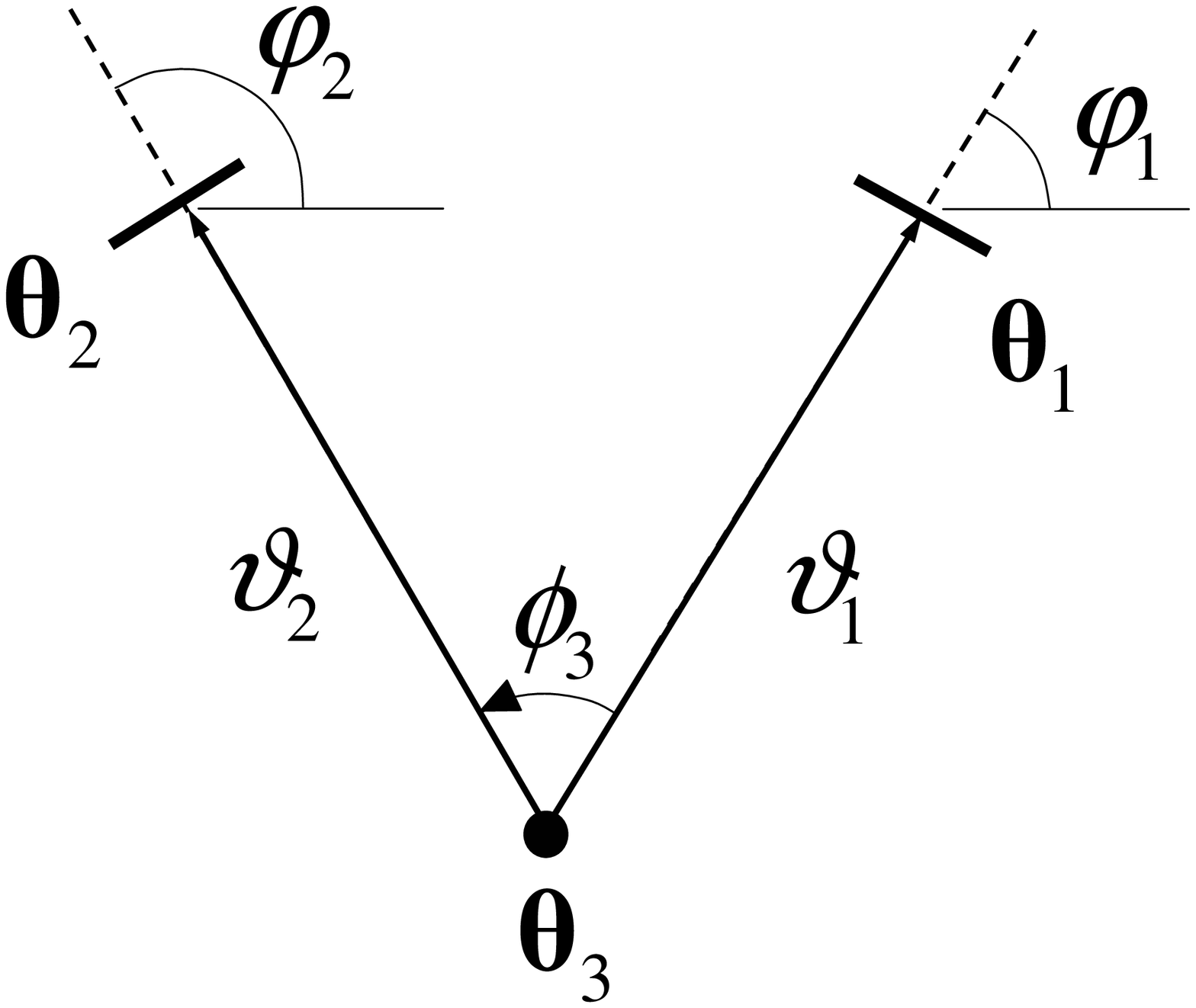,width=65mm,angle=0}
  \end{center}
  \caption{\label{fig:corrsketch}
    \verB{Illustration of the parametrisation of the lens-lens-shear
      three-point correlator $\widetilde{\cal
        G}(\vartheta_1,\vartheta_2,\phi_3)$ (\emph{top panel}), and
      the lens-shear-shear correlation function
      $\widetilde{G}_\pm(\vartheta_1,\vartheta_2,\phi_3)$
      (\emph{bottom panel}). These statistics are employed to estimate
      the aperture statistics in Fig. \ref{fig:apstatsketch}. The
      figure is copied from SW05.}}
\end{figure}

\subsection{Aperture statistics estimators}
\label{sect:estimators}

\verB{To obtain the third-order moments of the galaxy-matter aperture
  statistics, we utilise the lens-lens-shear correlation function
  $\cal G$ in the case of $\ave{{\cal N}^2M_{\rm ap}}$ and the
  lens-shear-shear correlation function $G_\pm$ for $\ave{{\cal
      N}M^2_{\rm ap}}$. This section provides only a brief description
  of this approach.  For a more details, its computationally
  optimised implementation as well as verification, we refer the
  reader to Sect. 3 of \citet{2008A&A...479..655S}.}

In practice, the aperture moments $\ave{{\cal N}^2M_{\rm ap}}$ or
$\ave{{\cal N}M^2_{\rm ap}}$ are not computed from the aperture mass
$M_{\rm ap}$ or aperture number counts $\cal N$ directly. The
information contained in the aperture statistics is also contained
inside two classes of three-point correlation functions (SW05), which
are relatively straightforward to estimate. Once the correlation
functions have been determined, they can be transformed to the
corresponding aperture statistics by an integral transformation. The
estimation process thus proceeds in two basic steps.  In the first
step, for $\ave{{\cal N}^2M_{\rm ap}}$ one estimates the source
tangential ellipticity relative to the midpoint connecting two lenses,
\begin{equation}
  \label{eq:G}
  \widetilde{{\cal G}}(\vartheta_1,\vartheta_2,\phi_3)=
  \frac{1}{\overline{n}_{\rm g}^2}
  \Ave{n_{\rm g}(\vec{\theta}_1)n_{\rm g}(\vec{\theta}_2)
    \gamma\!\left(\vec{\theta}_3;\frac{\varphi_1+\varphi_2}{2}\right)}\;.
\end{equation}
The meaning of the notation is illustrated in the left panel of
Fig. \ref{fig:corrsketch}.  For $\ave{{\cal N}M^2_{\rm ap}}$ one
estimates the correlation of the ellipticities of two sources relative
to the line connecting the sources as a function of separation from
one lens (right panel),
\begin{equation}
  \label{eq:Gpm}
  \widetilde{G}_\pm(\vartheta_1,\vartheta_2,\phi_3)=
  \frac{1}{\overline{n}_{\rm g}}
  \Ave{\gamma(\vec{\theta}_1;\varphi_1)\gamma^\pm(\vec{\theta}_2;\varphi_2)
    n_{\rm g}(\vec{\theta}_3)}\;.
\end{equation}
Here and in the following equations a superscript ``$\pm$'' as in
$\gamma^\pm$ means $\gamma$ for $\gamma^-$ (in case of
$\widetilde{G}_-$) and the complex conjugate $\gamma^\ast$ for
$\gamma^+$ (in case of $\widetilde{G}_+$).

Both correlation functions are estimated inside \emph{bins of similar
  triangles}, i.e., lens-source triples within a configuration of
comparable side lengths $\vartheta_{1,2}$ and opening angles $\phi_3$,
by summing over all relevant galaxy triplets. Any triple of three
galaxy positions $\vec{\theta}_i,\vec{\theta}_j,\vec{\theta}_k$ that
meets the criteria of a relevant triangle is flagged by
$\Delta_{ijk}^{\vartheta_1\vartheta_2\phi_3}=1$ and
$\Delta_{ijk}^{\vartheta_1\vartheta_2\phi_3}=0$ otherwise. For this
study, we utilise 100 logarithmic bins for both $\vartheta_1$ and
$\vartheta_2$, and 100 linear bins for the opening angle $\phi_3$. For
estimating $\widetilde{\cal G}$ we utilise
\begin{eqnarray}
\label{eq:gestimator}
\lefteqn{\widetilde{\cal G}^{\rm est}(\vartheta_1,\vartheta_2,\phi_3)=}\\
&&\nonumber
  \!\!\!\!\frac{
    -\sum\limits_{i=1}^{N_{\rm d}}
    \sum\limits_{j=1}^{N_{\rm d}}
    \sum\limits_{k=1}^{N_{\rm s}}
    w_k\,
    \epsilon_k\,\e^{-\i(\varphi_i+\varphi_j)}
    \big[1+\omega(|\vec{\theta}_i-\vec{\theta}_j|)\big]
    \Delta_{ijk}^{\vartheta_1\vartheta_2\phi_3}}
  {\sum\limits_{i=1}^{N_{\rm d}}
    \sum\limits_{j=1}^{N_{\rm d}}
    \sum\limits_{k=1}^{N_{\rm s}}w_k\Delta_{ijk}^{\vartheta_1\vartheta_2\phi_3}}
\;,
\end{eqnarray}
and for $\widetilde{G}_\pm$ the estimator
\begin{eqnarray}
  \label{eq:gpmestimator}
  \lefteqn{\widetilde{G}^{\rm
      est}_\pm(\vartheta_1,\vartheta_2,\phi_3)=}\\
  \nonumber&&
  \frac{
    \sum\limits_{i=1}^{N_{\rm d}}
    \sum\limits_{j=1}^{N_{\rm s}}
    \sum\limits_{k=1}^{N_{\rm s}}
    w_j\,w_k\,\epsilon_j\epsilon_k^\pm\e^{-2\i\varphi_j}\e^{\pm2\i\varphi_k}
    \Delta_{ijk}^{\vartheta_1\vartheta_2\phi_3}}{
    \sum\limits_{i=1}^{N_{\rm d}}
    \sum\limits_{j=1}^{N_{\rm s}}
    \sum\limits_{k=1}^{N_{\rm s}}w_j\,w_k\Delta_{ijk}^{\vartheta_1\vartheta_2\phi_3}}\;,
\end{eqnarray}
where $N_{\rm d}$ and $N_{\rm s}$ are the number of lenses and sources,
$w_i$ are statistical weights of sources, $\varphi_i$ are polar angles
of the position vectors of galaxies with respect to the coordinate
origin, $\epsilon_i$ are the source ellipticities, and
\begin{equation}
  \omega(|\vec{\Delta\theta}|)=
  \Ave{\kappa_{\rm g}(\vec{\theta})
    \kappa_{\rm g}(\vec{\theta}+\vec{\Delta\theta})}
\end{equation}
is the angular two-point clustering of the lenses
\citep[e.g.][]{peebles80}. In this paper, the angular clustering of
lenses is estimated by means of the estimator in \citet{las93} prior
to the estimation of $\widetilde{\cal G}$ and then interpolated. 
\verB{Sources are weighed by the inverse-variance uncertainty in the
 \lensfit ellipticity measurement \citep{METAL2012}.}

In a second step, we transform the estimates of $\widetilde{\cal G}$
and $\widetilde{G}_\pm$ to the aperture statistics by devising the
transformation integrals Eqs. (63), (57), and (59) in SW05.
\verB{There is no need to remove the unconnected terms in the
  correlation functions. As shown in SW05 (Sect. 7.2. therein), the
  transformation from ${\cal G}$ to $\ave{{\cal N}^2M_{\rm ap}}$
  yields the same result when $\widetilde{\cal G}$ is taken instead of $\cal
  G$. Therefore, the integral transformation automatically ignores
  unconnected second-order terms in the triple correlator,
  resulting in an aperture statistics that are only determined by pure
  (connected) third-order correlation terms. The same holds true for
  $\widetilde{G}_\pm$ and $\ave{{\cal N}M^2_{\rm ap}}$.}

\subsection{Relation to 3D galaxy-matter bispectra}
\label{sect:3dbispectra}

The aperture statistics are directly connected to the angular
cross-bispectra of the projected matter and lens distribution:
\begin{eqnarray}
  \label{eq:n2mapbi}
  \lefteqn{\ave{{\cal N}^2M_{\rm ap}}(\theta_1;\theta_2;\theta_3)=}\\
  \nonumber&&
  \!\!\!\!\!\!\!\!\int\frac{\d^2\ell_1}{(2\pi)^2}
  \int\frac{\d^2\ell_2}{(2\pi)^2}
  \tilde{u}(\ell_1\theta_1)
  \tilde{u}(\ell_2\theta_2)
  \tilde{u}(|\vec{\ell_1}+\vec{\ell}_2|\theta_3)
  b_{{\rm gg}\kappa}(\vec{\ell_1},\vec{\ell_2})\;,
  \\
  \label{eq:nmap2bi}
  \lefteqn{\ave{{\cal N}M^2_{\rm ap}}(\theta_1;\theta_2;\theta_3)=}\\
  \nonumber&&
  \!\!\!\!\!\!\!\!\int\frac{\d^2\ell_1}{(2\pi)^2}
  \int\frac{\d^2\ell_2}{(2\pi)^2}
  \tilde{u}(\ell_1\theta_1)
  \tilde{u}(\ell_2\theta_2)
  \tilde{u}(|\vec{\ell_1}+\vec{\ell}_2|\theta_3)
  b_{\kappa\kappa{\rm g}}(\vec{\ell_1},\vec{\ell_2})\;,
\end{eqnarray}
where the angular galaxy-galaxy-matter bispectrum is
\begin{equation}
  \ave{
    \tilde{\kappa}_{\rm g}(\vec{\ell}_1)
    \tilde{\kappa}_{\rm g}(\vec{\ell}_2)
    \tilde{\kappa}(\vec{\ell}_3)}=
  (2\pi)^2\delta^{(2)}_{\rm D}(\vec{\ell}_1+\vec{\ell}_2+\vec{\ell}_3)
  b_{{\rm gg}\kappa}(\vec{\ell_1},\vec{\ell_2})
\end{equation}
and the angular matter-matter-galaxy bispectrum is
\begin{equation}
  \ave{
    \tilde{\kappa}(\vec{\ell}_1)
    \tilde{\kappa}(\vec{\ell}_2)
    \tilde{\kappa}_{\rm g}(\vec{\ell}_3)}=
  (2\pi)^2\delta^{(2)}_{\rm D}(\vec{\ell}_1+\vec{\ell}_2+\vec{\ell}_3)
  b_{\kappa\kappa{\rm g}}(\vec{\ell_1},\vec{\ell_2})\;.
\end{equation}
For statistically homogeneous random fields, the triple correlators on
the left-hand side of the previous two equations can only be
non-vanishing when $\vec{\ell}_1+\vec{\ell}_2+\vec{\ell}_3=\vec{0}$,
which is reflected by the 2D Dirac delta functions $\delta^{(2)}_{\rm
  D}(\vec{x})$ on the right-hand sides. Owing to homogeneity, the
bispectra thus depend only on two independent arguments $\vec{\ell}$,
for which we arbitrarily choose $\vec{\ell}_1$ and
$\vec{\ell}_2$. This automatically implies
$\vec{\ell}_3=-(\vec{\ell}_1+\vec{\ell}_2)$. In addition the
statistical isotropy implies that the bispectra are solely functions
of the moduli of $\vec{\ell}_{1,2}$ and the angle enclosed by both
wave vectors.

As can be seen from Eqs. \Ref{eq:n2mapbi}, \Ref{eq:nmap2bi}, the
aperture statistics are a locally filtered version of the bispectrum
because the exponential $u$-filter is relatively localised in
$\ell$-space with a filter maximum at \mbox{$\ell_{\rm
    max}=\sqrt{2}/\theta$}. By means of the filtering, the
aperture statistics basically becomes a band power bispectrum version
of $b_{{\rm gg}\kappa}$ or $b_{\kappa\kappa\rm g}$. Hence the aperture
statistics Eqs. \Ref{eq:n2map}, \Ref{eq:nmap2} measure two different
angular galaxy-matter band power cross-bispectra.

By virtue of the Limber approximation
\citep{1992ApJ...388..272K,bas01} the angular bispectra and thereby
the aperture statistics Eqs. \Ref{eq:n2mapbi}, \Ref{eq:nmap2bi} can
directly be related to the 3D cross-bispectrum of the matter and lens
distribution (SW05) as primary physical quantities that are assessed
by the statistics:
\begin{eqnarray}
  \label{eq:biggk}
 \lefteqn{b_{{\rm gg}\kappa}(\vec{\ell}_1,\vec{\ell}_2)=}\\
 \nonumber&&
 \frac{3\Omega_{\rm m}}{2D_{\rm H}^2}
 \int_0^{\chi_{\rm h}}\d\chi\,
 \frac{g(\chi)p_{\rm f}^2(\chi)}{f^3_{\rm K}(\chi)a(\chi)}
 B_{\rm ggm}\big(\frac{\vec{\ell}_1}{f_{\rm K}(\chi)},
 \frac{\vec{\ell}_2}{f_{\rm K}(\chi)},\chi\big)\;,\\
 \label{eq:bikkg}
 \lefteqn{b_{\kappa\kappa\rm g}(\vec{\ell}_1,\vec{\ell}_2)=}\\
 \nonumber&&
 \frac{9\Omega^2_{\rm m}}{4D_{\rm H}^4}
 \int_0^{\chi_{\rm h}}\d\chi\,
 \frac{g^2(\chi)p_{\rm f}(\chi)}{f^2_{\rm K}(\chi)a^2(\chi)}
 B_{\rm mmg}\big(\frac{\vec{\ell}_1}{f_{\rm K}(\chi)},
 \frac{\vec{\ell}_2}{f_{\rm K}(\chi)},\chi\big)\;,
\end{eqnarray}
where the 3D bispectra are determined by the Fourier transforms of the
matter density contrast, $\tilde{\delta}_{\rm m}(\vec{k},\chi)$, and
galaxy number density contrast, $\tilde{\delta}_{\rm
  g}(\vec{k},\chi)$, at radial distance $\chi$, namely
\begin{eqnarray}
  \lefteqn{\ave{
      \tilde{\delta}_{\rm g}(\vec{k}_1,\chi)
      \tilde{\delta}_{\rm g}(\vec{k}_2,\chi)
      \tilde{\delta}_{\rm m}(\vec{k}_3,\chi)}=}\\
  \nonumber&&
  (2\pi)^3\delta^{(3)}_{\rm D}(\vec{k}_1+\vec{k}_2+\vec{k}_3)
  B_{\rm ggm}(\vec{k}_1,\vec{k}_2,\chi)\;,\\
  \lefteqn{\ave{
      \tilde{\delta}_{\rm m}(\vec{k}_1,\chi)
      \tilde{\delta}_{\rm m}(\vec{k}_2,\chi)
      \tilde{\delta}_{\rm g}(\vec{k}_3,\chi)}=}\\
  \nonumber&&
  (2\pi)^3\delta^{(3)}_{\rm D}(\vec{k}_1+\vec{k}_2+\vec{k}_3)
  B_{\rm mmg}(\vec{k}_1,\vec{k}_2,\chi)\;.
\end{eqnarray}
The vector $\vec{k}$ is the comoving wave number of modes entering the
triple correlator. As before with the angular bispectra, the spatial
bispectra are also isotropic, i.e., they are only functions of
$|\vec{k}_1|$, $|\vec{k}_2|$ and the angle spanned by $\vec{k}_1$ and
$\vec{k}_2$.

To refine the previous RCS1 measurement in \citet{2008A&A...479..655S}
for different galaxy populations, we focus on equally-sized apertures
with $\theta_1=\theta_2=\theta_3$ only. This leads us to the short
hand notations $\ave{{\cal N}^2M_{\rm ap}}(\theta):=\ave{{\cal
    N}^2M_{\rm ap}}(\theta;\theta;\theta)$, likewise for $\ave{{\cal
    N}M_{\rm ap}^2}$. Due to the action of the $u$-filter in the
Eqs. \Ref{eq:n2mapbi} and \Ref{eq:nmap2bi} this picks up mainly
bispectrum contributions from equilateral triangles
$|\vec{\ell}_1|=|\vec{\ell}_2|=|\vec{\ell}_1+\vec{\ell}_2|$, albeit
also mixing in signal from other triangles because of the finite width
of the $u$-filter in $\ell$-space.

\subsection{Systematics indicators}

The gravitational shear of distant galaxy images is produced by small
fluctuations $\delta\phi$ in the intervening gravitational
potential. To lowest order in $\delta\phi/c^2$ this is expected to
only produce curl-free shear fields (B-modes vanish). Current surveys
do not have the power to measure higher-order effects, such that we
expect these to be undetectable in our data. Shear-related correlation
functions, or aperture moments involving the aperture mass, hence
vanish after rotation of all sources by $45^\circ$, i.e., after
$\gamma_{\rm c}(\vec{\theta})\mapsto-\i\,\gamma_{\rm
  c}\vec(\vec{\theta})$. Translated into data analysis, a $45^\circ$
rotation of the source ellipticities should result in a measurement
that is statistically consistent with the experimental noise
\citep[e.g.][]{hss07}. We use this as a necessary (but not sufficient)
indicator for the absence of systematics in the data.

The estimator $\widetilde{G}^{\rm est}_\pm$ in
Eq. \Ref{eq:gpmestimator} incorporates two sources with two uniquely
different possibilities to probe systematics: rotating the
ellipticities $\epsilon_j$ and $\epsilon_k$ of both sources results in
the so-called B-mode channel of $\ave{{\cal N}M^2_{\rm ap}}(\theta)$,
denoted here by $\ave{{\cal N}M_\perp^2}(\theta)$, and the P-mode
channel, $\ave{{\cal N}M_\perp M_{\rm ap}}(\theta)$, if only either
$\epsilon_j$ or $\epsilon_k$ are rotated. As pointed out by
\citet{schneider03}, a P-mode is a signature of a parity-invariance
violation in the shear data, which in a parity-invariant universe can
only be generated by systematics in the PSF correction pipeline, or in
the algorithm for the statistical analysis of the data. Non-vanishing
B-modes, on the other hand, can have a physical cause. For example,
they can be associated with the intrinsic clustering of sources
\citep{SvWM02}, intrinsic alignment correlations of physically close
sources or intrinsic shape-shear correlations \citep[][and references
therein]{Heymans06}. Especially the latter two are a concern for this
analysis, as these effects are known to affect the E-mode channel of
the aperture statistics, which is the prime focus of this
work. However, currently it is unclear by how much this really affects
G3L. We discuss in the following Sect. \ref{sect:sysdiscuss} that the
influence of these systematics can be suppressed by separating lenses
and sources in redshift, which is carried out in our analysis.

Since the estimator $\widetilde{\cal G}^{\rm est}$ in
Eq. \Ref{eq:gestimator} involves one source, there is only a single
systematics indicator of $\ave{{\cal N}^2M_{\rm ap}}(\theta)$, which
is a parity violation indicator, a P-mode channel. In the following we
will denote these statistics as $\ave{{\cal N}^2M_\perp}(\theta)$. As
shown in SW05, the B- and P-modes of the statistics can be computed
from $\widetilde{\cal G}$ and $\widetilde{G}_\pm$ directly by
utilising an alternative integral kernel in the transformation from
correlation functions to aperture statistics; see their Sect. 7.1 and
7.2.

\subsection{Reduction of II- and GI-contributions}
\label{sect:sysdiscuss}

One possible source of systematics are correlations with intrinsic
ellipticities $\epsilon_{\rm s}$ of sources. A correlation between
$\epsilon_{\rm s}$ of different sources (II-correlations) or between
$\epsilon_{\rm s}$ and a fluctuation in the mass density field
generating shear (GI-correlations) is known to contribute to the shear
correlation functions
\citep[e.g][]{2004PhRvD..70f3526H,Heymans06,2011A&A...527A..26J}. For
a discussion of intrinsic alignments in CFHTLenS see also
\citet{HETAL2012}.  We argue here that selecting lenses and sources
from well separated distances ideally removes contaminations by II- or
GI-correlations in the G3L statistics.

Consider the galaxy number density contrasts $\kappa_{{\rm g},1}$ and
$\kappa_{{\rm g},2}$ in two arbitrary line-of-sight directions
$\vec{\theta}_1$ and $\vec{\theta}_2$, respectively, and a source
ellipticity $\epsilon_{\rm s}+\gamma$ in a third direction
$\vec{\theta}_3$. The shear $\gamma$ and $\epsilon_{\rm s}$ are
rotated in direction of the mid-point between the two lenses according
to the definition of $\cal G$. If lenses and sources are well
separated in distance, then their properties are statistically
independent. The lens-lens-shear correlator measures
\begin{eqnarray}
  {\cal G}&=&
  \ave{\kappa_{{\rm g},1}\kappa_{{\rm g},2}(\gamma+\epsilon_{\rm
      s})}\\
  \nonumber&=&
  \ave{\kappa_{{\rm g},1}\kappa_{{\rm g},2}\gamma}+
  \ave{\kappa_{{\rm g},1}\kappa_{{\rm g},2}}\ave{\epsilon_{\rm s}}\\
  \nonumber&=&
  \ave{\kappa_{{\rm g},1}\kappa_{{\rm g},2}\gamma}\;,
\end{eqnarray}
free of any systematic contribution from the intrinsic shape
$\epsilon_{\rm s}$, if $\epsilon_{\rm s}$ is statistically independent
of the lens number density fluctuation $\kappa_{\rm g}$, i.e.,
\begin{equation}
  \ave{\kappa_{{\rm g},1}\kappa_{{\rm g},2}\epsilon_{\rm s}}=
  \ave{\kappa_{{\rm g},1}\kappa_{{\rm g},2}}\ave{\epsilon_{\rm s}}\;,
\end{equation}
vanishing due to $\ave{\epsilon_{\rm s}}=0$.

Now, consider a lens number density contrast $\kappa_{\rm g}$ in one
direction and the ellipticities $\epsilon_{{\rm s},i}+\gamma_i$ of two
source images $i=1,2$ in two other directions. The ellipticities are
rotated in direction of line connecting the sources in accordance with
the definition of $G_\pm$. The triple correlator measures
\begin{eqnarray}
  G_\pm&=&
  \ave{\kappa_{\rm g}(\gamma^\pm_1+\epsilon^\pm_{{\rm
        s},1})(\gamma_2+\epsilon_{{\rm
        s},2})}\\
  \nonumber&=&
  \ave{\kappa_{\rm g}\gamma^\pm_1\gamma_2}+
  \ave{\kappa_{\rm g}\epsilon^\pm_{{\rm s}_,1}\gamma_2}+
  \ave{\kappa_{\rm g}\gamma^\pm_1\epsilon_{{\rm s},2}}+
  \ave{\kappa_{\rm g}\epsilon_{{\rm s},1}^\pm\epsilon_{{\rm s},2}}\\
  \nonumber&=&
  \ave{\kappa_{\rm g}\gamma^\pm_1\gamma_2}+
  \ave{\kappa_{\rm g}\epsilon^\pm_{{\rm s}_,1}\gamma_2}+
  \ave{\kappa_{\rm g}\gamma^\pm_1\epsilon_{{\rm s},2}}+
  \ave{\kappa_{\rm g}}\ave{\epsilon^\pm_{{\rm s},1}\epsilon_{{\rm
        s},2}}\\  
  \nonumber&=&
  \ave{\kappa_{\rm g}\gamma^\pm_1\gamma_2}+
  \ave{\kappa_{\rm g}\epsilon^\pm_{{\rm s}_,1}\gamma_2}+
  \ave{\kappa_{\rm g}\gamma^\pm_1\epsilon_{{\rm s},2}}\;.
\end{eqnarray}
The last term in the third line vanishes because $\kappa_{\rm g}$ in
the foreground is independent of the intrinsic shape of the sources in
the background and because of $\ave{\kappa_{\rm g}}=0$. The latter
follows from the definition of density fluctuations $\kappa_{\rm
  g}$. 

The last two terms in the last line are less clear. For example in
$\ave{\kappa_{\rm g}\epsilon_{{\rm s},1}^\pm\gamma_2}$, $\gamma_2$
could be correlated with both $\epsilon_{{\rm s},1}$ (GI signal, if
source 2 is behind source 1) \emph{and} $\kappa_{\rm g}$ (GGL
signal). However, on the level of accuracy of the Born approximation
that is used in Eq. \Ref{eq:kappa}, the shear $\gamma_2$ is linear in
the matter density contrast $\delta_{\rm m}$ up to the distance of
source 2. We can, therefore, split the contributions to $\gamma_2$
into three parts $\gamma_2=\gamma_\kappa+\gamma_\epsilon+\gamma_{\rm
  rest}$, namely (i) in contributions from matter within correlation
length to the lens, $\gamma_\kappa$, (ii) matter within correlation
distance to source 1, $\gamma_\epsilon$, and (iii) the rest
$\gamma_{\rm rest}$, which is neither correlated with $\kappa_{\rm g}$
nor with $\epsilon_{{\rm s},1}$. In this case we find
\begin{equation}
  \ave{\kappa_{\rm g}\epsilon^\pm_{{\rm s}_,1}\gamma_2}=
  \ave{\kappa_{\rm g}\gamma_\kappa}\ave{\epsilon^\pm_{{\rm s},1}}+
  \ave{\kappa_{\rm g}}\ave{\epsilon^\pm_{{\rm s}_,1}\gamma_\epsilon}+
  \ave{\kappa_{\rm g}}\ave{\epsilon^\pm_{{\rm s}_,1}}\ave{\gamma_{\rm rest}}\;.
\end{equation}
All three terms vanish owing to $\ave{\kappa_{\rm
    g}}=\ave{\epsilon^\pm_{{\rm s},1}}=0$. A similar rational shows
that also $\ave{\kappa_{\rm g}\gamma_1^\pm\epsilon_{{\rm s},2}}$
vanishes to lowest order, such that we expect to find in the weak
lensing regime
\begin{equation}
  G_\pm=\ave{\kappa_{\rm g}\gamma_1^\pm\gamma_2}\;.
\end{equation}

\subsection{Magnification of lenses}

\begin{figure}
  \begin{center}
    \psfig{file=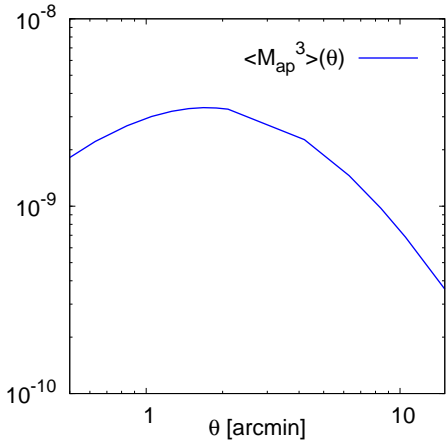,width=60mm,angle=-90}
  \end{center}
  \caption{\label{fig:map3} Prediction of the third-order moment of
    the aperture mass $\ave{M^3_{\rm ap}}$ for sources at redshift
    $z_{\rm s}=0.4$.}
\end{figure}

Another conceivable systematic effect is through cosmic magnification
\citep{1989ApJ...339L..53N,bas01} that is generated by matter density
fluctuations in front of lenses. To lowest order, foreground matter
density fluctuations with lensing convergence $\kappa_<$
(Eq. \ref{eq:kappa}) integrated to the \emph{lens distance} modify the
observed clustering of lenses on the sky above a certain flux limit
$f_{\rm lim}$ according to
\begin{equation}
  \kappa^\prime_{\rm g}=\kappa_{\rm g}+\lambda\kappa_<+{\cal O}(\kappa_<^2)\;,
\end{equation}
compared to the unmagnified lens number density $\kappa_{\rm
  g}$. Here, we have $\lambda:=2(\nu-1)$ with \mbox{$\bar{n}_{\rm
    g}(>f_{\rm lim})\propto f_{\rm lim}^{-\nu}$} being the mean number
density of lenses with flux greater than $f_{\rm lim}$. Normally
$\nu-1$ is of the order of unity \citep{2010MNRAS.401.2093V} or
smaller. Likewise the shear distortion $\gamma=\gamma_<+\gamma_>$,
Eq. \Ref{eq:kappagamma}, into the same l.o.s. direction contains a
contribution $\gamma_<$ related to $\kappa_<$, and $\gamma_>$ that is
the shear originating from matter fluctuations beyond the
foreground. This in combination produces as additional contribution to
${\cal G}=\ave{\kappa^\prime_{{\rm g},1}\kappa^\prime_{{\rm
      g},2}(\gamma_<+\gamma_>)}$ the term
$\lambda^2\ave{\kappa_{<,1}\kappa_{<,2}\gamma_<}$ and to
$G_\pm=\ave{\kappa^\prime_{\rm
    g}(\gamma_{<,1}+\gamma_{>,1})(\gamma_{<,2}+\gamma_{>,2})}$ the
term $\lambda\ave{\kappa_<\gamma_{<,1}\gamma_{<,2}}$.

These terms are basically third-order cosmic shear correlations or, in
terms of the aperture statistics, related to the $\ave{M_{\rm ap}^3(\theta)}$
statistics \citep{2005A&A...431....9S}. Third-order shear correlations
have been measured \citep{2003A&A...397..405B,
  2003ApJ...592..664P,2004MNRAS.352..338J,2011MNRAS.410..143S}, and
$\ave{M_{\rm ap}^3(\theta)}$ has been found
\citep{2004MNRAS.352..338J,2011MNRAS.410..143S} to be of the order of
$\lesssim10^{-7}$ for aperture scales of $\theta\sim1^\prime$ and
sources at $z_{\rm s}\sim1.0$. As this includes contributions from the
entire integrated matter up to $z_{\rm s}$, whereas the G3L
magnification effect only contributions from the matter integrated up
to the lens redshifts $z_{\rm d}\sim0.4$, we consider this an
empirical upper limit for the magnification effect. In
Fig. \ref{fig:map3}, we show a prediction of $\ave{M^3_{\rm
    ap}(\theta)}$ with sources at $z_{\rm s}=0.4$ for a WMAP7-like
cosmology based on the theory described in
\citet{2011MNRAS.410..143S}. This result implies that the impact of
lens magnification on the G3L aperture statistics is smaller than
$\lesssim10^{-8}$.


\section{Data}
\label{sect:data}

\begin{figure}
  \begin{center}
    \psfig{file=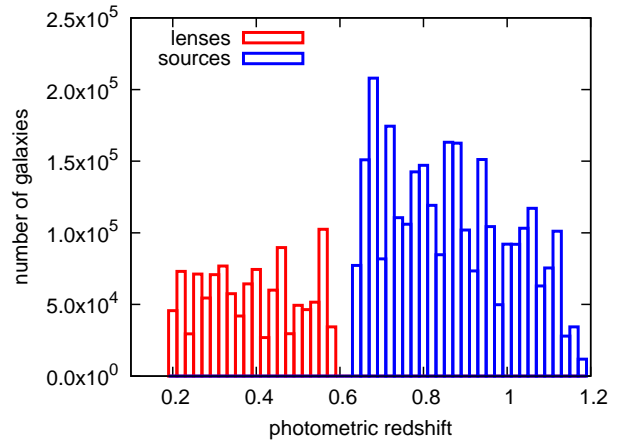,width=60mm,angle=-90}
  \end{center}
  \caption{\label{fig:pofz} Total number of lenses (red) and sources
    (blue) in the catalogue between \mbox{$0.2\le z_{\rm photo}<1.2$}
    and \mbox{$17.5\le i'<22.5$} for lenses or \mbox{$17.5\le
      i'<24.7$} for sources. The figures comprise all galaxies,
    complying with the selection cuts, contained within all 172
    pointings. For the G3L analysis, the lens sample is further
    subdivided in luminosity, stellar mass bins and photometric
    redshift, while sources are rejected for a photo-$z$ of $z_{\rm
      photo}<0.65$.}
\end{figure}

\begin{table*}
  \caption{\label{tab:samples} Selection criteria of lens
    samples and source sample for the G3L analysis applied to the
    samples in Fig. \ref{fig:pofz}, following
    \citet{2006MNRAS.368..715M} for the lenses. The luminosity bins
    (L), stellar mass bins (sm) and galaxy type bins (ETG: early-type
    galaxies; LTG: late-type galaxies) are again subdivided by 
    \mbox{$0.2\le z_{\rm photo}<0.44$} (``low-$z$'') and
    \mbox{$0.44\le z_{\rm photo}<0.6$} (``high-$z$''). Sources
    attributed no statistical weight $w$ by \lensfit are  not used in
    the source sample.  The galaxy numbers are for  all pointings of
    which the final analysis discards roughly 
    25\%. Luminosities and stellar masses assume $h=0.73$. 
    (1) $\bar{z}$: mean redshift, $\sigma_z$: r.m.s. variance of
    $p(z)$; (2) and (3): best-fit parameters of
    $\omega(\theta)=A_\omega(\theta/1^\prime)^{-\lambda}+\rm IC$
    within \mbox{$0^\prime\!.2\le\theta<10^\prime$}; (4):
    sample completeness; (5): mean $r$-band luminosity; (6): mean
    stellar mass in units of $10^{10}\,M_\odot$.}
  \center
  \begin{tabular}{lcrrrrrrr}
    Sample & Selection & \#Galaxies & $\bar{z}\pm\sigma_z^{(1)}$ 
    & $A_\omega/0.1^{(2)}$ & $\lambda^{(3)}$ &$f_c^{(4)}$ & $\ave{M_{r}}^{(5)}$& $\ave{M_\ast}^{(6)}$\\\hline\\
    L1 low-$z$& $-18\le M_{r}<-17$ & 36,372 & $0.22\pm0.16$ &
    $2.40\pm0.29$ & $0.45\pm0.11$ & $0.14$ & -17.75 & 0.04\\
    L1 high-$z$  &       "        & -- & -- & -- & -- & -- & -- & --\\
    L2 low-$z$& $-19\le M_{r}<-18$ & 157,306 & $0.28\pm0.15$ &
    $1.91\pm0.23$ & $0.35\pm0.05$ & $0.45$ & -18.60 & 0.10 \\
    L2 high-$z$   &       "               & -- & -- & -- & -- & -- &
    -- & --\\
    L3 low-$z$& $-20\le M_{r}<-19$ & 220,329 & $0.34\pm0.14$ &
    $1.41\pm0.12$ & $0.43\pm0.05$ & $0.81$ & -19.52 & 0.26 \\
    L3 high-$z$  &       "               & 75,902 & $0.48\pm0.11$ &
    $1.63\pm0.18$ & $0.54\pm0.08$ & $0.42$ & -19.72 & 0.29 \\
    L4 low-$z$& $-21\le M_{r}<-20$ & 149,190 & $0.34\pm0.12$ &
    $1.63\pm0.07$ & $0.53\pm0.03$ & $0.95$ & -20.50 & 0.91\\
    L4 high-$z$  &       "               & 185,286 & $0.51\pm0.10$ &
       $1.62\pm0.08$ & $0.69\pm0.04$ & $0.82$ & -20.53 & 0.98\\
    L5 low-$z$& $-22\le M_{r}<-21$ & 88,916 & $0.34\pm0.11$ &
    $2.19\pm0.14$ & $0.60\pm0.05$ & $0.98$ & -21.48 & 3.09\\
    L5 high-$z$  &         "             & 134,369 & $0.51\pm0.09$ &
       $2.06\pm0.05$ & $0.74\pm0.02$ & $0.99$ & -21.49 & 3.06\\
    L6 low-$z$ & $-23\le M_{r}<-22$ & 31,373 & $0.35\pm0.10$ &
    $3.02\pm0.24$ & $0.65\pm0.07$ & $0.99$ & -22.40 & 8.56\\
    L6 high-$z$  &         "             & 55,315 & $0.52\pm0.08$ &
       $2.50\pm0.10$ & $0.92\pm0.04$ & $1.00$ & -22.42 & 8.11\\
     \\
    sm1 low-$z$& $0.5\le M_\ast/10^{10}M_\odot<1.0$ &  78,181 & $0.34\pm0.12$
     & $2.41\pm0.34$ & $0.43\pm0.09$ & $0.94$ & -20.49 & 0.71 \\
    sm1 high-$z$   &          "                       & 69,784 & $0.50\pm0.10$ & 
     $1.72\pm0.33$ & $0.58\pm0.15$  & $0.77$ & -20.66 & 0.73\\
    sm2 low-$z$& $1.0\le M_\ast/10^{10}M_\odot<2.0$ &  61,650 & $0.34\pm0.11$ &
    $3.75\pm0.82$ & $0.36\pm0.11$ & $0.98$ & -20.98 & 1.42\\
    sm2 high-$z$  &           "                      & 82,411 & $0.51\pm0.09$ & 
     $2.39\pm0.07$ & $0.60\pm0.07$ & $0.90$ & -20.99 & 1.45\\
    sm3 low-$z$& $2.0\le M_\ast/10^{10}M_\odot<4.0$ &  48,632 & $0.34\pm0.10$
    & $3.47\pm0.31$ & $0.51\pm0.07$ & $0.99$ & -21.46 & 2.85\\
    sm3 high-$z$   &             "                    &  81,305 & $0.51\pm0.08$ &
    $2.44\pm0.13$ & $0.72\pm0.05$ & $0.98$ & -21.45 & 2.85\\
    sm4 low-$z$& $4.0\le M_\ast/10^{10}M_\odot<8.0$ & 33,218 & $0.35\pm0.09$
    & $4.05\pm0.39$ & $0.59\pm0.08$ & $0.99$ & -21.91 & 5.60\\
    sm4 high-$z$  &             "                    &  57,049 & $0.51\pm0.08$ &
    $2.72\pm0.11$ & $0.77\pm0.04$ & $0.99$ & -22.00 & 5.59\\
    sm5 low-$z$& $8.0\le M_\ast/10^{10}M_\odot<16.0$ & 15,527 & $0.36\pm0.08$
    & $5.00\pm0.41$ & $0.70\pm0.07$ & $1.00$ & -22.40 & 10.86\\
    sm5 high-$z$   &             "                    & 27,598 & $0.51\pm0.08$ &
    $3.56\pm0.24$ & $0.81\pm0.07$ & $1.00$ & -22.81 & 10.88\\
    sm6 low-$z$& $16.0\le M_\ast/10^{10}M_\odot<32.0$ &  4,605 & $0.36\pm0.07$ 
    & $6.58\pm0.50$ & $1.51\pm0.07$ & $1.00$ & -23.00 & 21.13\\
    sm6 high-$z$  &             "                    &  7,121 & $0.52\pm0.07$ &
    $4.18\pm0.78$ & $1.58\pm0.16$ & $1.00$ & -23.22 & 20.90\\
    sm7 low-$z$& $32.0\le M_\ast/10^{10}M_\odot<64.0$ &   526 & $0.38\pm0.06$
    & $8.89\pm1.37$ & $1.64\pm0.15$ & $1.00$ & -23.60 & 40.81\\
    sm7 high-$z$   &               "                  &   775 & $0.52\pm0.07$ &
    $5.61\pm1.30$ & $1.28\pm0.21$ & $1.00$ & -23.67 & 38.52\\
        \\
    ETG low-$z$& $0\le$\texttt{T\_B}$<2~|~-23\le M_{r}<-21$ & 89,359 &
    $0.34\pm0.10$ & $3.43\pm0.08$ & $0.68\pm0.02$ & $0.99$ & -21.88 & 5.91 \\
    ETG high-$z$   &      "        & 137,144 & $0.51\pm0.08$ & $2.90\pm0.09$ &
        $0.83\pm0.03$ & $1.00$ & -21.91 & 5.74\\
    LTG low-$z$& $2\le$\texttt{T\_B}$<6~|~-23\le M_{r}<-21$ & 30,926 & $0.35\pm0.13$
    & $0.70\pm0.13$ & $0.87\pm0.18$ & $0.96$ & -21.64 & 1.73\\
    LTG high-$z$   &       "       & 52,527 & $0.51\pm0.10$ & $1.33\pm0.16$ &
        $0.78\pm0.11$ & $0.99$ & -21.73 & 2.05\\
     \\
    SOURCES & $0.65\le z_{\rm photo}<1.2~|~w>0$ & 2,926,894 &
    $0.93\pm0.26$ & -- & -- & -- & -- & --
   \end{tabular}
\end{table*}

\begin{figure*}
  \begin{center}
    \psfig{file=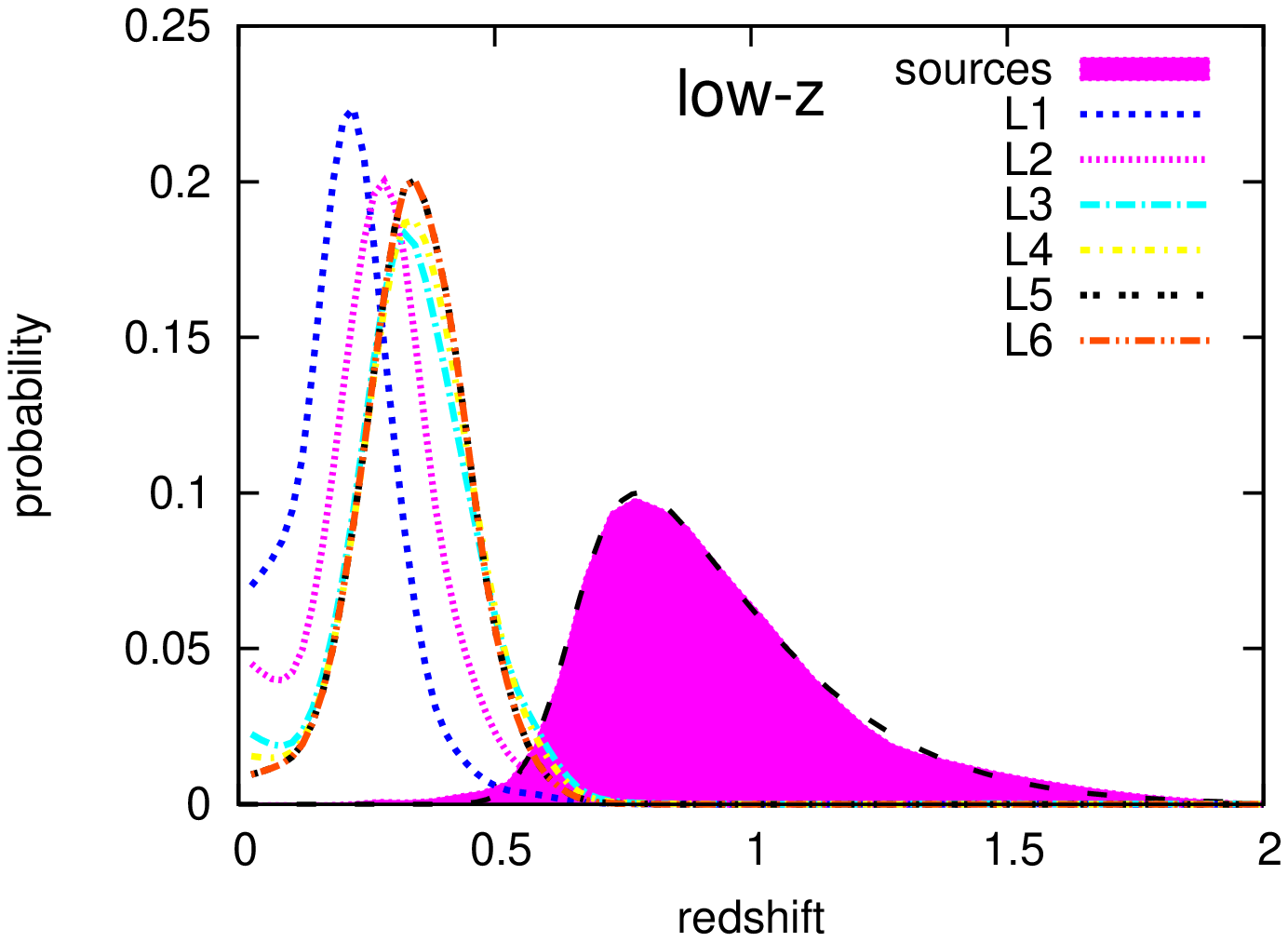,width=55mm,angle=-90}
    \psfig{file=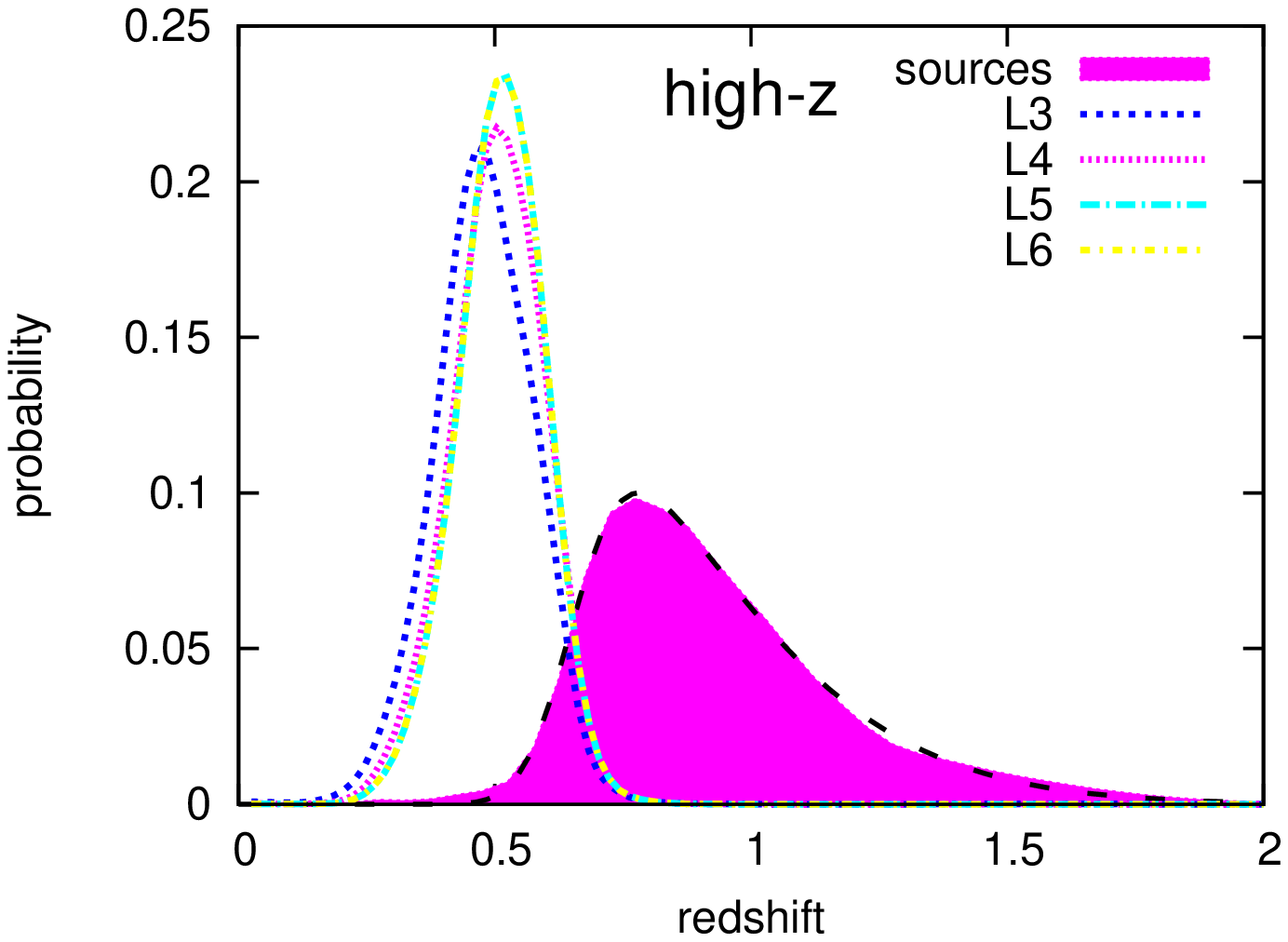,width=55mm,angle=-90}
    \\
    \psfig{file=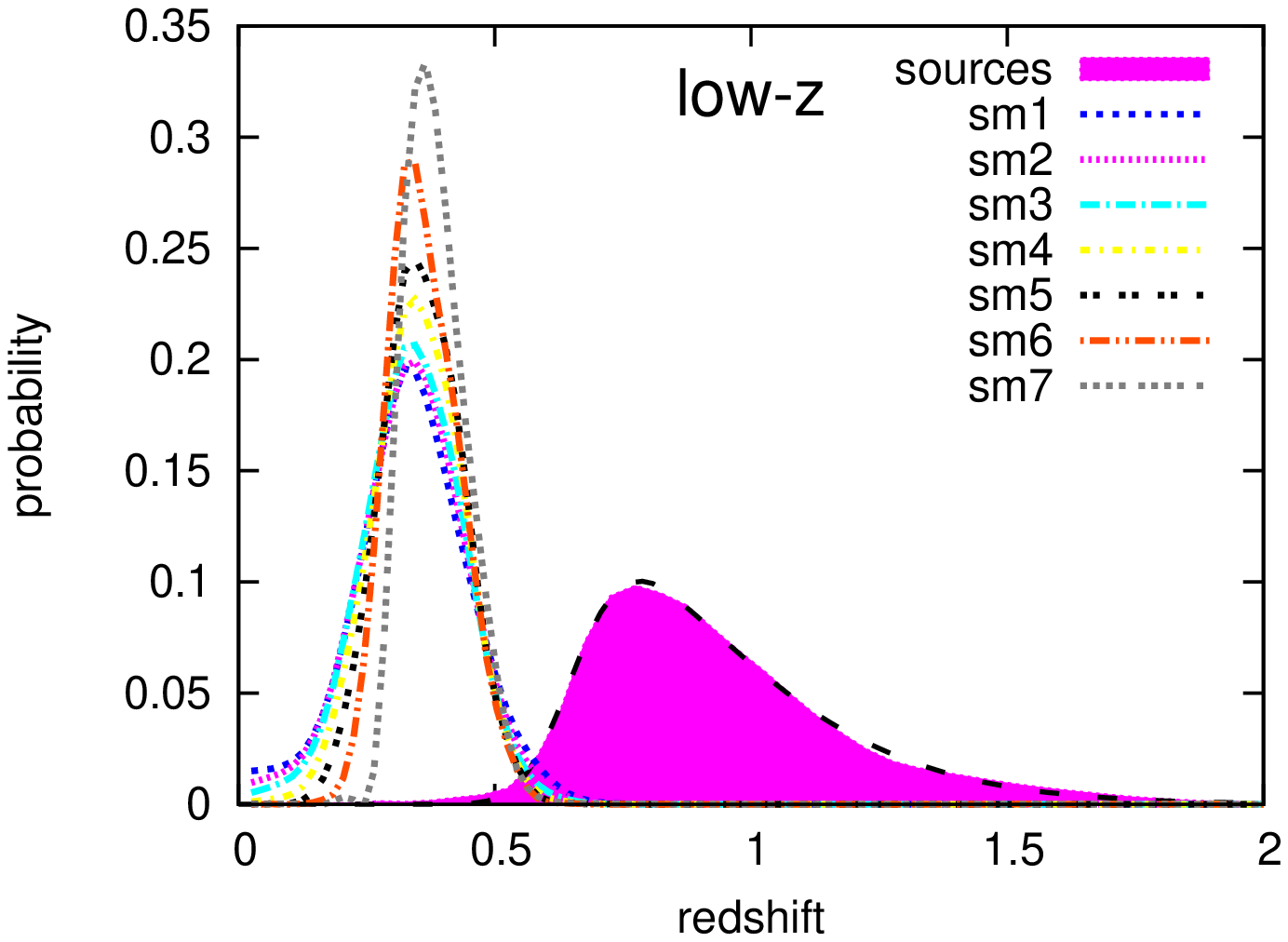,width=55mm,angle=-90}
    \psfig{file=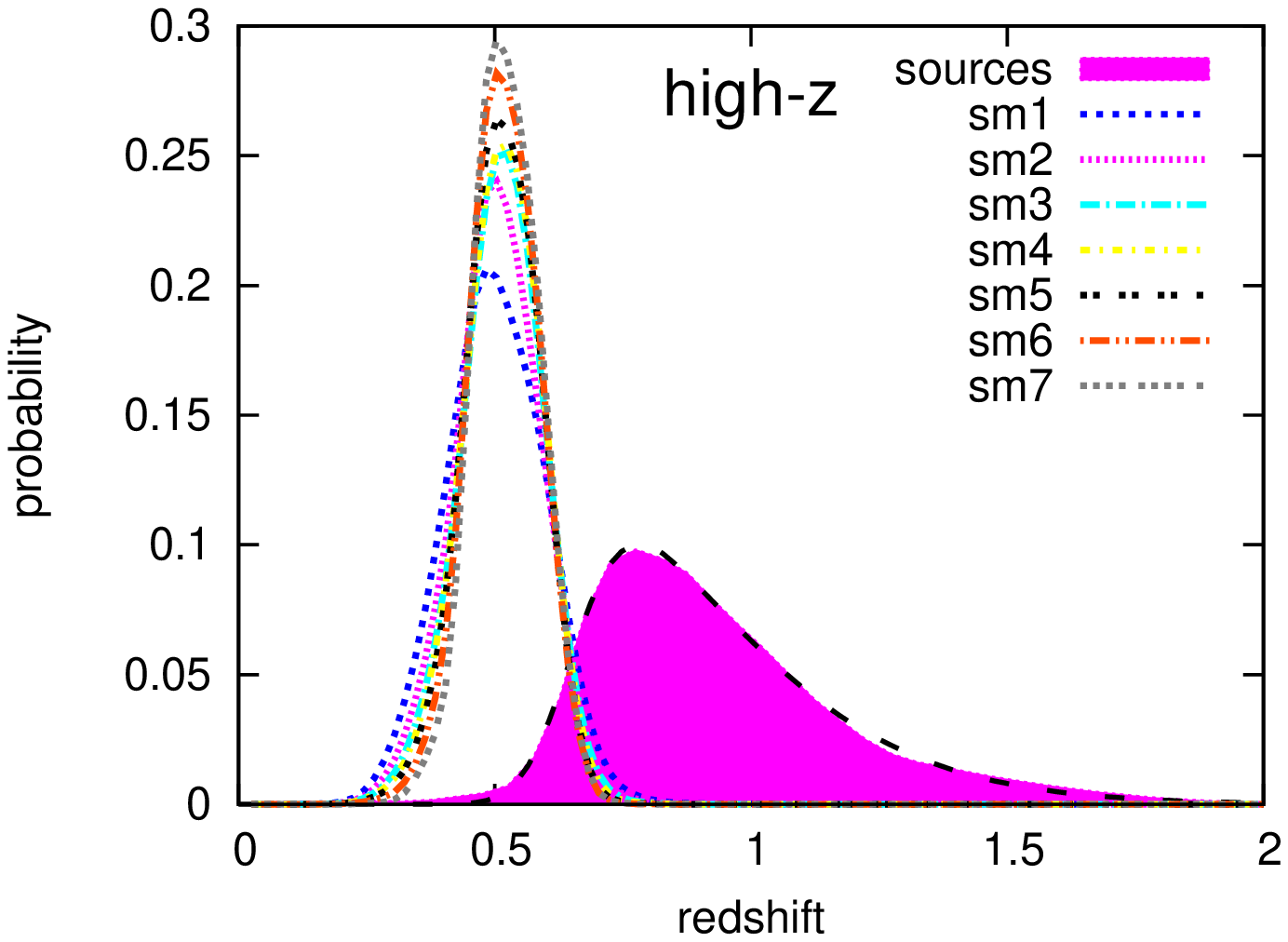,width=55mm,angle=-90}
    \\
    \psfig{file=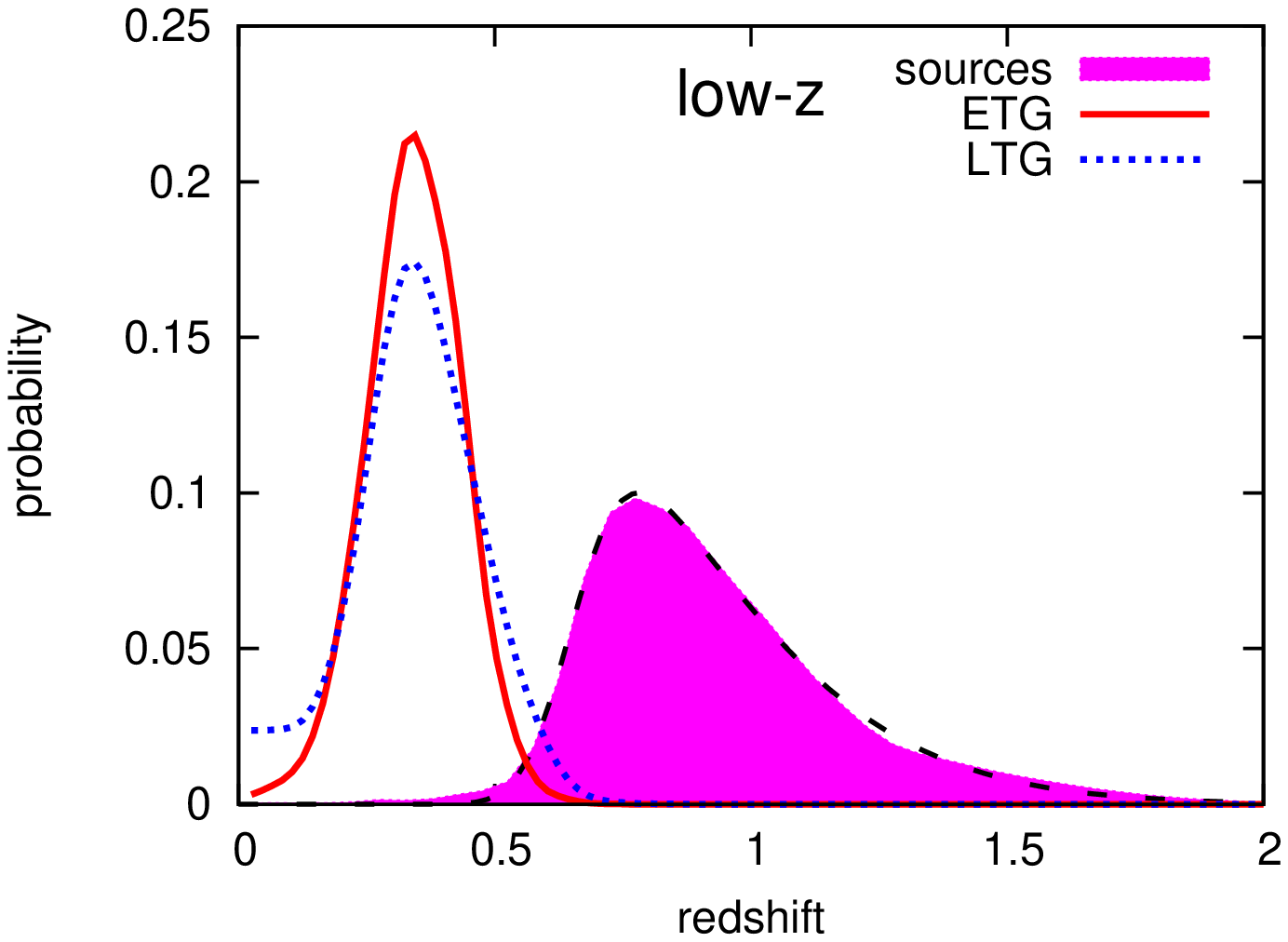,width=55mm,angle=-90}
    \psfig{file=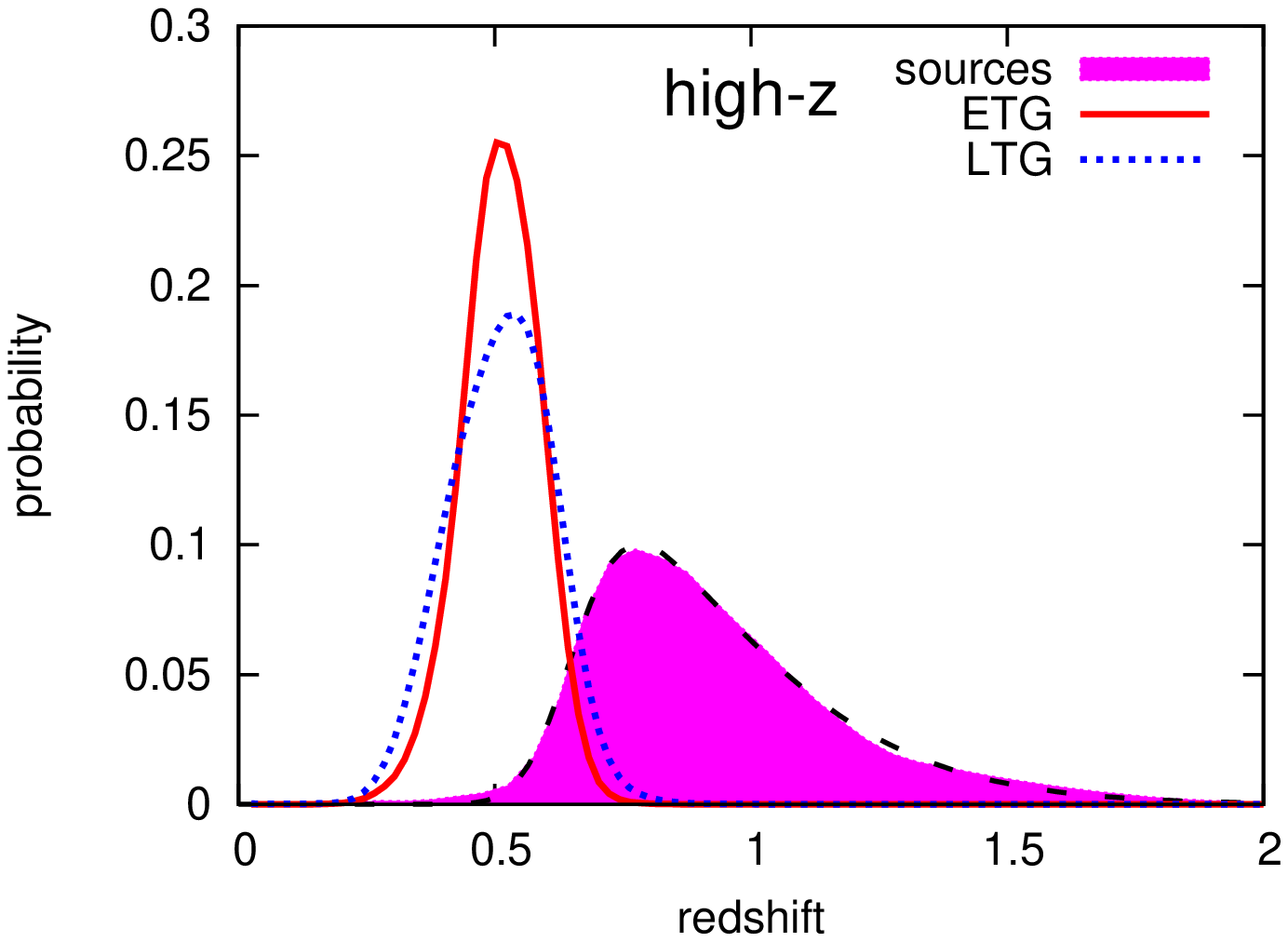,width=55mm,angle=-90}
  \end{center}
  \caption{\label{fig:subpofz} In contrast to Fig. \ref{fig:pofz},
    this figure shows the full \texttt{BPZ} redshift posterior of the
    various samples.  Low-$z$ lenses are selected from $z_{\rm
      photo}\in[0.2,0.44]$, high-$z$ lenses from $z_{\rm
      photo}\in[0.44,0.6]$, and sources from $z_{\rm
      photo}\in[0.65,1.2]$. The dashed black line enclosing the source
    p.d.f. is a parametrised fit, see Sect. \ref{sect:sourcesample}
    for best-fit parameters.}
\end{figure*}

\subsection{Object selection and photometric redshifts}

This work uses the full CFHTLenS data set, which originates from the
CFHTLS-Wide Survey. \verB{The CFHTLS-Wide imaged 171 MegaCam
(mounted on the CFHT) pointings in the five broad-band filters $u^*$,
$g'$, $r'$, $i'$, and $z'$.}  During the observation campaign of CFHTLS,
the $i'$-band filter was replaced by a new filter with a slightly
different transmission curve. For some of the pointings only the
updated $i'$-band filter magnitudes are available, which are treated
as the old filter magnitudes in the analysis. For details, see
\cite{EETAL2012}.

CFHTLenS has an effective area (different pointings partly overlap) of
about 154 square-degrees with high-quality photometric redshifts down
to $i'\approx 24.7$.  The data set and the extraction of our
photometric redshift catalogue are described in
\citet{2012MNRAS.tmp.2386H}. Our data processing techniques and
recipes are described in \citet{2009A&A...493.1197E} and
\citet{EETAL2012}. As primary selection criterion, we select sources
brighter than \mbox{$i'<24.7$} and lenses brighter than
\mbox{$i'<22.5$}. This will be further subdivided in the following by
using photometric redshifts (Fig. \ref{fig:pofz}) and, in the case of
lenses, $M_r$ rest frame magnitudes, stellar masses or SED information
(details below). 43 pointings out of 171 exhibit a significant PSF
residual signal, according to the detailed tests in Sect. 4.2 of
\citet{2012arXiv1210.0032H}, and are therefore discarded for the
analysis ($\sim25\%$ area); 129 pointings are included in the
analysis. This leaves a total effective survey area of $\sim120\,\rm
deg^2$ that is eventually used in the analysis.
\verB{Of this area an additional $\sim20\%$ percent is lost due to
  masking.} The analysis is performed on individual fields
which allows us to use field-to-field variances of the measurements to
estimate the covariance of measurement errors directly from the data.

\subsection{Lens samples}
\label{sect:lenssample}

To guarantee a high reliability of the photo-$z$ estimates for the
lenses, a magnitude cut of $i'\le22.5$ is applied. A detailed account
and tests of the CFHTLenS photo-$z$ pipeline can be found in
\citet{2012MNRAS.tmp.2386H}. Based on the galaxies endowed with
photometric redshifts, three classes of lens samples are selected
(Table \ref{tab:samples}):
\begin{itemize}

\item A luminosity or L-sample class, which consists of six distinct
  rest-frame $M_r$ bins \citep[SDSS
  $r$-filter;][]{2000AJ....120.1579Y}, labelled L1 to L6. The same
  formal luminosity bin limits as in \citet{2006MNRAS.368..715M} or
  \citet{VETAL2012} are applied, although we do not automatically
  expect equivalent completeness of the samples. To quantify the
  completeness, we introduce the $f_{\rm c}$ parameter below.

\item A stellar mass or sm-sample class, which is also further
  subdivided using seven distinct stellar mass bins. Again, we are
  guided by \citet{2006MNRAS.368..715M} for compiling this sample
  class. The sm class has sub-classes with labels sm1-sm7.

\item A galaxy type class using the \texttt{T\_B} parameter in
  \texttt{BPZ} \citep{ben00}, which provides the most likely galaxy
  SED for a given galaxy and its estimated photo-$z$; see
  \citet{EETAL2012} for more details. \texttt{T\_B}=2 as division
  line, we separate early-type galaxies (``ETG''), which have
  \texttt{T\_B}$<2$, from late type galaxies
  (``LTG'').\footnote{Within \texttt{BPZ} values of \texttt{T\_B}
    denote best-fitting galaxy templates: 1=CWW-Ell, 2=CWW-Sbc,
    3=CWW-Scd, 4=CWW-Im, 5=KIN-SB3, and 6=KIN-SB2. Note that the
    templates are interpolated, such that fractional numbers occur.}
  \verB{In order to define a volume-limited sample of ETG and LTG, we
    select only luminous galaxies with restframe luminosities $-23\le
    M_r<-21$. With this luminosity cut, ETG and LTG are actually
    subsamples of L5 and L6 combined.}
\end{itemize}
\verB{The stellar masses of the lenses are determined from the galaxy
  multi-colour data as described in Sect. 2.1 of
  \citet{VETAL2012}. The estimators assume a
  \citet{2003PASP..115..763C} star initial mass function.}

All three classes are further split into two photo-$z$ bins: a
``low-$z$'' bin with \mbox{$0.2\le z_{\rm photo}<0.44$} and a
``high-$z$'' bin with \mbox{$0.44\le z_{\rm photo}<0.60$}. As redshift
estimators we use the maximum probability redshifts of the redshift
posterior provided by \texttt{BPZ}. The redshift boundaries give
comparable numbers of lenses prior to attributing them to one of the
three lens classes (Fig. \ref{fig:pofz}). Not counting the high-$z$ L1
and L2 samples, which have too faint limits to contain
lenses\footnote{Actually, we find a few galaxies in the high-$z$ L1/L2
  samples. These are probably extreme outliers with greatly inaccurate
  redshift estimates.}, we have in total 28 lens subsamples.

The true redshift distribution of a lens sample is not identical to
the distribution of their photometric redshifts due to the errors in
the photo-$z$ estimators.
\verB{For a magnitude cut of \mbox{$i'<22.5$}, the errors are
  approximately \mbox{$\sigma_{\rm z}\lesssim 0.04(1+z)$} with a
  \mbox{$\sim3\%$} outlier rate \citep{2012MNRAS.tmp.2386H}. We
  combine the posterior redshift probability distribution functions
  (p.d.f.) of all lenses given by \texttt{BPZ}, see
  Fig. \ref{fig:subpofz}, to quantify the redshift uncertainties of
  complete lens samples.} The depicted redshift probability
distributions will be utilised when normalising the G3L aperture statistics.

To help the comparison of our G3L results to future studies, we also
quote the angular clustering and completeness of the lens samples. The
results are listed in Table \ref{tab:samples}, the details are
described in Appendices \ref{sect:clustering} (clustering) and
\ref{sect:completeness} (completeness). In short, for the angular
clustering of lenses, we approximate the angular galaxy two-point
correlation function by a power law over the angular range
\mbox{$0^\prime\!\!.2\le\theta<10^\prime$}.  For each lens sample with
the photo-$z$ bin $[z_1,z_2]$, we quote the completeness factor
$f_{\rm c}$ that expresses the average $V(z_1,z_{\rm max})/V(z_1,z_2)$
of all lenses in the sample; $V(z_1,z_2)$ is the light cone volume
between redshift $z_1$ and $z_2$, and $z_{\rm max}\le z_2$ is the
maximum redshift up to which a lens is still above the flux limit
$i^\prime=22.5$. A small $f_{\rm c}$ is a sign of a strong
incompleteness because many galaxies similar to those observed near
$z_1$ are presumably missing at higher redshifts. Due to the magnitude
limit, samples containing a substantial portion of faint galaxies are
most affected by incompleteness, most notably L1 and L2. As expected,
the completeness drops if one moves from the low-$z$ to the high-$z$
bin in almost all cases. The few minor exceptions, L5 for instance,
are probably due to shot noise in the $f_{\rm c}$ estimator. We
conclude that L4-L6, sm3-sm7 and ETG/LTG are the most complete,
volume-limited samples for our study ($f_{\rm c}>0.80$ for both
low-$z$ and high-$z$). In Table \ref{tab:samples} we also quote the
average absolute $r$-band flux of the samples, listed as magnitude
$\ave{M_r}$ and the their average stellar mass $\ave{M_\ast}$.
 
\subsection{Source sample}
\label{sect:sourcesample}

All details concerning the galaxy shape measurement (employing the
\lensfit algorithm; \citealt{2008MNRAS.390..149K},
\citealt{2007MNRAS.382..315M}, and \citealt{METAL2012}), CFHTLenS
source catalogue generation, and the discussion of shear systematics
are presented in \citet{2012arXiv1210.0032H} and \citet{METAL2012}.
We account for the multiplicative shear bias by employing the Miller
et al. (2012) normalisation scheme adjusted to our estimators (see
Appendix \ref{sect:mbias}).

In order to reduce the level of undesired II- and GI-correlations in
the measurements, we attempt to separate sources and lenses by
redshift, utilising photometric redshifts as estimators. As a
compromise between accurate redshift estimates and a large numbers of
sources, we apply a magnitude limit $i'<24.7$ to the \lensfit shear
catalogue and select sources between \mbox{$0.65\le z_{\rm
    photo}<1.2$}. As for the lenses, the true redshift distribution is
derived from the combined posterior redshift p.d.f. of individual
sources, shown in every panel of Fig. \ref{fig:subpofz} in comparison
to the redshift distribution of the lens samples. The individual
posteriors are weighted with the source weight that is also used in
the lensing analysis. The source redshift p.d.f. is well fitted by a
broken exponential distribution
\begin{equation}
  p_{\rm b}(z)\propto
  \left\{
  \begin{array}{ll}
    \exp{(-p_0(z_0-z)^{p_1})} & {\rm if}~z\le z_0,\\
    \exp{(-p_2(z-z_0)^{p_3})} & {\rm otherwise}
  \end{array}
  \right.
\end{equation}
with fit parameters $p_0=91.14$, $p_1=2.623$, $p_2=4.093$, $p_3=1.378$
and $z_0=0.794$ (dashed black lines). With our selections we find
about $3\times10^6$ sources with mean redshift $\bar z\approx0.93$. As
can be seen in Fig. \ref{fig:subpofz}, the overlap of the various
$p_{\rm f}(z)$ and the source $p_{\rm b}(z)$ is small but not entirely
vanishing, mainly at $z=0.5-0.7$ for the high-$z$ and at $z\sim0.6$
for the low-$z$ samples. The typical overlapping area of the redshift
probability distribution functions (visible in Fig. \ref{fig:subpofz})
is $\sim12\%$ for the high-$z$ samples and $\sim4\%$ for the low-$z$
samples.


\section{Results}
\label{sect:results}

\subsection{Measurements and their uncertainties}
\label{sect:measurement}

In order to obtain measurements for the lensing aperture statistics,
we use the method outlined in Sect. \ref{sect:estimators}. As the
binning grid for $\widetilde{\cal G}^{\rm est}$ and
$\widetilde{G}_\pm^{\rm est}$, $100$ log-bins ranging between 9 arcsec
and 50 arcmin are set up for $\vartheta_{1/2}$, $100$ linear bins are
used for the opening angle $\phi_3$, yielding overall $10^6$ bins with
bin widths $\Delta\phi_3=3.6\,\rm deg$ and
$\Delta\ln{\vartheta}=0.058$. All measurements are performed
separately on every individual pointing, out of 129 square pointings
with roughly $1\,\rm deg^2$ each. Adjacent pointings partly overlap,
however, which reduces the area that is actually used. In our study,
we crop the pointings to remove the overlap.  For the final result,
individual estimates are combined by averaging the individual
$\widetilde{\cal G}^{\rm est}$ and $\widetilde{G}_\pm^{\rm est}$
weighted by the number of triangles within each bin.

Finally, the combined estimates are transformed to the aperture
statistics by the integral transformations discussed in SW05. In his
way, the aperture statistics between
\mbox{$0^\prime\!\!.5\le\theta_{\rm ap}\le10^\prime$} for ten aperture
scale radii are computed. As addressed in \citet{2008A&A...479..655S},
the transformation from $\tilde{\cal G}$ or $\tilde{G}_\pm$ to
aperture statistics becomes biased towards small and large aperture
radii due to an insufficient sampling of the correlation functions. A
similar transformation bias is also known for the aperture mass
statistics \citep{kse06}.  For the small separations, the bias depends
in detail on the mean number density of the galaxies, most crucially
the lenses, and the clustering of the lenses, which in combination
determines the sampling of the correlation functions by small
triangles.  By comparison to simulated data, we made sure that this
bias is negligible (below $\sim10\%$) within the range of
\mbox{$1^\prime\lesssim\theta\lesssim10^\prime$} in our case (see
Fig. 1 in \citealt{2012arXiv1204.2232S} for an illustration of the
transformation bias).  The variance of the measurements across all 129
pointings is used to estimate the covariance of measurement errors
(Jackknifing; Appendix \ref{sect:clustering}). The inverse covariance
matrix is estimated from the pointing-to-pointing covariance according
to the method in \citet{2007A&A...464..399H}.

\begin{figure*}
  \begin{center}
    \psfig{file=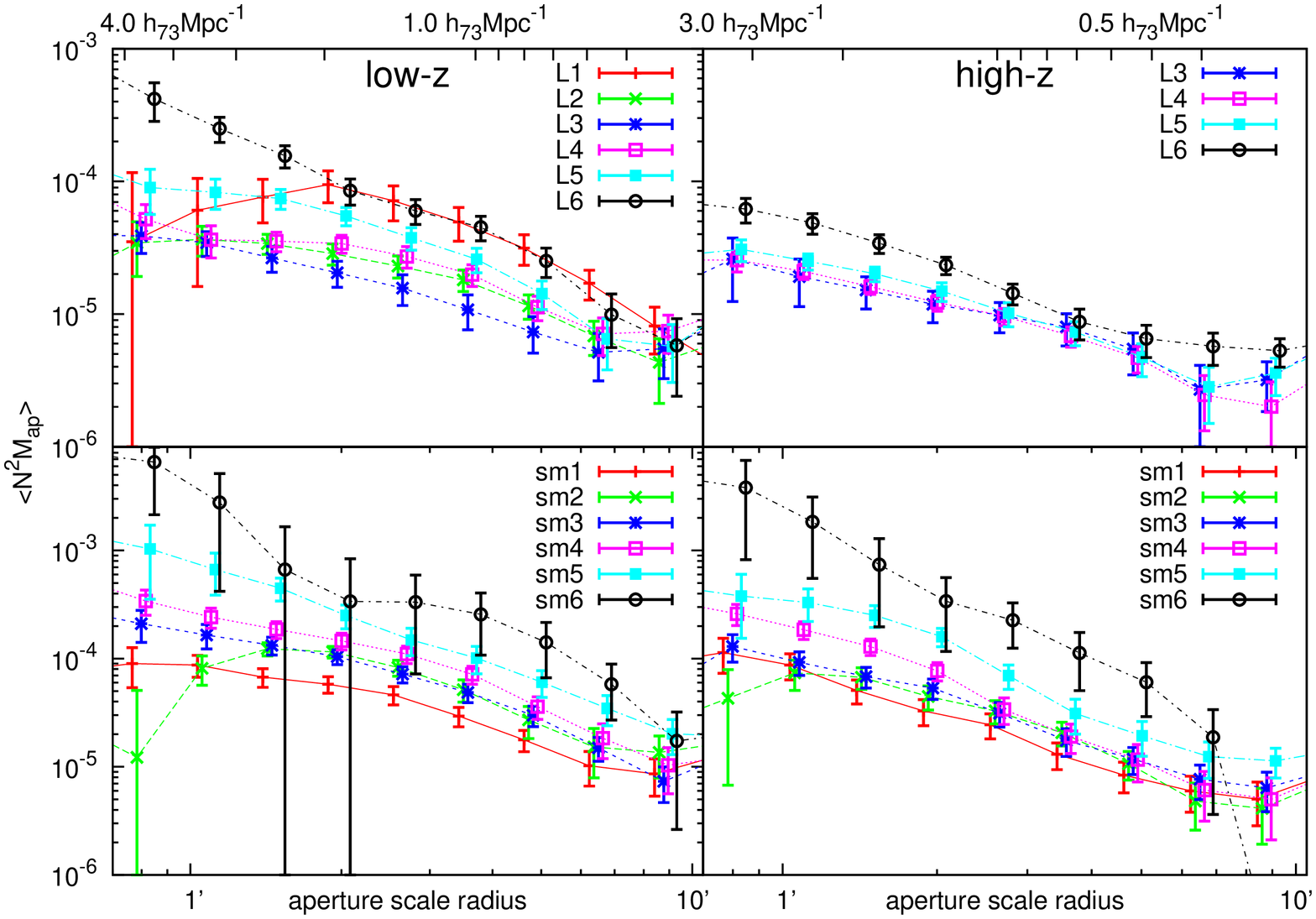,width=107mm,height=170mm,angle=-90}
    \psfig{file=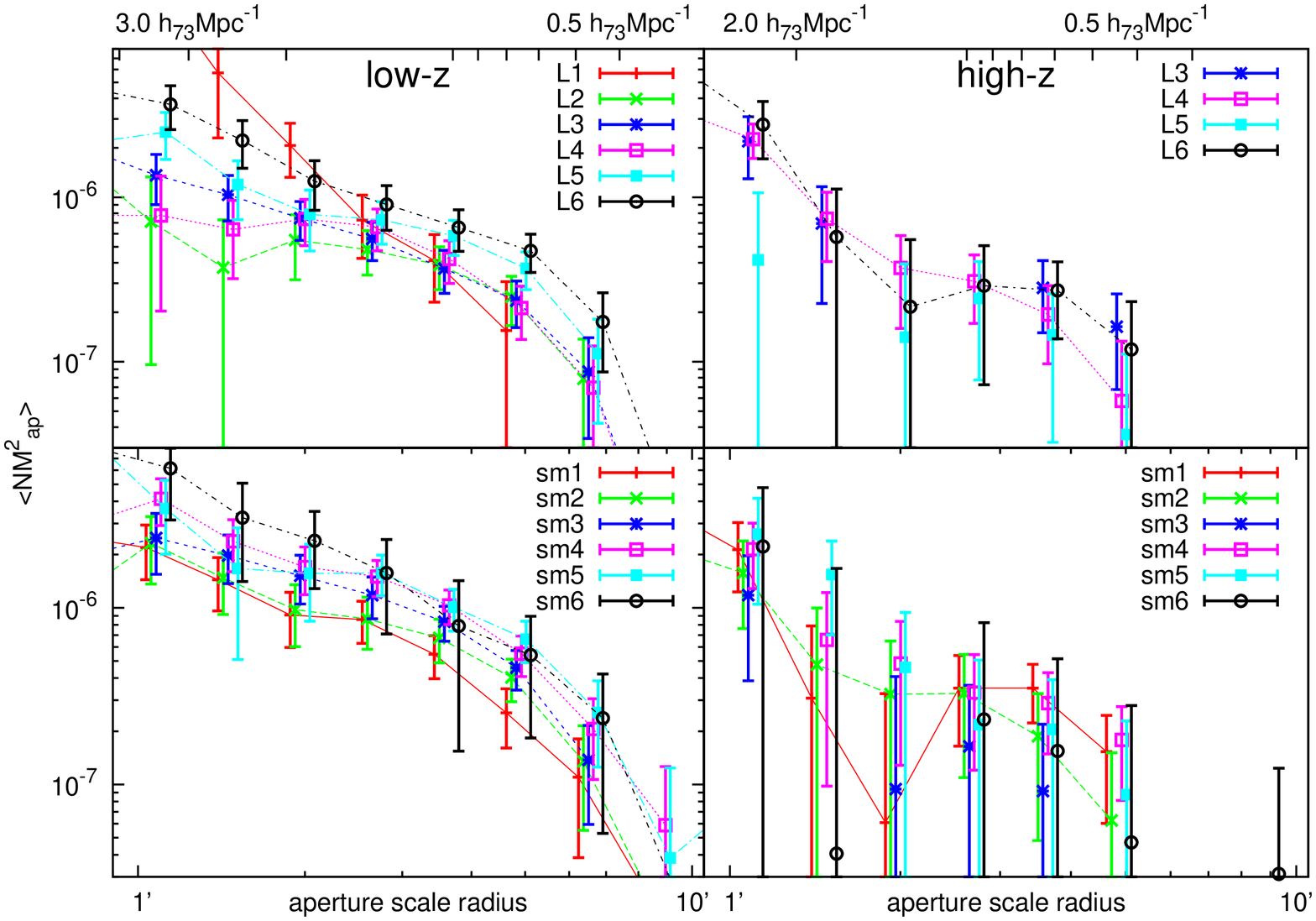,width=107mm,height=170mm,angle=-90}
  \end{center}
  \caption{\label{fig:n2map2} Measurements of the E-mode aperture
    statistics $\ave{{\cal N}^2M_{\rm ap}}(\theta)$ (top figure) and
    $\ave{{\cal N}M^2_{\rm ap}}(\theta)$ (bottom figure) as a function
    of aperture scale radius $\theta$. The left column depicts
    measurements for the low-$z$ bin, the right column the high-$z$
    bin. Different lines refer to different lens samples (Table
    \ref{tab:samples}). Note that the values get biased for
    $\theta\lesssim1^\prime$ due to the transformation bias. Error
    bars indicate the $1\sigma$ standard deviation of the mean of all
    pointings considered. Missing data points are outside the plotting
    range but consistent with zero. Numbers at the top indicate the
    effective scale of the statistics according to the maximum in the
    $u$-filter. L1-L6: $M_{r}$-luminosities increasing from -17.8
    mag to -22.4 mag; sm1-sm6: increasing stellar masses from
    $7\times10^{9}\,M_\odot$ to $2\times10^{11}\,M_\odot$.}
\end{figure*}

\begin{figure*}
  \begin{center}
    \psfig{file=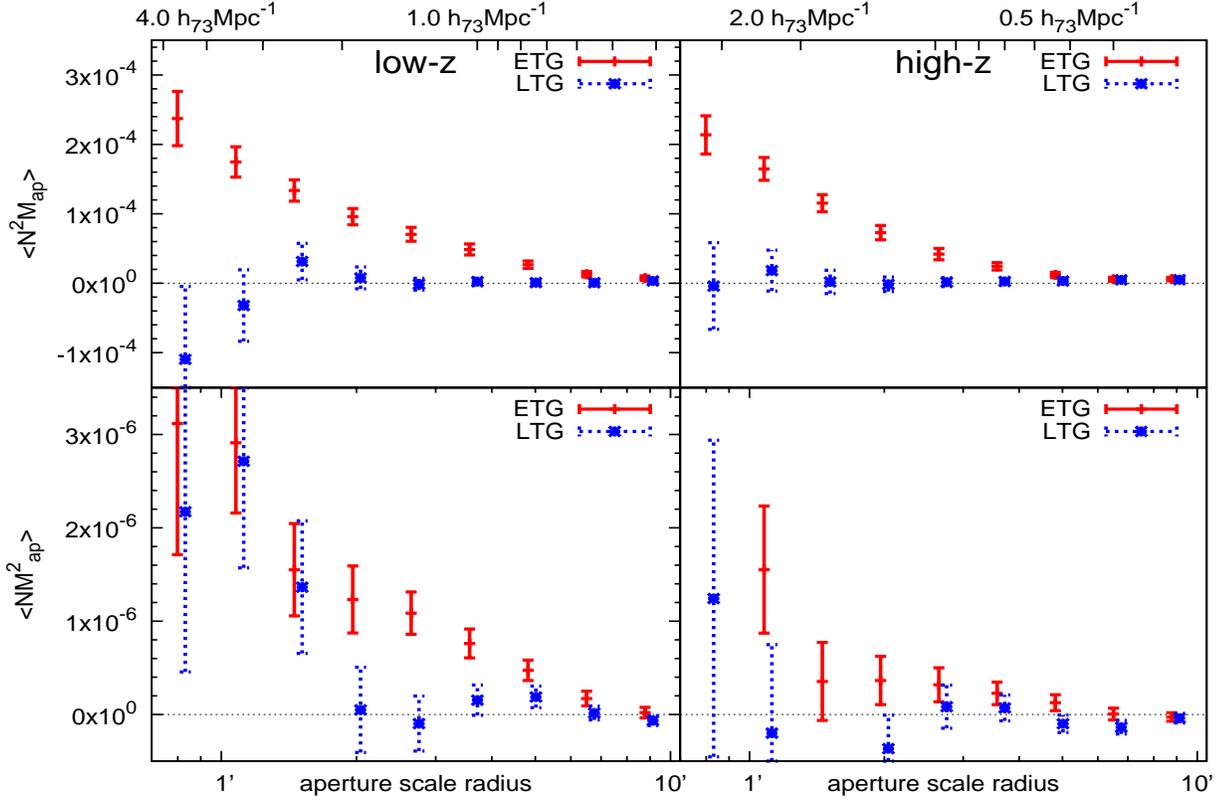,width=107mm,height=170mm,angle=-90}
  \end{center}
  \caption{\label{fig:mapstatgaltype} Aperture statistics results for
    the late-type and early-type lens samples. The left column shows
    the low-$z$ redshift bin, the right column the high-$z$ redshift
    bin. Top row and bottom row correspond to $\ave{{\cal N}^2M_{\rm
        ap}}$ and $\ave{{\cal N}M^2_{\rm ap}}$, respectively. Numbers
    at the top indicate the effective scale according to the maximum
    in the $u$-filter. ETG: early-type galaxies; LTG: late-type
    galaxies.}
\end{figure*}

\begin{table*}
  \caption{\label{tab:nulltest} 
    Results of the null tests for the
    different channels of the statistics. Quoted values are the 
    reduced $\Delta\chi^2$ with d.o.f=7. for the E-modes and P-mode,
    $\ave{{\cal N}^2M_\perp}$, and d.o.f.=14 for the combined
    B/P-mode, $\ave{{\cal N}M^2_\perp}$ and $\ave{{\cal
        N}M_\perp M_{\rm ap}}$. E-modes consistent with a null signal 
    are in \textbf{bold}, P- or P/B-modes inconsistent 
    with a null signal are \underline{underlined}. 
    Adopted confidence levels are $95\%$. Only data points within the
    range \mbox{$1^\prime\!\!.5\le\theta<10^\prime$} were
    used for this test.}
\center
  \begin{tabular}{lrrrrrrrr}
    & \multicolumn{4}{c}{$0.22\le z_{\rm photo}<0.44$} 
    & \multicolumn{4}{c}{$0.44\le z_{\rm photo}<0.60$}\\
    & \multicolumn{2}{c}{$\ave{{\cal N}^2M_{\rm ap}}$}&
    \multicolumn{2}{c}{$\ave{{\cal N}M^2_{\rm ap}}$}  &
    \multicolumn{2}{c}{$\ave{{\cal N}^2M_{\rm ap}}$}&
    \multicolumn{2}{c}{$\ave{{\cal N}M^2_{\rm ap}}$}\\
    Sample  & E-mode & P-mode & E-mode & P/B-mode &
    E-mode & P-mode & E-mode & P/B-mode\\
     \hline\\
L1 & 3.22	&	1.27	&	2.64	&	1.24 &
      --      &        --     &        --     &        -- \\
L2 & 7.32	&	1.57	&	2.31	&	\underline{1.75} &
      --      &        --     &        --     &        -- \\
L3 & 3.79	&	0.66	&	4.93	&	0.75 &
     3.21	&	0.94	&	2.38	&	1.22\\
L4 & 9.26	&	0.58	&	2.62	&	1.14 &
     7.66	&	0.89	&	2.38	&	0.79\\
L5 & 6.72	&	1.00	&	4.31	&	0.80 &
     7.08	&	1.95	&	\textbf{0.81} &	1.06\\
L6 & 7.76	&	0.99	&	4.92	&	0.74 &
     7.07	&	1.50	&	\textbf{1.74}	&	0.41\\
     \\
sm1 & 6.41	&	1.06	&	4.06	& 1.55 &
      5.18      &	0.74	&	\textbf{1.90}  &  0.68\\
sm2 & 12.43	&	0.59	&	3.55	&	1.11 &
      5.97      &	0.40	&	\textbf{0.95}	&	0.83\\
sm3 & 7.07	&	0.99	&	4.05	&	0.76 & 
      5.62	&	0.65	&	\textbf{0.18} 	& 1.19 \\
sm4 & 7.64	&	1.86	&	4.83	&	0.58 &
      5.93	&	0.27	&	2.28	&	1.12\\
sm5 & 3.76	&	1.27	&	4.95    &	0.52 &
      6.74	&	1.18	&	\textbf{0.78}	&	0.96\\
sm6 & \textbf{0.65}	&	0.91	& \textbf{1.15}	&	1.00 &
      2.33	&	0.68	&	\textbf{0.64}	&	1.42\\
sm7 & \textbf{0.93}	&	1.16	&	\textbf{0.66}	&	1.61 &
      \textbf{0.45}	&	0.42	&	\textbf{1.52}	& \underline{1.87}\\
\\
ETG & 12.50	&	0.87	&	6.62	&	0.65 &
      15.24	&	1.74	&	\textbf{1.21}	&	1.06 \\
LTG & \textbf{0.77}	&	0.85	&	\textbf{1.58}	&	1.52 &
      \textbf{0.90}	&	1.20	&	\textbf{1.45}	&	0.37
  \end{tabular}
\end{table*}

\subsection{E-mode measurements}

Fig. \ref{fig:n2map2} summarises the E-mode results for the luminosity
and stellar mass bins of the $\ave{{\cal N}^2M_{\rm ap}}$ (top) and
$\ave{{\cal N}M^2_{\rm ap}}$ statistics (bottom). Due to the
incompleteness in the samples, L1 and L2 are empty in the higher
redshift bin and hence are missing in the corresponding
plots. Likewise, due to the small number of lenses and correspondingly
large error bars, also the data points of sm7 are missing. The signal
dependence on galaxy type is displayed separately in
Fig. \ref{fig:mapstatgaltype}. For aperture radii greater than $\sim2$
arcmin the measurements seem to be well approximated by power laws,
which will be determined below. Below roughly 2 arcmin there are
indications of deviations from the power-law behaviour at smaller
radii in several cases, e.g., $\ave{{\cal N}^2M_{\rm ap}}$ of low-$z$
L1/L4, or $\ave{{\cal N}M_{\rm ap}^2}$ of high-$z$ L4/L6.

The result of $\ave{{\cal N}^2M_{\rm ap}}$ of the late-type galaxies
(LTG) stand out as being the only one that is completely consistent
with zero despite relatively small error bars. Therefore, the excess
mass around late-type galaxy pairs vanishes within the statistical
uncertainties. In strong contrast to that, the corresponding signal of
the early-type galaxy (ETG) sample is highly significant. From the LTG
signal upper limit we estimate the ETG signal to be greater by a
factor of at least $\sim10$. This confirms the prediction of
\citet{2012arXiv1204.2232S} that is based on galaxy population
synthesis models.

The low-$z$ sample L1, with the fewest number of lenses, presumably is
affected by the transformation bias. This can be seen by the clear
drop of the data points for $\ave{{\cal N}^2M_{\rm ap}}$ below
$\theta_{\rm ap}\sim2\,\rm arcmin$ compared to a power-law behaviour
at larger scales.

\subsection{Systematics tests}

General tests for systematics on the level of shear catalogue
generation are to be found in \citet{2012arXiv1210.0032H}. We only use
CFHTLenS pointings that passed the therein described tests for cosmic
shear applications. To further test for systematics in our
measurements, we check for the consistency of the aperture statistics
B- and P-modes with a null signal. The details of this test and,
moreover, G3L measurements within separate CFHTLenS fields (W1-W4) are
presented in Appendix \ref{sect:systematics}. The null test also
allows to quantify the significance of the signal in the E-mode
channels of the statistics. Table \ref{tab:nulltest} summarises the
tests for all statistics and galaxy samples.

In summary, we find that B/P-modes in the aperture statistics are
consistent with zero between 1 arcmin and 10 arcmin. When looking at
the combined L1-L6 sample, separate measurements within the survey
fields W1-W4 agree well for both the low-$z$ and the high-$z$ redshift
bin. This demonstrates the internal consistency of the data, and that
the observed signals do not originate from a single, possibly peculiar
field. As to the E-mode channels of the statistics, we find for
$\ave{{\cal N}^2M_{\rm ap}}$ highly significant signals ($95\%$
confidence) for all low-$z$ samples, except for sm6 and sm7, and most
high-$z$ lens samples. Sm6 and sm7 pose exceptions because they
contain relatively small numbers of galaxies. Apart from the high-$z$
L3, L4, and sm4, all high-$z$ sample measurements of $\ave{{\cal
    N}M^2_{\rm ap}}$ are consistent with zero, whereas their low-$z$
counterparts are mostly significant.  As $\ave{{\cal N}M^2_{\rm ap}}$
involves a three-point correlation function with two sources and one
lens, the noise level of this measurement is naturally higher than for
$\ave{{\cal N}^2M_{\rm ap}}$.

\section{Interpretation}
\label{sect:interpretation}

\begin{figure}
  \begin{center}
    \psfig{file=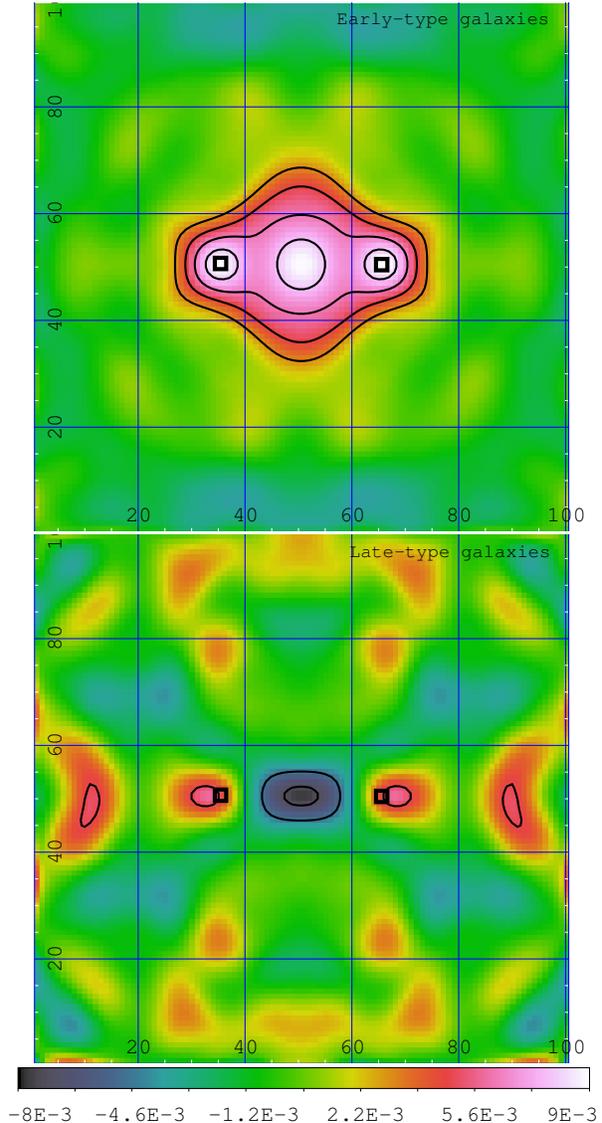,width=80mm,angle=0}
  \end{center}
  \caption{\label{fig:excessmap}
    \verB{Visualisation of the $\cal G$-measurement. Shown is the
      excess mass $\ave{\kappa_{\rm g}(\vec{\theta}_1)\kappa_{\rm
          g}(\vec{\theta}_2)\kappa(\vec{\theta}_3)}$ (intensity scale)
      as function of $\vec{\theta}_3$ around early-type (\emph{top})
      and late-type galaxy pairs (\emph{bottom}) with mean angular
      separations $|\vec{\theta}_1-\vec{\theta}_2|$ between 40 and 60
      arcsec; one map pixel corresponds to 1.67 arcsec ($x$- and
      $y$-axis labels). The lens positions
      $\vec{\theta}_1,\vec{\theta}_2$ are indicated as boxes, the
      contours show the S/N levels $2,3,\ldots$ (positive excess mass)
      and -3,-2 (negative excess mass). To increase the
      signal-to-noise the low-$z$ and high-$z$ maps of the ETG and LTG
      samples have been combined. A smoothing with a Gaussian kernel
      of r.m.s.-size $6.7$ arcsec has been applied to map.}}
\end{figure}

\subsection{Lens-pair excess mass}

\verB{Although aperture statistics and G3L-correlators essentially
  contain the same information, we would like to show our $\cal
  G$-measurements for at least the ETG and LTG samples. As outlined in
  \citet{2008A&A...479..655S}, the G3L correlation function $\cal G$
  can conveniently be interpreted as a convergence map (excess mass
  map) once the separation of the two lenses is fixed; the E-mode in
  $\cal G$ is a series of such maps for varying lens-lens
  separations. After a rotation, the correlator $\widetilde{\cal G}$
  is a stacked shear field around the lens pair, from which we
  subtract off the GGL signal around individual the lenses to determine
  the connected part $\cal G$.  To obtain the
  excess mass maps for the ETG and LTG in Fig. \ref{fig:excessmap} we
  transform this stacked shear field to a convergence map utilising
  the algorithm in \citet{kas93}. For these maps, we consider
  relatively small lens-lens separations between 40 and 60 arcsec as
  in \citet{2008A&A...479..655S}, and we combine the maps of the
  low-$z$ and high-$z$ samples; lens-lens-source triangles are
  rescaled inside the map such that lenses are always at the same
  position in the map (boxes). We also exploit the parity invariance
  of the maps by averaging the left and right half of the map, thereby
  increasing the signal-to-noise, see \citet{2008A&A...479..655S} for
  details.}

\verB{The ETG map contains more significant structure and higher
  convergence values compared to the LTG map, which has only a weak
  signal. Qualitatively, the excess mass of the ETG sample is
  concentrated between the lens pair, whereas the LTG lenses seem to
  possess a small halo of excess mass around the individual lenses and
  a convergence trough between them. The latter implies that the
  average convergence about a LTG pair (both lenses at similar
  distance) at given separation is lower than the sum of convergence
  around two mean individual late-type galaxies. We will study these
  maps in more detail in a forthcoming paper and focus on the aperture
  statistics for the remainder of this paper.}

\begin{figure*}
  \begin{center}
    \psfig{file=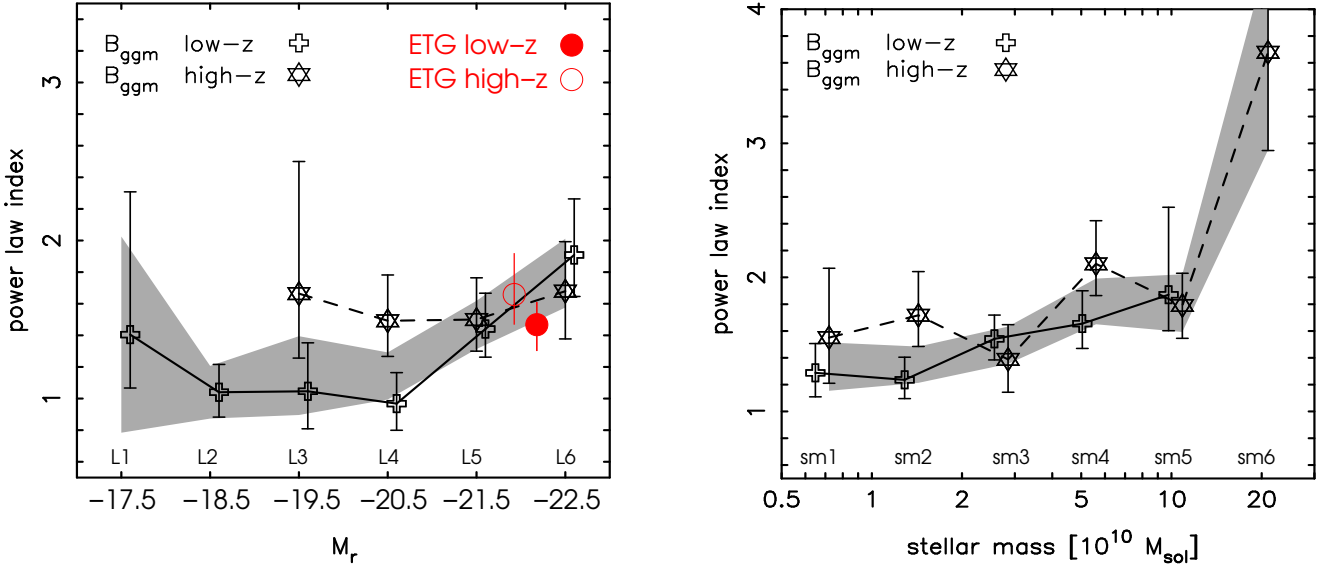,width=175mm,angle=0}
  \end{center}
  \caption{\label{fig:slopes} Dependence of the power-law index
    $\alpha$ in the normalised aperture statistics $\ave{{\cal
        N}^2M_{\rm ap}}$ between
    \mbox{$1^\prime\!\!.5\le\theta<10^\prime$} with $M_r$-band
    magnitude (left panel) and stellar mass of lenses (right panel).
    Table \ref{tab:apstatfits} lists the power law indices of all
    statistics. Only fits to measurements with a $95\%$ confidence
    detection are plotted.  The shaded region highlights the $68\%$
    credibility region of the combined low-$z$ (open crosses) and
    high-$z$ (open stars).  Also shown are the slopes for the
    early-type galaxy sample in the corresponding magnitude range
    (left panel).  For clarity, these data points are offset about
    their actual mean $\ave{M_r}=-21.88(-21.91)$ for the low-$z$
    (high-$z$) sample.}
\end{figure*}

\begin{table*}
  \caption{\label{tab:apstatfits}  
    Power-law fits
    $\ave{{\cal N}^nM^m_{\rm
        ap}}(\theta)=A_0(\theta/1^\prime)^{-\alpha}$ to the measured
    aperture statistics in Fig. \ref{fig:n2map2};
    $A_0$ is the signal amplitude at an aperture scale radius of
    $\theta=1\,\rm  arcmin$. The fit considers only data within
    $\theta\in[1^\prime\!\!.5,10^\prime]$. Quoted errors bracket a
    $68\%$ credibility region about the median.}
  \center
  \begin{tabular}{crrrrrrrr}
    & \multicolumn{4}{c}{$0.20\le z_{\rm photo}<0.44$}
    & \multicolumn{4}{c}{$0.44\le z_{\rm photo}<0.60$}\\
    & 
    \multicolumn{2}{c}{$\ave{{\cal N}^2M_{\rm ap}}$} & 
    \multicolumn{2}{c}{$\ave{{\cal N}M^2_{\rm ap}}$} & 
    \multicolumn{2}{c}{$\ave{{\cal N}^2M_{\rm ap}}$} & 
    \multicolumn{2}{c}{$\ave{{\cal N}M^2_{\rm ap}}$}\\
    Sample & 
    $A_0/10^{-5}$ & \multicolumn{1}{c}{$\alpha$} & $A_0/10^{-7}$ & \multicolumn{1}{c}{$\alpha$} & 
    $A_0/10^{-5}$ & \multicolumn{1}{c}{$\alpha$} & $A_0/10^{-7}$ & \multicolumn{1}{c}{$\alpha$}\\\hline\\ 
    L1 & $10.57_{-4.90}^{+5.12}$ &  $1.40_{-0.34}^{+0.90}$
       & $54.38_{-20.06}^{+25.92}$ & $2.59_{-0.59}^{+0.94}$ 
       &  --              & --
       &  --              & -- \\
    L2 & $5.15_{-0.99}^{+1.05}$ & $1.04_{-0.16}^{+0.18}$
       & $9.48_{-11.37}^{+8.30}$ & $2.11_{-0.73}^{+1.59}$
       &  --              & --
       &  --              & -- \\
    L3 & $3.66_{-0.98}^{+1.04}$ & $1.05_{-0.24}^{+0.31}$ 
       & $28.22_{-6.63}^{+7.95}$  & $1.87_{-0.29}^{+0.44}$
       & $8.56_{-2.75}^{+3.78}$ & $1.66_{-0.41}^{+0.84}$ 
       & $41.07_{-21.95}^{+27.18}$ & $3.84_{-1.06}^{+0.71}$  \\
    L4 & $5.10_{-0.95}^{+1.04}$ & $0.97_{-0.17}^{+0.20}$
       & $16.76_{-6.88}^{+8.95}$  & $1.95_{-0.51}^{+1.14}$
       & $6.85_{-1.34}^{+1.50}$ &  $1.49_{-0.23}^{+0.29}$
       & $31.90_{-13.26}^{+18.06}$ & $3.30_{-0.87}^{+0.95}$ \\
    L5 &  $12.30_{-2.33}^{+2.47}$ & $1.44_{-0.18}^{+0.23}$
       &  $34.56_{-10.48}^{+14.41}$ & $1.99_{-0.37}^{+0.80}$
       &  $9.15_{-1.70}^{+1.80}$  & $1.50_{-0.20}^{+0.26}$
       &  $-8.89_{-18.56}^{+14.23}$ & $3.58_{-1.33}^{+0.88}$  \\
    L6 &  $35.54_{-7.80}^{+9.15}$ & $1.91_{-0.26}^{+0.35}$ 
       &  $62.85_{-18.24}^{+25.33}$ & $2.10_{-0.36}^{+0.58}$
       &  $19.89_{-5.48}^{+6.28}$ & $1.68_{-0.30}^{+0.31}$ 
       &  $42.07_{-21.14}^{+29.43}$ & $3.43_{-1.02}^{+0.93}$  \\
       \\
   sm1 & $11.95_{-2.42}^{+2.57}$ & $1.29_{-0.18}^{+0.22}$
       & $42.70_{-12.91}^{+19.94}$  & $2.16_{-0.46}^{+0.95}$ 
       & $9.34_{-2.94}^{+4.01}$ & $1.55_{-0.34}^{+0.52}$ 
       & $28.69_{-20.40}^{+28.52}$ & $3.55_{-1.26}^{+0.91}$ \\
   sm2 & $19.91_{-2.96}^{+3.09}$ & $1.24_{-0.14}^{+0.17}$
       & $38.15_{-12.24}^{+16.87}$ & $1.90_{-0.39}^{+0.82}$
       & $14.48_{-3.22}^{+3.63}$ & $1.72_{-0.23}^{+0.33}$
       & $25.94_{-18.47}^{+22.36}$ & $3.50_{-1.08}^{+0.89}$ \\
   sm3 & $24.49_{-4.91}^{+5.22}$ &  $1.54_{-0.16}^{+0.18}$
       & $45.06_{-14.01}^{+16.95}$ & $2.09_{-0.36}^{+0.85}$
       & $11.11_{-3.00}^{+3.28}$ & $1.39_{-0.24}^{+0.26}$
       & $-4.83_{-19.77}^{+16.48}$ & $3.52_{-1.30}^{+0.91}$\\
   sm4 & $34.96_{-6.40}^{+6.88}$ & $1.65_{-0.19}^{+0.25}$
       & $61.43_{-16.56}^{+19.67}$ &$1.95_{-0.29}^{+0.51}$ 
       & $26.16_{-5.50}^{+5.84}$ & $2.10_{-0.24}^{+0.32}$
       & $31.46_{-16.94}^{+22.60}$ & $3.10_{-0.88}^{+1.04}$\\
   sm5 & $87.19_{-25.14}^{+33.76}$ & $1.87_{-0.27}^{+0.65}$
       & $77.89_{-22.99}^{+26.49}$ & $1.96_{-0.29}^{+0.45}$
       & $47.34_{-13.26}^{+14.86}$  & $1.79_{-0.24}^{+0.24}$
       & $63.93_{-35.21}^{+46.24}$ & $3.61_{-1.03}^{+0.82}$\\
   sm6 & $197.06_{-347.35}^{+313.36}$ & $3.60_{-1.02}^{+0.84}$
       & $124.38_{-64.46}^{+83.56}$ & $3.21_{-0.91}^{+1.00}$
       & $400.32_{-189.83}^{+212.89}$ &  $3.68_{-0.73}^{+0.72}$ 
       & $53.51_{-68.50}^{+88.69}$ & $3.71_{-1.25}^{+0.81}$\\
   sm7 & $315.06_{-274.91}^{+307.91}$ & $3.26_{-1.32}^{+1.04}$
       & $230.78_{-198.85}^{+265.46}$ & $3.81_{-1.19}^{+0.74}$ 
       & $-136.21_{-483.19}^{+521.62}$ & $3.14_{-1.15}^{+1.08}$  
       & $407.91_{-276.37}^{+307.45}$ &  $4.09_{-1.11}^{+0.57}$ \\
       \\
   ETG & $23.27_{-2.97}^{+3.05}$  & $1.49_{-0.13}^{+0.16}$
       & $41.94_{-9.29}^{+10.22}$ & $1.71_{-0.21}^{+0.29}$ 
       & $21.51_{-2.74}^{+2.85}$ & $1.72_{-0.16}^{+0.19}$
       & $19.73_{-12.41}^{+16.79}$ & $3.03_{-1.03}^{+1.12}$ \\
   LTG & $10.82_{-10.46}^{+13.05}$  &$3.64_{-1.16}^{+0.82}$ 
       & $59.49_{-35.02}^{+42.99}$ &  $3.86_{-1.04}^{+0.70}$
       & $3.25_{-4.51}^{+5.85}$ &  $2.81_{-1.52}^{+1.36}$
       & $-26.87_{-23.95}^{+18.66}$ & $3.41_{-1.19}^{+0.94}$
  \end{tabular}
\end{table*}

\subsection{Power-law fits to measurements}

For aperture scale radii larger than \mbox{$\theta\gtrsim2^\prime$},
our measurements are reasonably consistent with a simple power
law. Therefore, we fit power laws $\ave{{\cal N}^nM_{\rm
    ap}^m}(\theta_{\rm ap})=A_0(\theta/1^\prime)^{-\alpha}$ to data
points within \mbox{$1^\prime\!\!.5\le\theta<10^\prime$} to quantify
the measured profiles of the statistics $n=2,m=1$ and $n=1,m=2$, see
Table \ref{tab:apstatfits}. The fit starts at 1.5 arcmin in order not
to be too strongly influenced by the transformation bias. Fits use the
Jackknife covariance matrices based on the measurements in the
different pointings, as in Eq. \Ref{eq:jackknife}. For the fit, a
multivariate Gaussian noise model for the measurement errors is
adopted.  The quoted values indicate the posterior median and a $68\%$
credibility region about the median for the amplitude $A_0$ and slope
$\alpha$. The posterior adopts a top-hat prior for the power-law
slope, only allowing values within $\alpha\in[0,5]$.

Fig. \ref{fig:slopes} depicts the dependence of the slope $\alpha$ on
the lens $M_r$ magnitude and stellar mass for $\ave{{\cal N}^2M_{\rm
    ap}}$ of all samples with at least a $95\%$ confidence detection
(Table \ref{tab:nulltest}). We find a clear trend towards steeper
slopes (steeper equilateral bispectra) for more luminous galaxies and
galaxies with higher stellar mass. Note that sm6 and sm7 contain on
average galaxies more luminous than these of L5 and L6 (Table
\ref{tab:samples}). The figure also depicts the measured slopes for
the ETG samples, which are consistent with the L-subsamples of
comparable $M_r$ luminosity (between L5 and L6). Slopes weakly
constrained by the data have posterior medians that are drawn towards
the centre of the top-hat prior, $\alpha=2.5$. This mainly applies to
the noisier $\ave{{\cal N}M^2_{\rm ap}}$ measurements, for which
reason they are not included in Fig. \ref{fig:slopes} but are listed
in Table \ref{tab:apstatfits}.

\subsection{Normalised measurements}
\label{sect:calibration}

The G3L aperture statistics are directly related to the 3D
matter-galaxy cross-bispectra and the redshift distribution of lenses
and sources (Sect. \ref{sect:3dbispectra}). The radial galaxy
distributions and the fiducial cosmology define a smoothing kernel in
radial and transverse direction. To disentangle, to lowest order, the
dependence of the signal on the physically relevant bispectrum from
the dependence on source or lens distribution, we introduce a
normalisation scheme.

\begin{figure*}
  \begin{center}
    \psfig{file=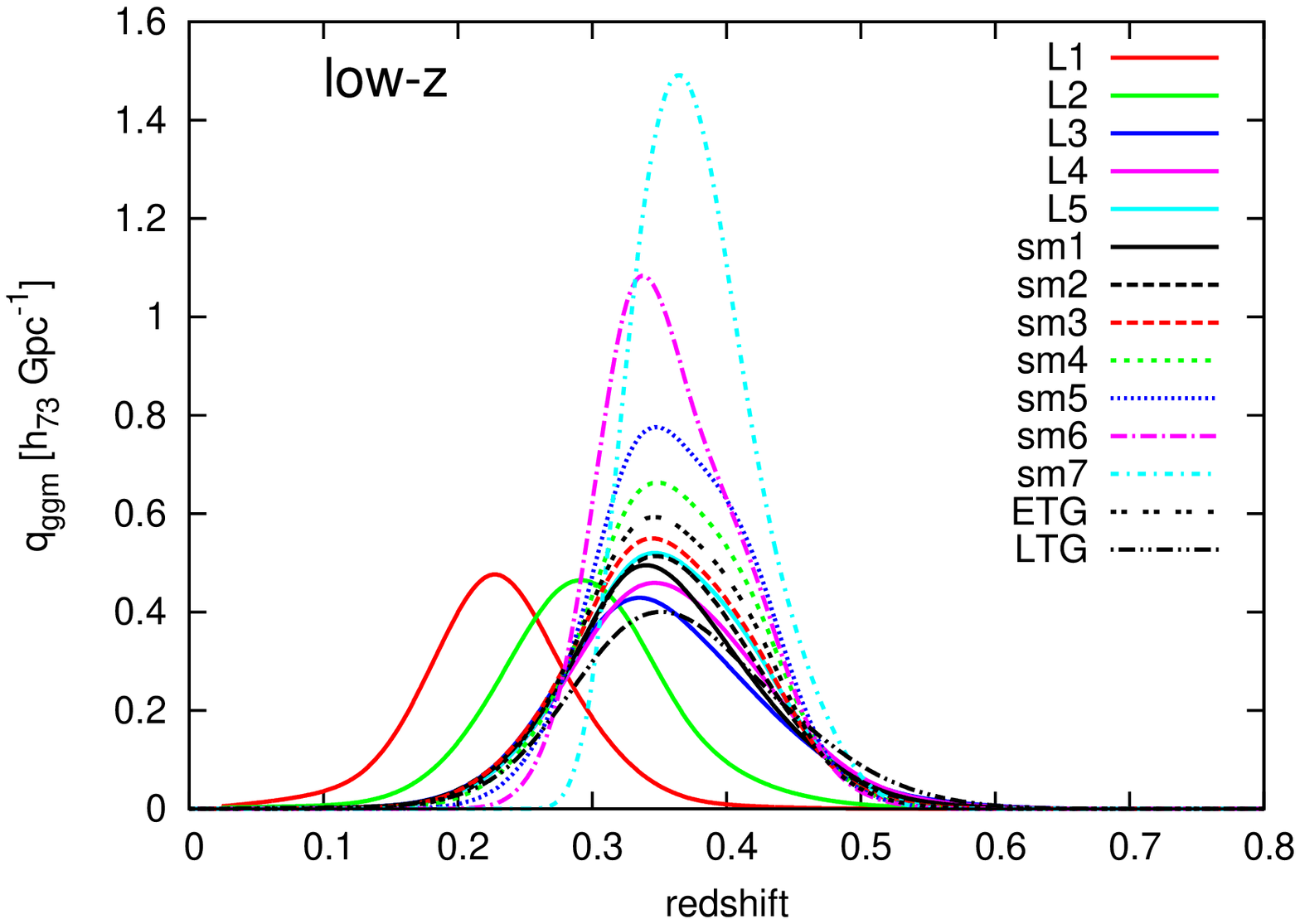,width=55mm,angle=-90}
    \psfig{file=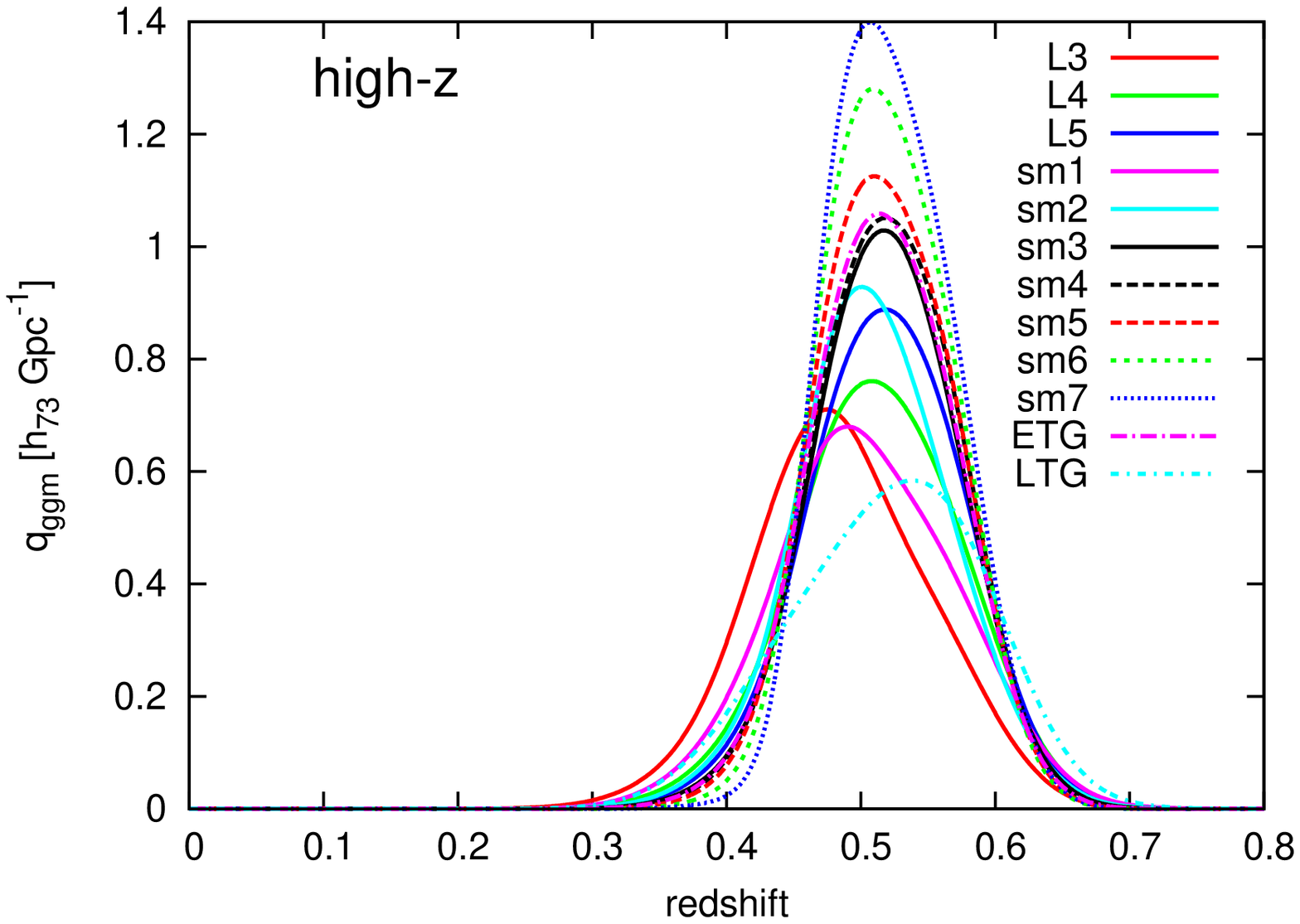,width=55mm,angle=-90}
    \\
    \psfig{file=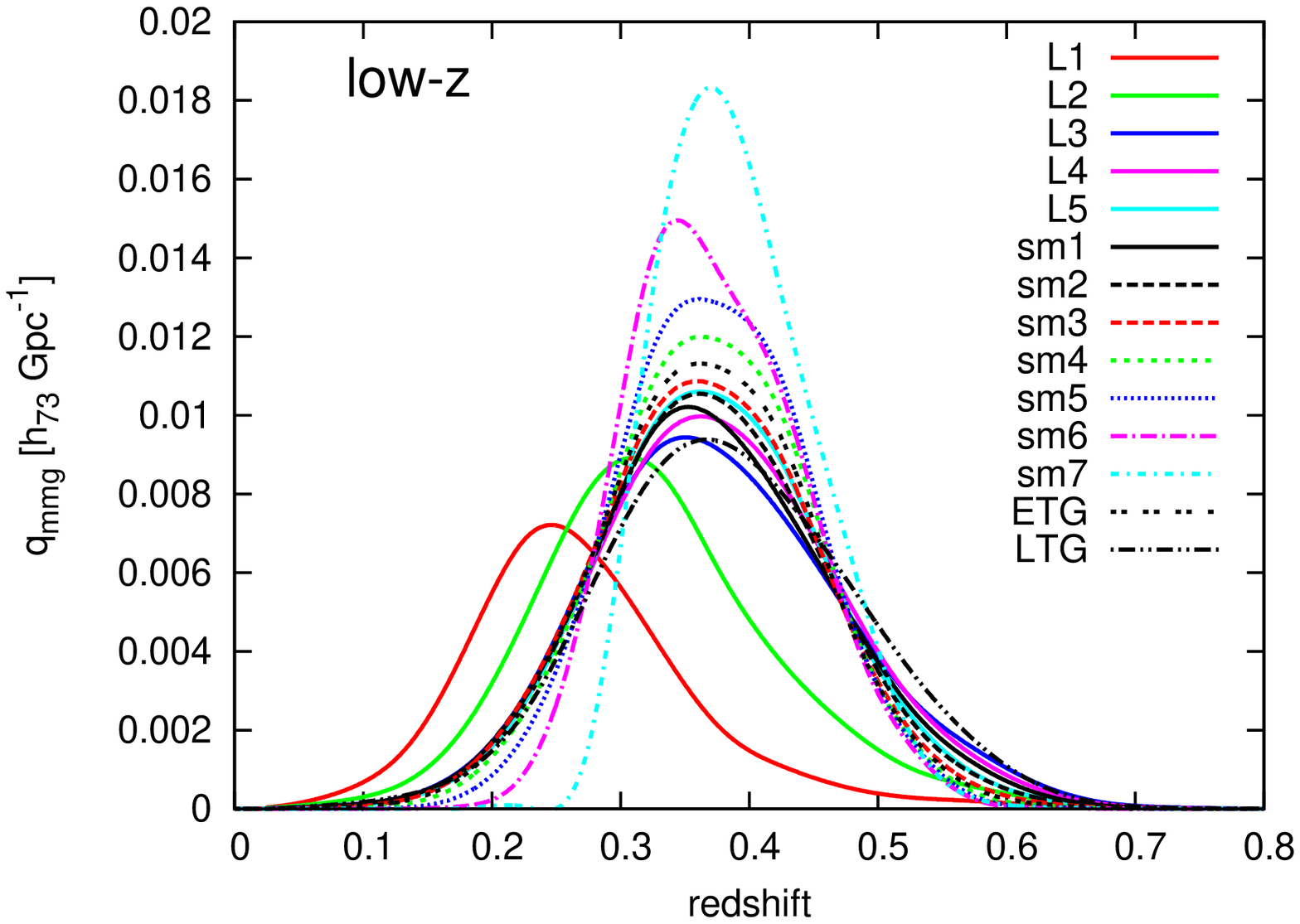,width=55mm,angle=-90}
    \psfig{file=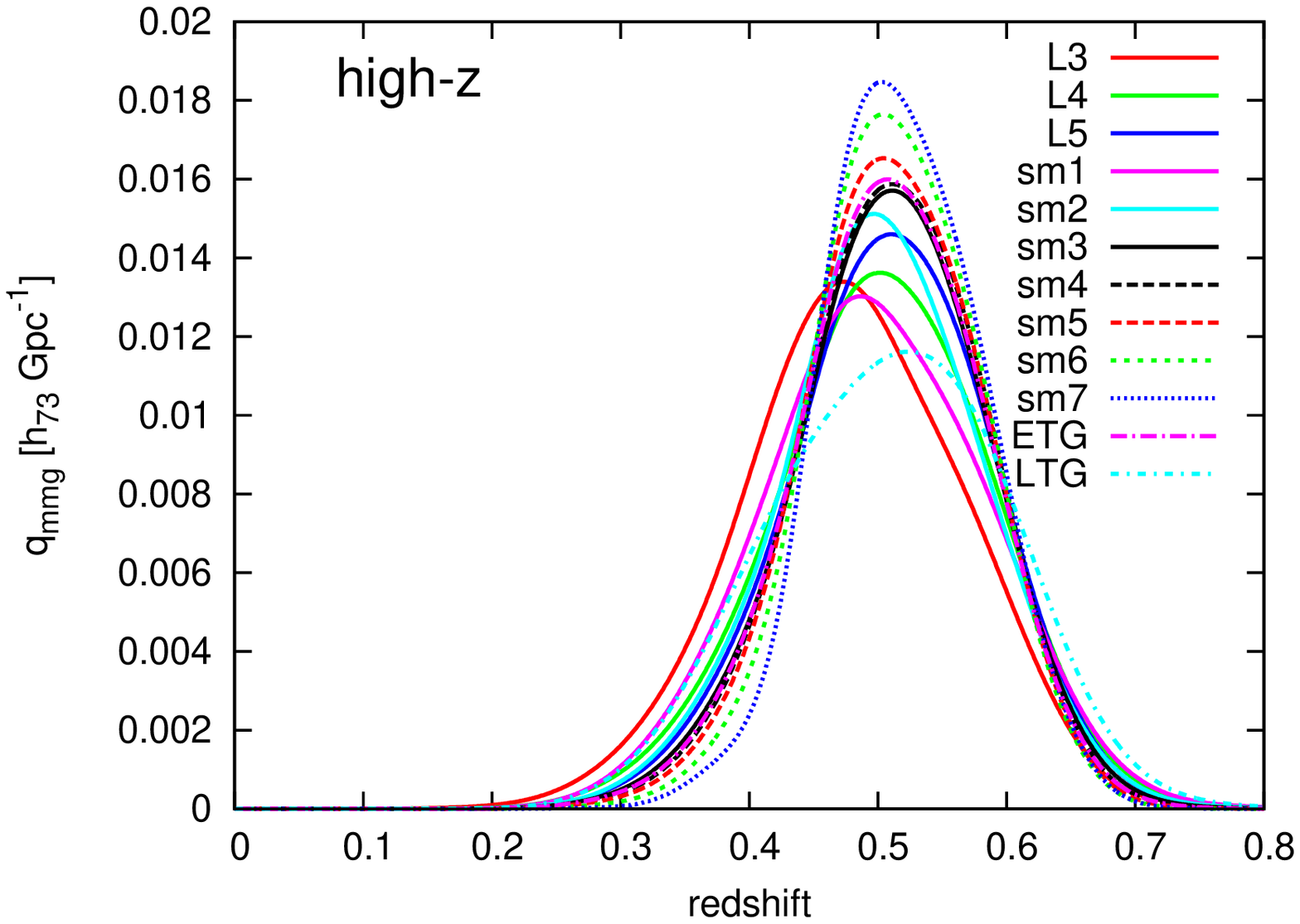,width=55mm,angle=-90}
  \end{center}
  \caption{\label{fig:sensitivity} 
    Radial weight of the matter-galaxy bispectrum for the different
    lens samples used in this study: $\ave{{\cal N}^2M_{\rm ap}}$ (top
    row) and $\ave{{\cal N}M^2_{\rm ap}}$ (bottom row). The left
    column is for lenses with $z_{\rm photo}\in[0.2,0.44]$ (low-$z$),
    the right column for $z_{\rm photo}\in[0.44,0.6]$ (high-$z$).
    L1-L6: $M_{r}$-luminosities increasing from -17.8 mag to -22.4
    mag; sm1-sm7: increasing stellar masses from
    $7\times10^{9}\,M_\odot$ to $4\times10^{11}\,M_\odot$; ETG:
    early-type galaxies; LTG: late-type galaxies.}
\end{figure*}

Combining Eq. \Ref{eq:n2mapbi} and Eq. \Ref{eq:biggk} shows that
$\ave{{\cal N}^2M_{\rm ap}}$ constitutes a radially and transversely
weighted average of the 3D bispectrum $B_{\rm
  ggm}(\vec{k}_1,\vec{k}_2,\chi)$, namely
\begin{eqnarray}
  \label{eq:calcalc}
 \lefteqn{\ave{{\cal N}^2M_{\rm ap}}(\theta_1;\theta_2;\theta_3)}\\
 &&\nonumber
 \!\!\!\!=\frac{3\Omega_{\rm m}}{2D_{\rm H}^2}\int_0^{\chi_{\rm H}}\!\!\!\d\chi
\frac{g(\chi)p_{\rm f}^2(\chi)}{f_{\rm K}^3(\chi)a^2(\chi)} 
\int\frac{\d^2\ell_1\d^2\ell_2\d^2\ell_3}{(2\pi)^6}\\
 &&\nonumber
 ~~\times\tilde{u}(\ell_1\theta_1)\tilde{u}(\ell_2\theta_2)\tilde{u}(\ell_3\theta_3)\\
 &&\nonumber
 ~~\times(2\pi)^2\delta_{\rm D}(\vec{\ell}_1+\vec{\ell}_2+\vec{\ell}_3)
 B_{\rm ggm}\big(\frac{\vec{\ell}_1}{f_{\rm K}(\chi)},
 \frac{\vec{\ell}_2}{f_{\rm K}(\chi)},\chi\big)\;.
\end{eqnarray}
By changing the integration variables as in $\vec{\ell}_i=f_{\rm
  K}(\chi)\vec{k}_i$ we write this integral as
\begin{eqnarray}
  \label{eq:apstatint}
  \lefteqn{\ave{{\cal N}^2M_{\rm ap}}(\theta_1;\theta_2;\theta_3)}\\
  &&\nonumber
  \!\!\!=:\int_0^{\chi_{\rm h}}\d\chi\,
  q_{{\rm ggm}}(\chi)\\
  &&\nonumber
  ~~~\times\overline{B}_{\rm
    ggm}\big(
  \frac{1}{f_{\rm K}(\chi)\theta_1},
  \frac{1}{f_{\rm K}(\chi)\theta_2},
  \frac{1}{f_{\rm K}(\chi)\theta_3},\chi\big)\;,
\end{eqnarray}
for which we introduce the $u$-filtered bispectrum
\begin{eqnarray}
  \label{eq:biband}
  \lefteqn{
    \overline{B}_{\rm ggm}\!\left(\frac{1}{R_1},\frac{1}{R_2},\frac{1}{R_3},\chi\right)}\\
  \nonumber&&
  \!\!\!\!\!\!\!\!\!:=
  \int\frac{\d^2k_1\d^2k_2}{(2\pi)^4D_{\rm H}^2}\,
  \big(\tilde{u}(k_1R_1)\tilde{u}(k_2R_2)\tilde{u}(|\vec{k}_1+\vec{k}_2|R_3)\\
  \nonumber&&
  ~~~~~~~~~~~~~~\times B_{\rm ggm}(\vec{k}_1,\vec{k}_2,\chi)\big)\;.  
\end{eqnarray}
As implied by \Ref{eq:apstatint}, the lensing aperture statistics is
basically the transversely $u$-filtered bispectrum $\overline{B}_{\rm
  ggm}$ averaged in radial direction by the kernel $q_{\rm
  ggm}(\chi)$. For equally-sized aperture radii
$\theta_1=\theta_2=\theta_3$, the $u$-filter gives most weight to the
equilateral bispectrum $B_{\rm ggm}(\vec{k}_1,\vec{k}_2,\chi)$ with
$k_1=k_2=|\vec{k}_1+\vec{k}_2|$, but also mixes in other triangle
configurations to some degree. The radial weighting kernel,
\begin{equation}
  \label{eq:qggm}
  q_{{\rm ggm}}(\chi):=
  \frac{3\Omega_{\rm m}}{2}
  \frac{g(\chi)p_{\rm f}^2(\chi)f_{\rm K}(\chi)}{a(\chi)}\;,
\end{equation}
is peaked at a radial distance $\chi_{\rm max}$ that is determined by
the redshift p.d.f. of lenses and sources (top row of
Fig. \ref{fig:sensitivity}). Therefore, most weight is given to the
bispectrum at distance $\chi_{\rm max}$. 

The kernel $q_{\rm ggm}(\chi)$ is not normalised, i.e.,
$\int\d\chi\,q_{\rm ggm}(\chi)\ne1$, such that the aperture statistics
assumes a value that depends not only on the underlying 3D bispectrum
$B_{\rm ggm}$ but also on the normalisation.  In order to make
measurements comparable for different lens and source samples, we
define a normalised statistics $\cal{B}_{\rm ggm}$ through the
relation
\begin{eqnarray}
  \label{eq:calbggm}
  \lefteqn{\ave{{\cal N}^2M_{\rm ap}}
  (\theta_1;\theta_2;\theta_3)}\\
  &&\nonumber
  ={\cal B}_{{\rm ggm}}\!\left(R_1,R_2,R_3\right)
  \int_0^{\chi_{\rm h}}\d\chi\,q_{{\rm ggm}}(\chi)
\end{eqnarray}
with $R_i:=f_{\rm K}(\chi_{\rm max})\theta_i$. We emphasise that by
this definition ${\cal B}_{\rm ggm}$ is not a deprojection of the
angular aperture statistics to the spatial 3D bispectrum. This would
involve the inversion of the $\chi$-integral. Instead we, in effect,
normalise the statistic by the area $\int\d\chi\,q_{\rm ggm}(\chi)$,
and we convert angular scales to projected physical scales through the
angular diameter distance $f_{\rm K}(\chi_{\rm max})$ at maximum
weight $q_{\rm ggm}(\chi_{\rm max})$. 

The same line of reasoning can be applied to the second G3L aperture
statistics for which we obtain
\begin{eqnarray}
  \label{eq:calbmmg}
  \lefteqn{\ave{{\cal N}M^2_{\rm ap}}
    (\theta_1;\theta_2;\theta_3)}\\
  &&\nonumber
  ={\cal B}_{{\rm mmg}}\!\left(R_1,R_2,R_3\right)
  \int_0^{\chi_{\rm h}}\d\chi\,q_{{\rm mmg}}(\chi)\;,
\end{eqnarray}
with its own radial filter
\begin{equation}
  \label{eq:qmmg}
  q_{{\rm mmg}}(\chi):=
  \frac{9\Omega^2_{\rm m}}{4D^2_{\rm H}}
  \frac{g^2(\chi)p_{\rm f}(\chi)f^2_{\rm K}(\chi)}{a^2(\chi)}\;.
\end{equation}
Examples of kernels $q_{\rm mmg}(\chi)$ relevant for this work are
depicted in the bottom row of Fig. \ref{fig:sensitivity}.

By definition the aperture statistics $\ave{{\cal N}^nM_{\rm ap}^m}$,
as moments of smoothed density contrasts on the sky, are
dimensionless. As $q_{\rm ggm}(\chi)$ has the dimension $\rm
[length^{-1}]$, Eq. \Ref{eq:qggm}, we deduce from
Eq. \Ref{eq:apstatint} that the $u$-filtered ${\cal B}_{\rm ggm}$ is
also dimensionless. This becomes also obvious from
Eq. \Ref{eq:calbggm} because the normalisation integral is
dimensionless. A similar argument applies to the dimensionless ${\cal
  B}_{\rm mmg}$.

\begin{figure*}
  \begin{center}
    \psfig{file=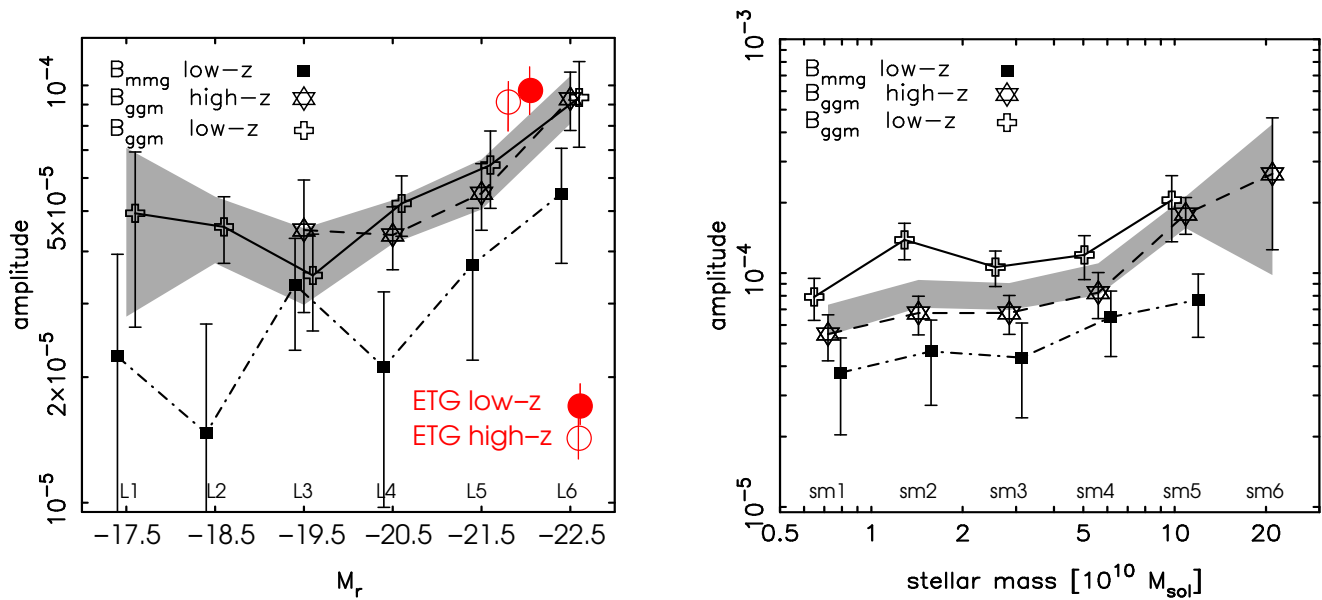,width=175mm,angle=0}
  \end{center}
  \caption{\label{fig:calibrateddata} \verB{Trends of the (significant)
    galaxy-matter bispectra ${\cal B}_{\rm mmg}(R)$ and ${\cal B}_{\rm
      ggm}(R)$ for $k=\sqrt{2}/R=1.03\,h_{73}\rm Mpc^{-1}$ with
    function of $M_r$-luminosity (left panel) and stellar mass (right
    panel). Shown are both redshift bins together.  The left panel
    also includes ${\cal B}_{\rm ggm}(R)$ data points of the
    early-type galaxy sample. These data points are somewhat offset
    with respect to their actual mean $\ave{M_r}=-21.88(-21.91)$
    (high-$z$ in brackets) for clarity.  The shaded area highlights
    the $1\sigma$ constraints of ${\cal B}_{\rm ggm}(R)$ for low-$z$ and
    high-$z$ combined.}}
\end{figure*}

In our analysis, we estimate the equilateral ${\cal B}_{\rm ggm}$ or
${\cal B}_{\rm mmg}$ amplitudes of all samples at a common comoving
length scale of \mbox{$R_i=R_{\rm 1Mpc}=1\,h_{100}^{-1}\rm Mpc$} (or
\mbox{$k\approx\sqrt{2}/R_{\rm 1Mpc}\sim1.4\,h_{100}\rm Mpc^{-1}$} for
the exponential $u$-filter). For this purpose, the power-law fits in
Table \ref{tab:apstatfits} to the aperture statistics are employed,
which essentially describe the data at the scales of interest, to
interpolate in the case of ${\cal B}_{\rm ggm}$
\begin{eqnarray}
  \lefteqn{{\cal B}_{\rm ggm}(R):={\cal B}_{\rm ggm}(R,R,R)}\\
  &&\nonumber
  \!\!\!\!=\underbrace{\frac{A_0}{\int_0^{\chi_{\rm h}}\d\chi\,q_{\rm ggm}(\chi)}  
  \left(\frac{f_{\rm K}(\chi_{\rm
        max})\times1^\prime}{h_{100}^{-1}\,\rm  Mpc}\right)^{+\alpha}}_{=:{\cal B}_{\rm ggm}(R_{\rm 1Mpc})}
  \left(\frac{R}{h_{100}^{-1}\rm Mpc}\right)^{-\alpha}\;.
\end{eqnarray}
We are quoting only the amplitude ${\cal B}_{\rm ggm}(R_{\rm 1Mpc})$
in the following. Likewise for the matter-matter-galaxy bispectrum
${\cal B}_{\rm mmg}$ we have
\begin{eqnarray}
  \lefteqn{{\cal B}_{\rm mmg}(R):={\cal B}_{\rm mmg}(R,R,R)}\\
  &&\nonumber
  \!\!\!\!=\underbrace{\frac{A_0}{\int_0^{\chi_{\rm h}}\d\chi\,q_{\rm mmg}(\chi)}  
  \left(\frac{f_{\rm K}(\chi_{\rm
        max})\times1^\prime}{h_{100}^{-1}\,\rm  Mpc}\right)^{+\alpha}}_{=:{\cal B}_{\rm mmg}(R_{\rm 1Mpc})}
  \left(\frac{R}{h_{100}^{-1}\rm Mpc}\right)^{-\alpha}\;.
\end{eqnarray}
To assess the uncertainty in the bispectrum amplitude, we marginalise
over the uncertainties in $A_0$, the aperture statistics amplitude at
1 arcmin, and $\alpha$, the power-law index, taking into account the
correlation of their errors. A value of $R_{\rm 1Mpc}$ corresponds to
an aperture scale radius between $2.5$ to $3.5\,\rm arcmin$ depending
on the mean redshift of the lens samples. The compiled results are
plotted in Fig. \ref{fig:calibrateddata} -- one of the main results of
our study -- to highlight the trends with $M_r$ magnitude and stellar
mass. As before, only measurements with highly significant detections
are included in the plot.

At the corresponding $M_r$ magnitude range, we include also the
normalised $\ave{{\cal N}^2M_{\rm ap}}$ amplitude of the ETG
sample. Their amplitude is somewhat higher in comparison to L5 and
L6. This can be explained by the fact that the LTG sample is included
in the L samples but not in the ETG sample of similar luminosity: the
LTG have a normalised amplitude considerably smaller than that of the
ETG.

\subsection{Third-order galaxy biasing}

SW05 introduced a set of third-order galaxy biasing parameters
$b_3,r_1,r_2$ to parametrise the galaxy-matter bispectra relative to
the matter bispectrum $B_{\rm mmm}$,
\begin{eqnarray}
  B_{\rm ggm}(\vec{k}_1,\vec{k}_2,\chi)&=&
  b_3^2r_2\,
  B_{\rm mmm}(\vec{k}_1,\vec{k}_2,\chi)\;,\\
  B_{\rm mmg}(\vec{k}_1,\vec{k}_2,\chi)&=&
  b_3r_1\,
  B_{\rm mmm}(\vec{k}_1,\vec{k}_2,\chi)\;.
\end{eqnarray}
The coefficients $b_3,r_1,r_2$ are also functions of
$\vec{k}_1,\vec{k}_2,\chi$, which has been omitted here to save
space. For galaxies faithfully tracing the underlying matter
distribution one finds \mbox{$r_1=r_2=b_3=1$} for all scales. This
parametrisation generalises the earlier similar second-order galaxy
bias parametrisation \citep[e.g.,
][]{2003MNRAS.346..994P,hvg02,1999ApJ...518L..69T} to the
third-order. Our normalised G3L measurements of ${\cal B}_{\rm ggm}(R)$
and ${\cal B}_{\rm mmg}(R)$ can be utilised to constrain the ratio
$r_1/(r_2b_3)$ for a physical scale $R$ by considering the combined
statistics
\begin{eqnarray}
  \label{eq:psi}
  \Psi(R)&:=&\frac{{\cal B}_{\rm mmg}(R)}{{\cal B}_{\rm
  ggm}(R)}\\
 \nonumber&=&
 \frac{\int_0^{\chi_{\rm h}}\d\chi\,q_{\rm ggm}(\chi)}
 {\int_0^{\chi_{\rm h}}\d\chi\,q_{\rm mmg}(\chi)}
 \frac{\ave{{\cal N}M_{\rm ap}^2}\left(R f^{-1}_{\rm K}(\chi_{\rm
       max})\right)}
{\ave{{\cal N}^2M_{\rm ap}}\left(R f^{-1}_{\rm K}(\chi_{\rm
       max})\right)}
\end{eqnarray}
for a lens sample. We assume here that both kernels $q_{\rm ggm}$ and
$q_{\rm mmg}$ peak at the same distance $\chi_{\rm max}$, which is
approximately valid for our study. This bias parameters in
$r_1/(r_2b_3)$ are smoothed in $k$ and $\chi$ with maximum weight at
$k\approx\sqrt{2}/R$ (equilateral triangles) and $\chi_{\rm max}$. The
exact smoothing kernels are given in Appendix \ref{sect:biassmooth}. A
deviation of $\Psi(R)$ from unity hence indicates a biased galaxy
population.

\begin{figure*}
  \begin{center}
    \psfig{file=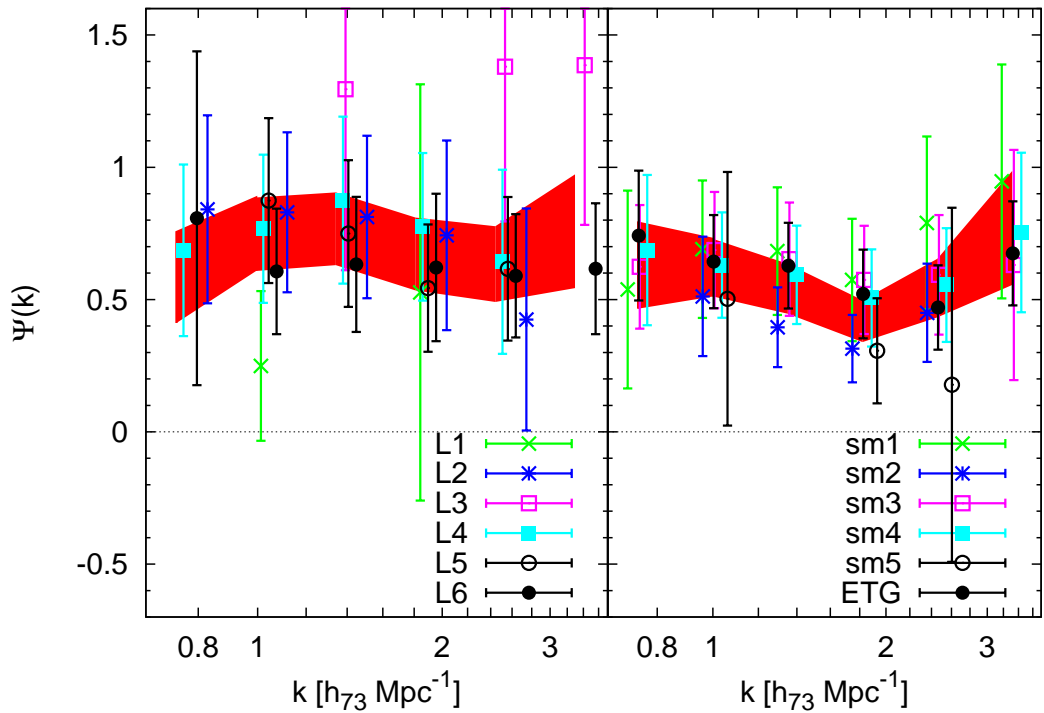,width=100mm,angle=-90}
  \end{center}
  \caption{\label{fig:bias} Results of the $\Psi$-statistics which
    probes the third-order galaxy biasing parameters of our lens
    samples.  Plotted is $r_1/(b_3r_2)$ as function of scale
    $k=\sqrt{2}/R=\sqrt{2}/(f_{\rm K}(\chi_{\rm max})\theta)$ for the
    angular range $1^\prime\le\theta<10^\prime$.  Used are only
    significant measurements in the low-$z$ samples. Luminosity
    samples are in the left panel, stellar mass samples and the ETG
    sample in the right panel. Data points with uncertainties larger
    than $\sigma_\Psi>0.8$ are not shown. The shaded area indicates
    the mean and standard deviation of the mean of combined
    samples. L1-L6: $M_{r}$-luminosities increasing from -17.8 mag to
    -22.4 mag; sm1-sm5: increasing stellar masses from
    $7\times10^{9}\,M_\odot$ to $10^{11}\,M_\odot$; ETG: early-type
    galaxies.}
\end{figure*}

We calculate the $\Psi(R)$ statistics for the samples L1-L6, samples
sm1-sm5, and the ETG sample (all low-$z$ only) for angular scales
between one and ten arcmin. The remaining measurements are too noisy
for useful constraints. Fig. \ref{fig:bias} summarises these novel
measurements. The error distributions of the ratios $\Psi(R)$ are
estimated by employing Monte-Carlo realisations of ${\cal B}_{\rm
  ggm}$ and ${\cal B}_{\rm mmg}$ (assumed Gaussian); depicted are the
mean and variances $\sigma_\Psi$ in the resulting
distributions. Alternatively, one could utilise the analytic
probability distribution function given in \citet{BIOMET1969HINKLEY}.

\section{Discussion}  
\label{sect:discussion}

\verB{We performed a G3L analysis of approximately 100 square degree
  of the CFHTLenS data set.} The data is endowed with photometric
redshifts of galaxies and 
\lensfit estimates of the PSF-corrected source ellipticities. For the
first time, the signal-to-noise of the lensing data is sufficient to
measure third-order galaxy-galaxy lensing as a function of lens
luminosity, stellar mass and galaxy type. The work of
\citet{2008A&A...479..655S}, analysing the RCS1 data, demonstrated
that G3L measurements are principally possible with contemporary
lensing surveys. This is confirmed by this study.
We further subdivided the lens samples in $M_r$-luminosities, stellar
masses, SED types, and a ``low-$z$'' (\mbox{$0.2\le z_{\rm
    photo}<0.44$}) and a ``high-$z$'' (\mbox{$0.44\le z_{\rm
    photo}<0.6$}) redshift bin by utilising the photometric redshifts
of the lenses. We presented the G3L measurements in terms of aperture
statistics that probes the angular bispectrum of the (projected)
matter-galaxy three-point correlations.  In one case
(``lens-lens-shear''), the measurements quantify correlations between
two lens positions and the surface matter density around the lens
pair; this can be interpreted as excess surface mass density about
galaxy pairs \citep{RNSIMON2011}. In the other case
(``lens-shear-shear''), it expresses correlations between a lens
position and the surface mass density in two different directions
close to the lens. The here adopted G3L aperture statistics has the
practical advantage to separate E- and B/P-modes from these
measurements, which is utilised to detect signatures of possible
systematics in the data. On this level, no significant G3L systematics
signals were detected.

\begin{figure}
  \begin{center}
    \psfig{file=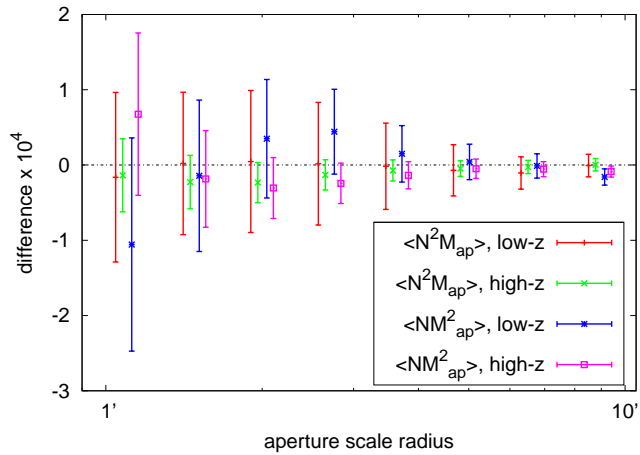,width=62mm,angle=-90}
  \end{center}
  \caption{\label{fig:IIGItest}
    \verB{Plotted are the differences of $\ave{{\cal N}^2M_{\rm
          ap}}(\theta)/\int\d\chi\,q_{\rm ggm}(\chi)$ for sources with
      \mbox{$z_{\rm photo}\ge0.65$} and \mbox{$z_{\rm photo}\ge0.8$},
      and the analogue for $\ave{{\cal N}M^2_{\rm
          ap}}(\theta)/\int\d\chi\,q_{\rm mmg}(\chi)$. For fixed lens
      samples (sm3-5 combined; either low-$z$ or high-$z$), the result
      is expected to be consistent with zero. Note that neighbouring
      error bars are correlated.}}
\end{figure}

To reduce the impact due to intrinsic alignments of sources, we
separated lens and source galaxy samples physically from each other by
exploiting the photometric redshifts in the survey. We showed that in
the ideal case of no radial overlap, neither II-correlations nor
GI-correlations contribute to the correlator. Owing to the uncertainty
in the galaxy redshifts, however, perfectly non-overlapping
distributions are hard to achieve.  We found that our low-$z$ lens
samples have still a small overlap of $\sim4\%$ with the source
sample, the high-$z$ samples a moderate overlap of $\sim12\%$ (overlap
of redshift probability distribution functions). Because of the small
overlap, at least for the low-$z$ samples we do not expect significant
contributions from intrinsic alignment correlations.
\verB{To test the degree of contamination by GI- and II-correlations,
  we compared the aperture statistics of the combined sm3-sm5 samples,
  both low-$z$ (\mbox{$\bar{z}_{\rm d}=0.35$}) and high-$z$
  (\mbox{$\bar{z}_{\rm d}=0.51$}), for two cases. In one case, we
  selected sources by \mbox{$0.65\le z_{\rm photo}<1.2$}
  (\mbox{$\bar{z}_{\rm s}=0.93$}) as before. In the second case, we
  were more conservative by selecting only sources within
  \mbox{$0.8\le z_{\rm photo}<1.2$} ($\bar{z}_{\rm s}=1.02$), thereby
  discarding about one third of our sources. However, the latter case
  reduced the overlap from $3.3\%$ ($10.9\%$) to $0.6\%$ ($2.5\%$) for
  the low-$z$ (high-$z$) sample. The statistics were normalised by
  $\int\d\chi q_{\rm ggm}(\chi)$ ($\int\d\chi q_{\rm mmg}(\chi)$) to
  compensate the signal change owing to the different source
  depths. The maximum signal increase is \mbox{$\sim30\%$} for
  $\ave{{\cal N}M_{\rm ap}^2}$ of the high-$z$ lens
  sample. Fig. \ref{fig:IIGItest} shows the difference in normalised
  statistics for fixed lens samples but varying source depths. Here we
  assumed that $f_{\rm K}(\chi_{\rm max})$ is identical for both
  compared signals, i.e., both signals were differenced at the same
  aperture scale radius. This assumption is accurate within a few
  percent here. As expected, the difference is consistent with zero,
  the level of GI/II-systematics in the statistics is therefore
  negligible within the measurement errors.}

As second possible source of systematics we identified the
magnification of the lens number densities by matter fluctuations in
front of the lenses. To first order this effect is comparable to the
aperture mass skewness $\ave{M_{\rm ap}^3(\theta)}$ associated with
sources at redshift $z_{\rm s}\sim0.4$. We estimated this effect to be
of the order of $\lesssim10^{-8}$ at $\theta\sim1$ arcmin.  We
therefore conclude in comparison with our measurements that in the
range $1^\prime\lesssim\theta\lesssim10^\prime$ the lens number
density magnification effect is negligible for both $\ave{{\cal
    N}^2M_{\rm ap}}$ and for $\ave{{\cal N}M^2_{\rm ap}}$. For a more
detailed investigation of systematics, however, realistic models of
intrinsic alignments and magnification for third-order correlations
are required, which are currently unavailable.

The observed aperture statistics depend on the redshift distributions
of lenses and sources. Results of the statistics would hence differ
when changing the source or lens redshift distribution, even if the
underlying comoving spatial 3D matter-galaxy bispectrum were unchanged
throughout the light-cone. In order to partially correct for this
effect, we employed a new technique that normalises the aperture
statistics with the lensing efficiency and relates angular scales to
effective spatial scales; this yields two kinds of galaxy-matter
bispectra ${\cal B}_{\rm ggm}(R)$ and ${\cal B}_{\rm mmg}(R)$,
originating from $\ave{{\cal N}^2M_{\rm ap}}$ and $\ave{{\cal N}M_{\rm
    ap}^2}$, respectively.  The normalised statistics obtained are
basically band bispectra due to the smoothing of the exponential
$u$-filter in $k$-space and the radial smoothing of the lensing
kernels. 
Only by means of our or similar normalisation schemes,
measurements for different lens samples or same galaxy selections at
different redshifts become comparable. In particular, the application
of our normalisation simplifies the comparison with results from
future studies. The problem of unnormalised measurements becomes
particularly obvious for $\ave{{\cal N}^2M_{\rm ap}}$ of the low-$z$
L1 sample in Fig. \ref{fig:n2map2} (top left panel) in comparison to
the L6 sample in the same panel: for \mbox{$\theta\gtrsim2$ arcmin}
both measurements are basically identical, although the normalisation
reveals that the lower luminosity galaxies have a smaller bispectrum
amplitude (Fig. \ref{fig:calibrateddata}, left panel). This effect
results from a completely different redshift distribution of the L1
lenses that, due to sample incompleteness, have a mean redshift of
$z\sim0.22$ instead of L6's $z\sim0.34$.

We estimated the measurement errors directly from the data by
Jackknifing the signal variance across the survey pointings. Ideally,
with statistically independent pointings this would properly account
for uncertainties due to source shape noise, sampling noise and cosmic
variance. However, the pointings are bundled together in large fields
W1-W4 with extensions of several degrees across the sky. This makes
pointings of the same field partly correlated.
Therefore, the cited uncertainties are probably somewhat too
optimistic in the sense that they underestimate the cosmic variance.

To refine the previous RCS1 measurement in \citet{2008A&A...479..655S}
for different galaxy populations and to investigate the dependences of
bispectra amplitudes on galaxy populations, we focussed here on
equally-sized apertures. This gives most weight to the equilateral
bispectra. We found that the aperture statistics are reasonably well
described by a power law on angular scales ranging from roughly two
arcmin to ten arcmin. On smaller angular scales, we observe evidence
for a change of slope, but we are also increasingly affected by the
transformation bias. For instance, $\ave{{\cal N}^2M_{\rm ap}}$ of the
fainter low-$z$ L4 sample clearly flattens below $\sim2$ arcmin.
Qualitatively, this behaviour is also observed in the semi-analytic
galaxy models studied in \citet{2012arXiv1204.2232S}, see their
Fig. 8. A similar change of slope, maybe also a steepening, is visible
for the $\ave{{\cal N}M^2_{\rm ap}}$ statistics of the luminosity
samples. A comparison of galaxy models to our measurements requires a
careful replication of the galaxy sample selections, their
uncertainties, and the survey incompleteness. Moreover, as concluded
in \citet{2012arXiv1204.2232S}, no reliable galaxy model is currently
available to predict the correct amplitude of G3L measurements -- or
to even double-check whether our results may be strongly effected by
galaxy selection effects. We hence postpone this task to a future
paper.

\begin{table}
  \caption{\label{tab:ETLTG} \verB{Values of the normalised galaxy-matter
    bispectra  ${\cal B}_{\rm ggm}(R)$ (top half) and ${\cal B}_{\rm
      mmg}(R)$ (bottom half) at $R=1\,h_{100}^{-1}\rm Mpc$
    for the late-type (LTG) and early-type (ETG) galaxy sample.}}
  \begin{center}
  \begin{tabular}{lcc}
    Sample & low-$z$ & high-$z$
    \\\hline\hline\\
    ETG & $+1.04_{-0.17}^{+0.17}\times10^{-4}$ &
    $+9.48_{-1.23}^{+1.19}\times10^{-5}$ \\
    LTG & $+4.37_{-5.01}^{+9.21}\times10^{-6}$ &
    $+1.07_{-1.02}^{+1.67}\times10^{-5}$
    \\\hline
    ETG & $+5.98_{-1.41}^{+1.35}\times10^{-5}$ &
    $+1.42_{-0.92}^{+1.37}\times10^{-5}$\\
    LTG & $+6.93_{-4.46}^{+9.91}\times10^{-6}$ &
    $-1.39_{-1.62}^{+0.95}\times10^{-5}$
  \end{tabular}
  \end{center}
\end{table}

The measurements of ${\cal B}_{\rm ggm}$ utilising ETG and LTG pairs
(subdivision of the combined L5/L6 sample) show that the excess mass
around pairs is a strong function of galaxy type. The high excess mass
signal of ETG is comparable to the strong signal of pairs in our
sm-samples with stellar masses of $\sim10^{11}\,M_\odot$, whereas the
excess mass of LTG is consistent with zero in our case (Table
\ref{tab:ETLTG}). A plausible explanation for this is the fact that
many early-type galaxies in the ETG sample
($\ave{M_\ast}\approx6\times10^{10}\,M_\odot$) are satellites in dense
cluster environments, whereas LTG are frequently field galaxies. This
was, for instance, found by the GGL study of
\citet{2006MNRAS.368..715M} in SDSS. The splitting into ETG from
over-dense and under-dense regions that was conducted in this study is
actually comparable to the lens-lens-shear correlation function
because the G3L correlator gives more weight to pairs in cluster
environments, simply because more pairs are found in these
regions. Recently, \citet{2012arXiv1204.2232S} studied the excess mass
for two state-of-the-art semi-analytic galaxy models. Although the G3L
amplitudes and colour distributions of the two considered models are
inconsistent, both models predict a large difference in $\ave{{\cal
    N}^2M_{\rm ap}}$ for $z_{\rm d}=0.17$ red and blue galaxies up to
a factor of $\sim10^2$ at $\theta\sim2\,\rm arcmin$. With our
uncertainties in the LTG signal, we estimate the difference to be at
least a factor $\sim10$, strongly confirming the previous prediction.

By forming ratios of normalised bispectra, our new statistic
$\Psi(R):={\cal B}_{\rm mmg}(R)/{\cal B}_{\rm ggm}(R)$ approximately
yields the ratio $r_1/(r_2b_3)$ of (smoothed) third-order biasing
parameters (SW05).  The details of the smoothing are determined by the
shapes of the $u$-filter, the peaked kernels $q_{\rm mmg}, q_{\rm
  ggm}$, and to some extent also the matter bispectrum (Appendix
\ref{sect:biassmooth}). Deviations of $\Psi$ from unity indicate
galaxies that not faithfully trace the underlying matter density
field, i.e., biased galaxies. This new technique for investigating
galaxy bias with lensing advances the methodology of
\citet{1998A&A...334....1V} and \citet{1998ApJ...498...43S} that focus
on second-order galaxy bias. The application of the latter found that
galaxies are generally biased tracers
\citep{2001ApJ...558L..11H,2007A&A...461..861S,2012ApJ...750...37J}. We
confirm this finding by employing third-order galaxy-matter
correlations.

\section{Conclusions}
\label{sect:conclusions}

\begin{itemize}
\item We detect G3L with unprecedented high significance in the
  CFHTLenS for galaxy populations of different luminosity, stellar
  mass, and SED type. This applies to both third-order
  galaxy-galaxy-matter correlations (${\cal B}_{\rm ggm}$) and
  galaxy-matter-matter correlations (${\cal B}_{\rm mmg}$).
\item We find that the (equilateral) galaxy-matter bispectra are,
  within the remaining statistical errors, reasonably well
  scale-invariant for the spatial (comoving) scales
  \mbox{$0.3\,{\rm Mpc}^{-1}\lesssim k\lesssim 2.2\,{\rm
      Mpc}^{-1}$}. On smaller scales, not included in our power-law
  fits, there are indications of deviations from the power-law shape
  in several cases.
\item The low-$z$ and high-$z$ counterparts of the same lens samples
  yield very similar bispectra amplitudes ${\cal B}_{\rm ggm}$
  (Fig. \ref{fig:calibrateddata}) and slopes
  (Fig. \ref{fig:slopes}). This points to little evolution of the
  bispectrum between redshift $z\sim0.3-0.5$, especially for our
  $M_{r}$-selected galaxies. There is, however, some evidence for a
  change in the amplitude of ${\cal B}_{\rm ggm}$ for stellar-mass
  selected galaxies below $\sim10^{11}\,\msol$: high-$z$ lenses show a
  lower amplitude than the low-$z$ lenses (right panel of
  Fig. \ref{fig:calibrateddata}). This implies an increase of excess
  mass about pairs of galaxies of fixed stellar mass with time, as,
  e.g., may be expected in a CDM scenario due to the continuous
  accretion of matter. Evolution trends of ${\cal B}_{\rm mmg}$ are
  unclear due to the measurement uncertainties in the high-$z$
  samples.
\item For ${\cal B}_{\rm ggm}$ the slope and the amplitude is a
  changing function of galaxy luminosity, stellar mass and galaxy
  type.  The amplitude change is also observed for ${\cal B}_{\rm
    mmg}$. Brighter or more massive galaxies (by stellar mass) exhibit
  a steeper bispectrum, which implies that the excess mass is more
  concentrated in these cases. Moreover, there is a clear trend
  towards higher amplitudes for both more luminous and more massive
  galaxies. This shows that more luminous or massive galaxies, or
  galaxy pairs in the case of ${\cal B}_{\rm ggm}$, inhibit denser
  environments than fainter or lighter galaxies.
\item We observe a strong signal for the excess mass around early-type
  galaxies (ETG) pairs. Late-type galaxy pairs (LTG), on the other
  hand, have a signal that is consistent with zero
  \verB{when studied as aperture statistics}. This remarkable
  observation is in excellent agreement with the recent prediction of
  \citet{2012arXiv1204.2232S} based on semi-analytic galaxy models.
  The measurement therefore suggests that virtually all signal in this
  magnitude range originates from ETG pairs, and possibly mixed pairs
  of ETG and LTG, rather than from LTG pairs. This can be explained by
  the fact that a large fraction of ETG are satellite galaxies in
  cluster.
  \verB{By explicitly mapping out the excess mass around LTG and ETG galaxy pairs we have also found that both maps are fundamentally different in their amplitudes as well as in their general appearance.}
\item
  \verB{The mismatch between ${\cal B}_{\rm ggm}$ and ${\cal B}_{\rm
      mmg}$ for the same lens galaxy sample immediately indicates galaxies biased
    with respect to the matter distribution.
  This mismatch is captured by the galaxy bias statistics $\Psi(R)$
  (Fig. \ref{fig:bias}) that shows for our low-$z$ samples values
  comparable for a wide range of galaxy luminosities and stellar
  masses.} Therein, we probe the non-linear regime on scales smaller
  than $k\sim0.8\,\rm Mpc^{-1}$. We find best constraints on $\Psi(R)$
  with the stellar mass samples, which has for all samples sm1-sm5 and
  scales combined (minimum-variance weighted) an average value of
  $\overline{\Psi}=0.51\pm0.07$.  This shows -- for the first time
  employing third-order lensing statistics -- that galaxies are biased
  tracer of the matter density field. Although $\Psi(R)$ indicates
  that the ratio $r_1/(r_2b_3)$ stays relatively constant with scale,
  with a possible shallow local minimum at $k\approx1.8\,\rm
  Mpc^{-1}$, the additionally observed change of the bispectrum
  amplitudes with galaxy luminosity or mass means that the individual
  bias parameters have to differ between the galaxy samples.
\item Finally, we emphasise that theory is lacking behind in
  interpreting the G3L measurements. Reliable model predictions
  \verB{, e.g., in the vein of \citet{2003MNRAS.340..580T},} are
  needed, not only to properly interpret the measurements, but also to
  gain a better understanding of systematics and to verify that
  selection effects in the data do not spoil the measurement.
\end{itemize}


\section*{Acknowledgements}

We thank Stefan Hilbert for useful discussions and Hananeh Saghiha
for verifying our correlator code output with hers on simulated data.

This work is based on observations obtained with MegaPrime/MegaCam, a
joint project of CFHT and CEA/IRFU, at the Canada-France-Hawaii
Telescope (CFHT) which is operated by the National Research Council
(NRC) of Canada, the Institut National des Sciences de l'Univers of
the Centre National de la Recherche Scientifique (CNRS) of France, and
the University of Hawaii. This research used the facilities of the
Canadian Astronomy Data Centre operated by the National Research
Council of Canada with the support of the Canadian Space Agency.  We
thank the CFHT staff for successfully conducting the CFHTLS
observations and in particular Jean-Charles Cuillandre and Eugene
Magnier for the continuous improvement of the instrument calibration
and the {\sc Elixir} detrended data that we used. We also thank
TERAPIX for the quality assessment and validation of individual
exposures during the CFHTLS data acquisition period, and Emmanuel
Bertin for developing some of the software used in this
study. CFHTLenS data processing was made possible thanks to
significant computing support from the NSERC Research Tools and
Instruments grant program, and to HPC specialist Ovidiu Toader.  The
early stages of the CFHTLenS project were made possible thanks to the
support of the European Commission’s Marie Curie Research Training
Network DUEL (MRTN-CT-2006-036133) which directly supported members of
the CFHTLenS team (C. Bonnett, L. Fu, H. Hoekstra, B.T.P. Rowe,
P. Simon, M. Velander) between 2007 and 2011 in addition to providing
travel support and expenses for team meetings.

T. Erben is supported by the Deutsche Forschungsgemeinschaft through
project ER 327/3-1 and, with P. Simon and P. Schneider, by the
Transregional Collaborative Research Centre TR33 - "The Dark
Universe". C. Heymans, H. Hoekstra \& B.T.P. Rowe acknowledge support from
the European Research Council under the EC FP7 grant numbers 240185
(CH), 279396 (H. Hoekstra+ES) \& 240672 (BR).  L. van Waerbeke
acknowledges support from the Natural Sciences and Engineering
Research Council of Canada (NSERC) and the Canadian Institute for
Advanced Research (CIfAR, Cosmology and Gravity
program). H. Hildebrandt is supported by the Marie Curie IOF 252760, a
CITA National Fellowship, and the DFG grant Hi 1495/2-1. H. Hoekstra
and E. Semboloni also acknowledge support from Marie Curie IRG grant
230924 and the Netherlands Organisation for Scientiﬁc Research grant
number 639.042.814.  T.D. Kitching acknowledges support from a Royal
Society University Research Fellowship.  Y. Mellier acknowledges
support from CNRS/INSU (Institut National des Sciences de l'Univers)
and the Programme National Galaxies et Cosmologie (PNCG).  L. Fu
acknowledges support from NSFC grants 11103012 and 10878003,
Innovation Program 12ZZ134 and Chen Guang project 10CG46 of SMEC, and
STCSM grant 11290706600 \& Pujiang Program 12PJ1406700.  M.J. Hudson
acknowledges support from the Natural Sciences and Engineering
Research Council of Canada (NSERC).  T. Schrabback acknowledges
support from NSF through grant AST-0444059-001, SAO through grant
GO0-11147A, and NWO. M. Velander acknowledges support from the
Netherlands Organization for Scientific Research (NWO) and from the
Beecroft Institute for Particle Astrophysics and Cosmology. C. Bonnett
is supported by the Spanish Science Ministry AYA2009-13936
Consolider-Ingenio CSD2007-00060, project 2009SGR1398 from Generalitat
de Catalunya and by the European Commission’s Marie Curie Initial
Training Network CosmoComp (PITN-GA-2009-238356).

{\small Author Contributions: All authors contributed to the
  development and writing of this paper.  The authorship list reflects
  the lead authors of this paper (P. Simon, T. Erben, and
  P. Schneider) followed by two alphabetical groups.  The first
  alphabetical group includes key contributers to the science analysis
  and interpretation in this paper, the founding core team and those
  whose long-term significant effort produced the final CFHTLenS data
  product.  The second group covers members of the CFHTLenS team who
  made a significant contribution to either the project, this paper or
  both.  C. Heymans and L. van Waerbeke co-led the CFHTLenS
  collaboration. }

\bibliographystyle{mn2e}
\bibliography{g3l.cfhtlens}

\appendix

\section{Multiplicative shear bias}
\label{sect:mbias}

\citet{METAL2012} discusses a calibration scheme for correlation
function estimators involving shear estimates from the \lensfit
pipeline. For details, we refer the reader to the mentioned article,
Sect. 8.3 and 8.4. Analogous to the calibration scheme of the
two-point shear-shear correlation function detailed therein, we divide
$\tilde{\cal G}^{\rm est}$, Eq. \Ref{eq:gestimator}, and
$\tilde{G}_\pm^{\rm est}$, Eq. \Ref{eq:gpmestimator}, by $1+K_{\cal
  G}(\vartheta_1,\vartheta_2,\phi_3)$ and
$1+K_{G_\pm}(\vartheta_1,\vartheta_2,\phi_3)$, respectively. Both
calibration factors are given by
\begin{eqnarray}
  \lefteqn{1+K_{\cal
    G}(\vartheta_1,\vartheta_2,\phi_3)}\\
\nonumber&&
=\frac{
  \sum\limits_{i=1}^{N_{\rm d}}
  \sum\limits_{j=1}^{N_{\rm d}}
  \sum\limits_{k=1}^{N_{\rm s}}
  w_k(1+m_k)\Delta_{ijk}^{\vartheta_1\vartheta_2\phi_3}}
{
  \sum\limits_{i=1}^{N_{\rm d}}
  \sum\limits_{j=1}^{N_{\rm d}}
  \sum\limits_{k=1}^{N_{\rm s}}
  w_k\Delta_{ijk}^{\vartheta_1\vartheta_2\phi_3}}\;,\\
  \nonumber
 \lefteqn{1+K_{G_\pm}(\vartheta_1,\vartheta_2,\phi_3)}\\
 \nonumber&&
 =\frac{
    \sum\limits_{i=1}^{N_{\rm d}}
    \sum\limits_{j=1}^{N_{\rm s}}
    \sum\limits_{k=1}^{N_{\rm s}}
    w_jw_k(1+m_j)(1+m_k)\Delta_{ijk}^{\vartheta_1\vartheta_2\phi_3}
  }{
    \sum\limits_{i=1}^{N_{\rm d}}
    \sum\limits_{j=1}^{N_{\rm s}}
    \sum\limits_{k=1}^{N_{\rm s}}
    w_jw_k\Delta_{ijk}^{\vartheta_1\vartheta_2\phi_3}}\;.
\end{eqnarray}
The multiplicative bias factors $m_i$, provided in the CFHTLenS
catalogue\footnote{Publicly available under
  \url{http://www.cadc-ccda.hia-iha.nrc-cnrc.gc.ca/community/CFHTLens/query.html}}
for each source, are functions of the source signal-to-noise and
angular size. Note that for non-vanishing values of \mbox{$1+m_i>0$},
the calibration is mathematically equivalent to employing the
transformation $\epsilon_i\mapsto\epsilon_i/(1+m_i)$ and $w_i\mapsto
w_i(1+m_i)$ in the estimators $\widetilde{\cal G}^{\rm est}$ and
$\widetilde{G}^{\rm est}_\pm$.

\section{Lens samples supplement}

\subsection{Angular clustering of lenses}
\label{sect:clustering}

The angular correlation function $\omega(\theta)$ of the lenses as a
function of separation $\theta$ is approximated by a simple power law
\citep{peebles80}:
\begin{equation}
  \label{eq:powerlaw}
  \omega(\theta)=
  A_\omega\,\left(\frac{\theta}{1^\prime}\right)^{-\lambda}+{\rm IC}\;,
\end{equation}
where $A_\omega$ is the clustering strength at a separation of one
arcmin, $\lambda$ the slope of the correlation function and the
constant offset IC, the integral constraint \citep{gp77}. We find this
fitting function to be a good description of $\omega(\theta)$ between
$0^\prime\!\!.2\lesssim\theta\lesssim10^\prime$. The estimator of
$\omega(\theta)$ by \citet{las93}, employed for this paper, requires
the repeated counting of galaxy pairs in separation bins for random
realisations of unclustered galaxy distributions, factoring in the
incompleteness of the survey. The figures quoted in Table
\ref{tab:samples} consider the masks of individual survey pointings,
but presuming the same survey completeness within regions that are not
masked out. For the final $\omega(\theta)$, all pair counts from all
individual survey fields are added so that, in effect, fields with
more galaxies attain a higher weight in the average. The binned
$\omega(\theta)$ is stored as data vector $\overline{\vec{d}}$.

\textbf{Pointing-to-pointing Jackknife sampling} To estimate the
statistical uncertainty of $\overline{\vec{d}}$, we prepare a set of
$N_{\rm p}$ Jackknife samples $\bar{\vec{d}}^{\rm jn}_i$, where
$\bar{\vec{d}}_i$ is the combined data vector omitting the $i$th
patch. The Jackknife covariance of the sample mean is then
\citep{1986ANNSTAT,2009MNRAS.396...19N}:
\begin{equation}
  \label{eq:jackknife}
  \mat{C}=
  \frac{1}{N_{\rm p}}
  \sum_{i=1}^{N_{\rm p}}
  \Delta\bar{\vec{d}}_i^{\rm jn}\left[\Delta\bar{\vec{d}}_i^{\rm
      jn}\right]^{\rm t}\;,
\end{equation}
where
\begin{equation}
  \Delta\bar{\vec{d}}_i^{\rm jn}:=
  (N_{\rm p}-1)(\bar{\vec{d}}-\bar{\vec{d}}_i^{\rm jn})\;.
\end{equation}
For Table \ref{tab:samples}, a power law fit is applied to the
ensemble average $\overline{\vec{d}}$ of all pointings, taking into
account the Jackknife covariance $\mat{C}$. The $\theta$-binning in
$\overline{\vec{d}}$ is also applied to the power law model,
Eq. \Ref{eq:powerlaw}, by averaging the model over the width of each
bin. Note that the inverse of $\mat{C}$, required for the likelihood
analysis of the model fit parameters, has to be corrected to obtain an
unbiased estimator of the inverse covariance
\citep{2007A&A...464..399H}. Similar corrections of inverse
covariances are applied throughout the paper without further
mentioning.

\subsection{Completeness of lens sample}
\label{sect:completeness}

Here we define a parameter that quantifies the completeness of our
lens samples. First we define the comoving volume $V(z_1,z_2)$
confined by the redshift boundaries \mbox{$z_1\le z<z_2$},
\begin{equation}
  V(z_1,z_2)=
  \Omega\int_{\chi(z_0)}^{\chi(z_1)}\d\chi\,f_{\rm K}^2(\chi)\;,
\end{equation}
where $\Omega$ is the solid angle of the patch field-of-view minus the
solid angle of mask regions, and
\begin{equation}
  \chi(z)=D_{\rm H}\int_0^z\frac{\d z^\prime}{E(z^\prime)}\;,
\end{equation}
where $E(z):=H(z)/H_0$ is the Hubble parameter $H(z)$ as a function of
redshift normalised to $H_0$.

Due to the incompleteness in our flux-limited survey, a galaxy is only
visible up to a certain redshift $z_{\rm max}$. In general, and
especially for faint galaxies, one can expect the limit $z_{\rm max}$
to be a complicated function of intrinsic galaxy properties and
position, survey instrumentation, survey conditions and the data
reduction pipeline. Nevertheless, here we take the simplistic view
that the main factor is the apparent $i'$-band magnitude of the lens
(extinction corrected), which is limited to $i'\le22.5$, such that our
lens samples are predominantly magnitude limited. We further assume
that a $K$-correction is negligible over the redshift bin $[z_1,z_2]$
of interest. Under these circumstances, one finds implicitly for
$z_{\rm max}$:
\begin{equation}
  D_{\rm L}(z_{\rm max})=10^{0.4(m_{\rm limit}-m)}
  D_{\rm L}(z)\;,
\end{equation}
where $z$ is the redshift of the galaxy, $m$ its $i'$-band magnitude,
and $m_{\rm limit}=22.5$ the asserted magnitude limit of the lens
catalogue. By $D_{\rm L}(z)=(1+z)f_{\rm K}\!\left(\chi(z)\right)$ we
denote the comoving luminosity distance as a function of redshift.

In order to quantify for Table \ref{tab:samples} the completeness of a
sample of $N_{\rm g}$ galaxies, we estimate over which fraction
$f_{\rm c}=V(z_1,z_{\rm max})/V(z_1,z_2)$ an observed galaxy in the
sample would be observable. We take the average of all volume
fractions of all lenses in a sample, 
\begin{equation}
  f_c=
  \frac{1}{N_{\rm g}}
  \sum_{i=1}^{N_{\rm g}}\int_{z_1}^{z_2}\d z\,p_i(z)
  \frac{V\big(z_1,{\rm MIN}(z_{{\rm max},i};z_2)\big)}{V(z_1,z_2)}\;,
\end{equation}
and marginalise over the uncertainties in the galaxy redshifts,
quantified by the p.d.f. $p_i(z)\d z$.  Importantly, $z_{{\rm max},i}$
denotes the maximum redshift at which the $i$th galaxy would still be
included within the galaxy catalogue, complying with all survey and
sample selection criteria.  A completeness parameter close to unity
means that essentially all galaxies in the sample are visible
throughout the entire volume, whereas $f_{\rm c}\ll1$ indicates a
significant fraction of galaxies that is only visible in a small
subvolume at lower redshift. Obviously, $f_{\rm c}$ is merely an
estimator (upper limit) for the sample completeness as galaxies
already not observed at redshift $z_1$ are not accounted for. Note
that the solid angle $\Omega$ cancels inside the expression for
$f_{\rm c}$ and is hence not needed.

\section{Systematics}
\label{sect:systematics}

\subsection{B/P-mode consistency with null}

\begin{figure}
  \begin{center}
    \psfig{file=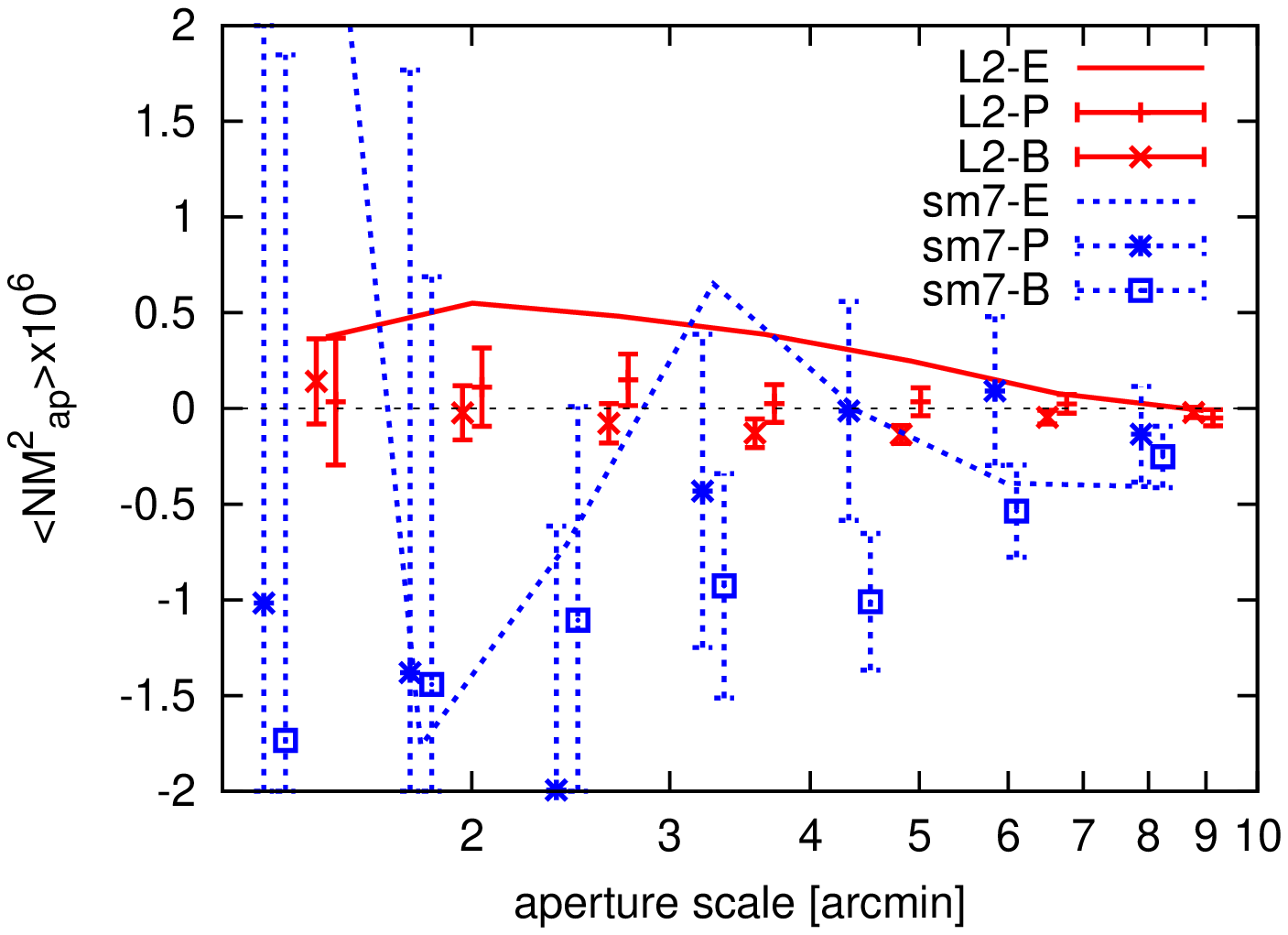,width=61mm,angle=-90}
  \end{center}
  \caption{\label{fig:systematics} Plots of $\ave{{\cal N}M_\perp
      M_{\rm ap}}$, $\ave{{\cal N}M^2_\perp}$ (L1 low-$z$ and sm7
    high-$z$) in comparison to the E-mode (lines). Both samples failed
    the 95\% confidence level null test for the P/B-modes, see Table
    \ref{tab:nulltest}.}
\end{figure}

As indicator of possible systematics in the estimators, we test
$\ave{{\cal N}^2M_\perp}$ and the combined $\ave{{\cal N}M_\perp
  M_{\rm ap}}$, $\ave{{\cal N}M^2_\perp}$ against the null
hypothesis. A null measurement should result in a
\begin{equation}
  \Delta\chi^2=\vec{d}^{\rm t}\mat{C}^{-1}\vec{d}\;,
\end{equation}
that is statistically consistent with a vanishing signal, with
$\vec{d}$ being a vector consisting of the measurements for the P-mode
($\ave{{\cal N}^2M_{\rm ap}}$) or both the P- and B-mode ($\ave{{\cal
    N}M^2_{\rm ap}}$). By $\mat{C}$ we denote the Jackknife covariance
of the measurements as obtained from the variance of B/P-mode
measurements in the pointings, as explained in
Sect. \ref{sect:clustering}. This covariance is larger than a null
hypothesis covariance as it possibly also contains power from
B/P-modes present in the data. A true null model would contain only
power from galaxy shape noise and sampling noise. The test results can
be found in Table \ref{tab:nulltest}. Measurements inconsistent with a
null signal ($95\%$ confidence) are underlined, thus for
$\Delta\chi^2\ge2.0$ ($1.68$) per degree of freedom for $\ave{{\cal
    N}^2M_{\rm ap}}$ ($\ave{{\cal N}M_{\rm ap}^2}$). In total, we find
two lens samples that fail the test; they are plotted in comparison to
their E-mode in Fig. \ref{fig:systematics}. In both cases the failures
are related to the $\ave{{\cal N}M_{\rm ap}^2}$ statistics and
significantly negative B-modes. Note that errors between neighbouring
bins are strongly correlated.

Finding two measurements out of in total 57 that fail the $95\%$ 
test is what we would expect as false positive rate. We therefore
conclude that the influence of systematics on the E-mode measurement
that reveal themselves via the P- or B-modes is likely to be small
compared to our measurement uncertainties. Note that the sm7 sample
that failed the null test is not used in the final analysis because
the corresponding E-mode signal is consistent with zero.

\begin{figure*}
  \begin{center}
    \psfig{file=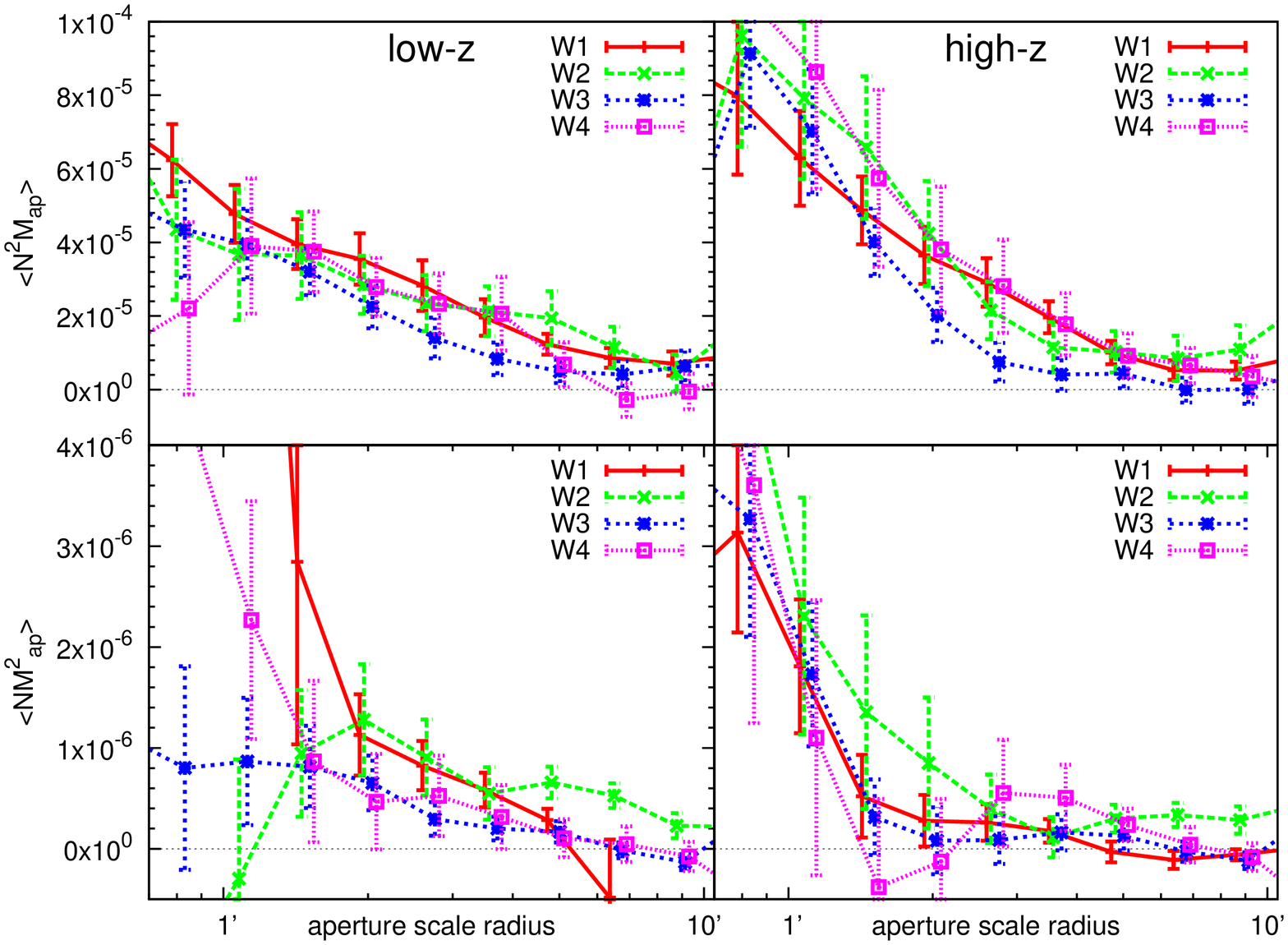,width=120mm,angle=-90}
  \end{center}
  \caption{\label{fig:mapstatW14} Measurements of the aperture
    statistics (top row: $\ave{{\cal N}^2M_{\rm ap}}$, bottom row:
    $\ave{{\cal N}M^2_{\rm ap}}$) for the combined L1-L6 sample
    (low-$z$ and high-$z$ redshift bin separately). The measurements
    are split for the four fields W1-W4. The error bars indicate only
    the pointing-to-pointing variance within the same field. The
    fields vary in size and thus the sizes of lens and source
    catalogues vary. Lines connect the data points to guide the eye.}
\end{figure*}

\subsection{Field dependence of the G3L signal}

The CFHTLS wide survey consists of four contiguous fields W1
($\sim72\,\rm deg^2$), W2 ($\sim33\,\rm deg^2$), W3 ($\sim49\,\rm
deg^2$) and W4 ($\sim25\,\rm deg^2$); the field areas do not include
masking or excluded fields due to significant PSF residuals. The
fields are well separated on the sky and were observed at different
times of the year. By splitting the measurements into W1-W4, we check
whether the G3L measurements are comparable for different subsets of
the data. To have a possibly large sample for this test, we combine
the signals of the samples L1 to L6 for each field, see
Fig. \ref{fig:mapstatW14}. Only measurements from pointings within the
same fields W1-W4 are combined, their statistical uncertainties
originate from the Jackknife technique (as in Eq.
\ref{eq:jackknife}). Therefore, the error bars do not include the
cosmic variance between the fields, which should be most prominent at
the larger angular scales.  

We find excellent agreement between the measurements, considering that
statistical uncertainties at larger scales are higher than indicated
and that errors between neighbouring angular bins are correlated. In
particular, this separation of data shows that the G3L signal does not
originate from singular fields that are extreme outliers compared to
the others. Since the uncertainties of the final combined measurements
are based on the pointing-to-pointing variance of the entire survey, a
possible systematic deviation of one field will be included as
systematic error inside the error bars.

\section{Third-order galaxy biasing}
\label{sect:biassmooth}

The values ${\cal B}_{\rm ggm}$ (${\cal B}_{\rm mmg}$) measure the
$u$-filtered bispectrum $\overline{B}_{\rm ggm}$ ($\overline{B}_{\rm
  mmg}$), radially smoothed with maximum weight at $\chi_{\rm
  max}$. The maximum weight of the $u$-filter in Fourier space is at
$k=\sqrt{2}/R$ for a given real space scale $R$.  From the definition
of the $\Psi$-statistics, Eq. \Ref{eq:psi}, from Eq. \Ref{eq:calcalc}
and from a similar equation for $\ave{{\cal N}M^2_{\rm ap}}$ it
follows that
\begin{eqnarray}
  \lefteqn{\Psi(R)}\\
  &&\nonumber
  \!\!\!\!\!\!\!=\frac{\int\d\chi\d^2k_1\d^2k_2\,F_{\rm
      mmg}(\vec{k}_1,\vec{k}_2,\chi;R)\,
    \left(b_3r_1\right)(\vec{k}_1,\vec{k}_2,\chi)}
  {\int\d\chi\d^2k_1\d^2k_2\,F_{\rm
      ggm}(\vec{k}_1,\vec{k}_2,\chi;R)\,
    \left(b^2_3r_2\right)(\vec{k}_1,\vec{k}_2,\chi)}\;,
\end{eqnarray}
where the smoothing kernels in $(\vec{k},\chi)$-space are
\begin{eqnarray}
  \label{eq:fkernels}
  \lefteqn{F_{\rm mmg}(\vec{k}_1,\vec{k}_2,\chi;R)}\\
  &&\nonumber
  \!\!\!\!:=\frac{q_{\rm mmg}(\chi)}{\int_0^{\chi_{\rm h}}\d\chi\,q_{\rm
        mmg}(\chi)}\times\\
  &&\nonumber
  ~\tilde{u}(k_1\Lambda)\tilde{u}(k_2\Lambda)\tilde{u}(|\vec{k}_1+\vec{k}_2|\Lambda)
  B_{\rm mmm}(\vec{k}_1,\vec{k}_2,\chi)\;,\\
  \nonumber
 \lefteqn{F_{\rm ggm}(\vec{k}_1,\vec{k}_2,\chi;R)}\\
  &&\nonumber
  \!\!\!\!:=\frac{q_{\rm ggm}(\chi)}{\int_0^{\chi_{\rm h}}\d\chi\,q_{\rm
        ggm}(\chi)}\times\\
  &&\nonumber
  ~\tilde{u}(k_1\Lambda)\tilde{u}(k_2\Lambda)\tilde{u}(|\vec{k}_1+\vec{k}_2|\Lambda)
  B_{\rm mmm}(\vec{k}_1,\vec{k}_2,\chi)
\end{eqnarray}
with \mbox{$\Lambda:=Rf_{\rm K}(\chi)/f_{\rm K}(\chi_{\rm max})$}. As
can be seen, the detailed weight within a band defined by the width of
the $u$-filter is also determined by the actual matter bispectrum
$B_{\rm mmm}$.

We can further exploit the statistical isotropy of the galaxy-matter
bispectra, which means that both $B_{\rm mmm}$ and the galaxy biasing
parameters $r_1,r_2,b_3$ are only functions of
$|\vec{k}_1|,|\vec{k}_2|,\phi$; $\phi$ is the angle spanned by
$\vec{k}_1$ and $\vec{k}_2$. The previous expressions therefore
simplify to
\begin{eqnarray}
  \lefteqn{\Psi(R)}\\
  &&\nonumber
  \!\!\!\!\!\!\!=\frac{\int\d\chi\d\phi\,\d k_1\d k_2k_1k_2\,F_{\rm
      mmg}(\ldots)\,
    \left(b_3r_1\right)(k_1,k_2,\phi,\chi)}
  {\int\d\chi\d\phi\,\d k_1\d k_2k_1k_2\,F_{\rm
      ggm}(\ldots)\,
    \left(b^2_3r_2\right)(k_1,k_2,\phi,\chi)}\;,
\end{eqnarray}
where
\begin{eqnarray}
  \lefteqn{F_{\rm mmg}(\ldots)}\\
  &&\nonumber
  \!\!\!\!:=\frac{q_{\rm mmg}(\chi)}{\int_0^{\chi_{\rm h}}\d\chi\,q_{\rm
        mmg}(\chi)}\times\\
  &&\nonumber
  ~\tilde{u}(k_1\Lambda)\tilde{u}(k_2\Lambda)\tilde{u}(|\vec{k}_1+\vec{k}_2|\Lambda)
  B_{\rm mmm}(k_1,k_2,\phi,\chi)\;,\\
  \nonumber
 \lefteqn{F_{\rm ggm}(\ldots)}\\
  &&\nonumber
  \!\!\!\!:=\frac{q_{\rm ggm}(\chi)}{\int_0^{\chi_{\rm h}}\d\chi\,q_{\rm
        ggm}(\chi)}\times\\
  &&\nonumber
  ~\tilde{u}(k_1\Lambda)\tilde{u}(k_2\Lambda)\tilde{u}(|\vec{k}_1+\vec{k}_2|\Lambda)
  B_{\rm mmm}(k_1,k_2,\phi,\chi)\;,
\end{eqnarray}
and
\begin{equation}
  |\vec{k}_1+\vec{k}_2|=
  \sqrt{k_1^2+k_2^2+2k_1k_2\cos{\phi}}\;.
\end{equation}
Note that for equilateral triangles we have
$k_1=k_2=|\vec{k}_1+\vec{k}_2|$ and thus $\cos{\phi}=-1/2$.

\label{lastpage}

\end{document}